\documentclass[journal,10pt,onecolumn,draftclsnofoot,]{IEEEtran}
\usepackage{graphicx}
\usepackage{float}
\usepackage{mathtools}
\usepackage{amsmath}
\usepackage{amsthm}
\usepackage{amssymb}
\usepackage{graphicx}
\usepackage{tikz,pgfplots}  
\usepackage{booktabs}
\usepackage{caption}
\usepackage{cuted}
\usepackage{verbatim}
\usepackage{xcolor}
\usepackage{mathtools}
\usepackage{dsfont}
\usepackage{mathrsfs}
\usepackage{bbm}
\usepackage{enumitem}
\usepackage{mathabx}
\pgfplotsset{compat=1.15}
\usetikzlibrary{arrows}
\usetikzlibrary{decorations.pathreplacing,calligraphy}
\usetikzlibrary{shapes.geometric}
\usetikzlibrary{graphs}
\usepackage{tikz-qtree}
\usepackage{filecontents}
\usepackage{csvsimple}
\usepackage{enumitem}
\usetikzlibrary{arrows.meta}
\usetikzlibrary{pgfplots.groupplots}
\usepackage[normalem]{ulem}
\usepackage{url}
\usepackage{algorithm}
\usepackage{algpseudocode}
\usepackage{cite} 
\usepackage{framed}
\usepackage{bm}
\usepackage{booktabs}
\usepackage{array}
\usepackage{longtable}
\usepackage{titlefoot}
\usepackage{cancel}

\date{}

\newtheorem{definition}{Definition}
\newtheorem{lemma}{Lemma}
\newtheorem{corollary}{Corollary}

\newtheorem{proposition}{Proposition}

\newtheorem{remark}{Remark}
\newtheorem{theorem}{Theorem}

\newcommand{\bupcnq}{{$\Pi_{n}^q$ }}

\newcommand{\supp}[1]{\textnormal{Supp(}#1\textnormal{)}}

\newcommand{\bikdjournal}[1]{}

\def\dc{d_{\textnormal{c}}}

\newcommand{\inpapproxcell}[1]{\hat{Q}_{\text{ip},#1}}
\newcommand{\outapproxcell}[1]{\hat{Q}_{\text{op},#1}}

\newcommand{\compositionclass}[1]{\mathscr{C}_{#1}}
\newcommand{\seqofsame}[1]{\cT_{#1}}

\newcommand{\setofweights}[2]{\cN_{#1,#2}}

\def\intdist{\hat{Q}_{\text{ip}}}

\def\cA{\mathcal{A}}
\def\cB{\mathcal{B}}
\def\cP{\mathcal{P}}
\def\cD{\mathcal{D}}
\def\cT{\mathcal{T}}
\def\cU{\mathcal{U}}

\def\cZ{\mathcal{Z}}

\def\cN{\mathcal{N}}

\def\cW{\mathcal{W}}

\def\vl{\boldsymbol\ell}

\def\vu{\mathbf{u}}

\def\vx{\mathbf{x}}

\def\vt{\mathbf{t}}
\def\vy{\mathbf{y}}
\def\vt{\mathbf{t}}
\def\vv{\mathbf{v}}
\def\vz{\mathbf{z}}
\def\vk{\mathbf{k}}

\def\vw{\mathbf{w}}

\def \cW{\mathcal{W}}

\def\vT{\mathbf{T}}
\def\vW{\mathbf{W}}
\def\qed{\hfill{$\blacksquare$}}

\def\tvtu{\tilde{\vt}_{\cellid}}

\def\vX{\mathbf{X}}
\def\vY{\mathbf{Y}}

\def\inpdist{Q_{\text{ip}}}
\def\opdist{Q_{\text{op}}}

\def\inpdistj{Q_{\text{ip},j}^{(i)}}

\def\oupdistj{Q_{\text{op},j}^{(i)}}
\def\oupdistk{Q_{\text{op},k}^{(i)}}

\def\outapproxj{\hat{Q}_{\text{op},j}^{(i)}}
\def\outapproxk{\hat{Q}_{\text{op},k}^{(i)}}

\def\inpdistapproxj{\hat{Q}_{\text{ip},j}^{(i)}}
\def\inpdistapproxk{\hat{Q}_{\text{ip},k}^{(i)}}

\def\indicator{\mathbbm{1}}

\def\outdist{Q_{\text{op}}}
\def\inpapprox{\hat{Q}_{\text{ip}}}
\def\outapprox{\hat{Q}_{\text{op}}}

\def\binpt{\text{B}_{(n,p)}(t)}

\def\multnv{\multinomial{n}{\vv}}

\newcommand{\multinomial}[2]{\text{M}_{(#1,#2)}}

\def\tv{d_{\text{TV}}}

\def\bnpmax{\overline{\text{B}}_{(n,p)}}

\def\upcn{\bupcnq}

\def\tvatom{\delta}

\def\bupcnq{\Pi^q_{n}}

\def\nupcn{\Sigma_{n,U}}

\def\nupcnc{\Sigma_{U}}

\def\dmc{U}
\def\dmcn{U^{(n)}}

\def\qww{q\text{-NCC}_{n,U}}
\def\qwwu{q\text{-NCC}_{U}}
\newcommand{\compositionof}[1]{\underbar{N}(#1)}

\def\dmcmatrix{U}
\def \kl{\cD_{\text{KL}}}

\def\Pperm{P_{\Pi}}
\def\Psigma{P_{\Sigma}}

\def\Pbww{P_{\text{BWW}}}
\def\Pqww{P_{q\text{-NC}}}

\def\weightqn{\cN_{q,n}}

\def\ve{\mathbf{e}}

\def\vZ{\mathbf{Z}}

\def\qwwi{q\text{-NCC}_{n_i,U}}

\def\tvatom{\alpha}

\def\Ssum{|\splus{\vu}{\vv}|+|\sminus{\vu}{\vv}|}

\newcommand{\splus}[2]{S_{(>)}\left(#1,#2\right)}
\newcommand{\sminus}[2]{S_{(<)}\left(#1,#2\right)}

\def\atomno{\nu(\vu,\vv)}
\def\atomnoi{\nu(\vu',\vv)}
\def\atomnoab{\nu(\vu_a,\vu_b)}

\def\cellid{\bm{l}}

\def\msgsize{L}

\begin{document}

\title{Identification Over Noisy Permutation Channels}
\setcounter{page}{1}
\author{
  \IEEEauthorblockN{Abhishek~Sarkar and Bikash~Kumar~Dey}\\
 \IEEEauthorblockA{Department of Electrical Engineering\\
                    Indian Institute of Technology Bombay\\
                    \{absarkar, bikash\}@ee.iitb.ac.in}
}
 \maketitle
 \pagenumbering{arabic}

\thispagestyle{empty}

\begin{abstract}

We study message identification over the noisy permutation channel. For discrete memoryless channels (DMCs), the number of identifiable messages grows doubly exponentially, and the maximum second-order exponent is same as the Shannon capacity of the DMC. We consider a $q$-ary noisy permutation channel where the transmitted vector is first permuted by a permutation chosen uniformly at random, and then passed through a DMC with strictly positive entries in its transition probability matrix $\dmcmatrix$. In an earlier work, we showed that over $q$-ary noiseless permutation channel, $2^{c_n n^{q-1}}$ messages can be identified if $c_n\rightarrow 0$, and a strong converse holds for $2^{c_n n^{q-1}}$ messages if $c_n\rightarrow \infty$. For the $q$-ary noisy permutation channel, we show that message sizes growing as $2^{R_n \left( \frac{n}{\log n}\right)^{(r-1)/2}}$, where $r$ be the rank of $\dmcmatrix$, are identifiable for any $R_n\rightarrow 0$. We also prove a strong converse result showing that for any sequence of identification codes with  
$$2^{\left(R_n n^{(q-1)/2}(\log n)^{1+\frac{(q-1)(q-2)}{2}}\right)},$$
messages, where $R_n \rightarrow \infty$, the sum of Type-I and Type-II error probabilities approaches at least $1$ as $n\rightarrow \infty$. Our proof of the strong converse uses the idea of channel resolvability. The permutation channel only preserves the composition vector or composition of the transmitted vector. Hence the effect of the noisy permutation channel is captured in the way the composition of the transmitted vector is changed, when passed through the DMC. This is represented by a new channel -- ``$q$-ary noisy composition channel ($\qww$)''. We propose a novel deterministic quantization scheme for quantization of a distribution over the set of all compositions by an $M$-type input distribution when the distortion is measured on the output distribution (over the $\qww$) in total variation distance. This plays a key role in the converse proof. We also study identification with deterministic encoder and decoder over, and proved tight achievability, weak converse, and strong converse.
\end{abstract}

\section{Introduction}\label{sec:intro}

\unmarkedfntext{Part of this work is accepted for presentation in ISIT 2025.}

The problem of message identification was introduced by JaJa~\cite{ja1985identification}, and  Ahlswede and Dueck~\cite{ahlswede1989identification, ahlswede1989identification_f}. In message identification over a channel, a decoder needs to determine whether a particular message of interest was transmitted. The decoder outputs an ``Accept'' or ``Reject'' depending on its decision. There are two types of decoding errors. Due to channel noise, a decoder may ``Reject'' a message, even when that very message was encoded. This error event is termed as a Type-I error. On the other hand, a decoder may ``Accept'' even when a different message was encoded. This is known as a Type-II error.
Ahlswede and Dueck \cite{ahlswede1989identification} demonstrated that with stochastic encoding, a doubly exponential number of messages can be identified over discrete memoryless channels (DMCs). The identification capacity of a channel, defined as the maximum second-order exponent of the number of identifiable messages, was shown to equal the Shannon capacity of the channel \cite{ahlswede1989identification, han1992new, han1993approximation, steinberg1998new, watanabe2021minimax,koga2002information}. Identification has since been extensively studied under various channel models \cite{ahlswede1995new, spandri2023information, rosenberger2023capacity, bracher2017identification, rosenberger2023identification, ahlswede1991identification, salariseddigh2021deterministic, labidi2021identification, ahlswede1999identification} and under various input constraints \cite{wiese2022identification, salariseddigh2021deterministic_it, salariseddigh2023deterministic}. 

A Permutation channel permutes/reorders the components of the transmitted vector. A noisy permutation channel~\cite{makur2020coding} additionally introduces noise to each component by passing through a DMC $\dmcmatrix$. Permutation channels are relevant in DNA-based storage and multipath routing in communication networks \cite{goldreich2012theory, gadouleau2010binary, walsh2009optimal, kovavcevic2015perfect, kovavcevic2018codes, akan2016fundamentals, pierobon2014fundamentals, Sima2023ErrorCF, sima2023robust, shomorony2022information, langberg2017coding,chee2016string, makur2020coding,tang2023capacity, tandon2019bee,tandon2020bee}. 

Identification over noiseless uniform permutation channel was studied in  \cite{sarkar2024identification, sarkar2024identificationbinary}. It was shown that for $q$-ary noiseless permutation channel, message sizes growing as $2^{\epsilon_nn^{q-1}}$ are identifiable under stochastic encoding for any $\epsilon_n\rightarrow 0$. A soft converse was proved -- for any $R>0$, there is no sequence of identification codes with message size growing as $2^{Rn^{q-1}}$ with a power-law decay ($n^{-\mu}$) of the error probability.  A strong converse was also proved, which showed that for any sequence of identification codes with message size $2^{R_n n^{q-1}}$, where $R_n \rightarrow \infty$, the sum of Type-I and Type-II error probabilities approaches at least $1$ as $n\rightarrow \infty$.
It was shown that under deterministic encoding, any message size up to $\binom{n+q-1}{q-1}$ is identifiable, and a strong converse holds above this message size. When block-wise and noiseless feedback is available, one can identify up to $2^{q^n}$ number of messages, and a strong converse was proved above this size.

In this paper, we consider a noisy permutation channel $\nupcn$. This channel consists of an $n$-block $q$-ary uniform permutation channel followed by a $q$-ary discrete memoryless channel (DMC) having a transition probability matrix $\dmcmatrix$.  The message is encoded into a $q$-ary vector $\vx$ which is first permuted by a permutation chosen uniformly at random from $S_n$, the set of all permutations of $\{1,\ldots,n\}$.
Following this, the components of the permuted vector pass through the DMC $\dmcmatrix$ to produce the final output vector.  Since permutation channels support reliable communication of a set of messages of size $N=n^R$ for some $R$, the communication rate is defined as $\frac{\log N}{\log n} $. It was proved that
for a noisy permutation channel consisting of a DMC $\dmcmatrix$ with strictly positive entries, the capacity equals $\frac{r-1}{2}$, where $r$ is the rank of  $\dmcmatrix$ \cite{makur2020coding, tang2023capacity}.

We prove that over a $q$-ary noisy permutation channel, $\msgsize=2^{\epsilon_n \left(\frac{n}{\log n}\right)^{(r-1)/2}}$ messages can be identified if $\epsilon_n \rightarrow 0$. We also prove a strong converse result when $U$ has strictly positive entries. We show that if the number of messages grows as  $$L=2^{\left(R_n n^{(q-1)/2}(\log n)^{1+\frac{(q-1)(q-2)}{2}}\right)},$$
for any $R_n \rightarrow \infty$, then the sum of the Type-I and Type-II error probabilities approaches $1$. Defining the identification rate as $\frac{\log\log L}{\log n}$, we see that the strong converse and achievability meet in terms of rate, if $U$ has full rank (i.e. $r=q$) and has strictly positive entries. The proof of the strong converse is based on ideas of channel resolvability \cite{han1992new,han1993approximation,koga2002information}. It is first noted that, due to the constituent permutation channel, the noisy permutation channel only passes some information on the composition/type of the transmitted vector. We relate the problem of identification over noisy permutation channel to that over a related channel, that we call the $q$-ary noisy composition channel ($\qww$) (see Lemma~\ref{lemma: noisy permutation to noisy type}). This new channel takes a composition as input and also outputs a composition. It captures the change in the composition of a vector when it passes through the DMC $\dmcmatrix$. Our proof of the strong converse crucially depends on the approximation of input distributions for the $\qww$ by $M$-type distributions so that the approximation causes small distortion (in TV distance) in the output distribution. We propose a deterministic quantization scheme for such input distributions that uses the structure of the channel.

We also study message identification with deterministic encoder and decoders. We show that $L=\left( \frac{n}{c\log n} \right)^{(r-1)/2}$ messages can be identifiable over $q$-ary noisy permutation channel, where $r$ is the rank of the constituent DMC $U$. We prove two converse results for deterministic identification over $q$-ary noisy permutation channels when $U$ has strictly positive entries. Our weak converse shows that there exists a $R'>0$ such that for all $R>R'$, and any sequence of ID codes with deterministic encoders and decoders and $L=Rn^{(q-1)/2}(\log n)^{(q-1)(q-2)/2}$ messages, the sum of Type-I and Type-II error probabilities is bounded away from $0$. Our strong converse results shows that for any sequence $R_n \rightarrow \infty$, and any sequence of ID codes with deterministic encoders and decoders and  $L= R_n n^{(q-1)/2}(\log n)^{(q-1)(q-2)/2}$ messages, the sum of Type-I and Type-II error probabilities approaches $1$.

Our quantization scheme for general $q$ has two stages. In the first stage, the input alphabet of the $\qww$, the set of compositions of $n$-length vectors over $[1:q]$, is partitioned into ``cubic'' cells and probability mass is shifted around inside each cell so that the probability mass at all but one point in each cell is a multiple of $1/M$. Hence, in the first stage, the required shifts of probability mass happen over a ``limited'' distance.  We show the existence of a Gray-like ordering of the cells, and this is used next stage. In the second stage, probability mass is shifted as required among the unquantized points in the cells, preferably among neighboring cells according to a Gray-like ordering of the cells.

Approximation or quantization of distributions under divergence metric and variational distance metric have been studied in~\cite{bocherer2016optimal}. More generally, covering of the space of distributions under divergence metric, when the ``centers'' of these ``divergence balls'' are not necessarily $M$-type distributions, has been studied in~\cite{tang2022divergence}. However, we are interested in similar questions of quantization or covering of the input distributions when there is a channel, and the metric is the variational distance between the output distributions over this channel. This is studied in channel resolvability~\cite{han1993approximation}. However, \cite{han1993approximation} used randomized quantization technique for the input distribution, where the quantization scheme does not depend on the channel. This was used in \cite{han1992new,han1993approximation,steinberg1998new,watanabe2021minimax} to prove the strong converse for identification. A direct application of the same approach for permutation channels only gives a strong converse for any positive second order rate. To prove the strong converse for a tighter message size, we use a channel-specific deterministic quantization scheme.

The organization of this paper is as follows.
In Section~\ref{sec:problem_setup}, we introduce notations and definitions related to the problem. The main results are presented in Section~\ref{sec:result}. The proof of  the achievability result (Theorem~\ref{thm:stoch}\ref{thm:ach_stoch}) is given in Section~\ref{sec:proof_ach_thm}. The proof of the strong converse (Theorem~\ref{thm:stoch}\ref{thm:converse_stoch}) is presented in Section~\ref{sec:proof_of_theorem_strong_converse_qary}. Three subsequent sections develop some techniques and results that are used in the proof of the strong converse. In Section~\ref{sec:multinomial}, we derive bounds on the multinomial distribution that are instrumental in proving the strong converse. In Section~\ref{sec:unit_change}, we derive bounds on the change in output distribution due to changes in the input distribution. A deterministic quantization algorithm is presented in Section~\ref{sec:algo}.
The proof of Theorem~\ref{thm:det} (on deterministic ID codes) is presented in Section \ref{sec:proof_of det_ID}.
We conclude in Section~\ref{sec:conclusion}.

\section{Notations and Problem setup}\label{sec:problem_setup}

\subsection{General notations and definitions}

Let $n$ and $q \ge 2$ be positive integers. For any two integers $i\leq j$, we denote $[i:j]:=\{i,i+1, \ldots, j\}$.  The set of all permutations, i.e. bijective maps, of the set $[1:n]$ is denoted by $S_n$.
For any set $\cA$, we denote $\cP(\cA)$ as the set of all distributions over $\cA$.
For any two distributions $P$ and $ Q$ over $\mathbb{Z}^l$, where $l$ is a positive integer, $P \ast Q$ denotes the convolution of $P$ and $Q$. 
For any positive integer $l \in \mathbb{Z}$, let $Q_1$ and $Q_2$ be two distributions over $\mathbb{Z}^l$. The total variation (TV) distance between $Q_1$ and $Q_2$ is defined as
\begin{align*}
    \tv (Q_1,Q_2) :=\frac{1}{2}\sum_{\vt\in \mathbb{Z}^l} |Q_1(\vt)-Q_2(\vt)|.
\end{align*}
For any real number $x$, $\lfloor x\rfloor$ denotes the greatest integer that is smaller than or equal to $x$.
For any real number $x$, and positive integer $M$, we define the following  function:
\begin{align}
\lfloor x \rfloor_{1/M} & := \frac{\lfloor Mx \rfloor}{M}.
\end{align}
This is the greatest multiple of $1/M$ that is smaller than or equal to $x$.
$\indicator\{\cdot\}$ denotes the indicator function, taking the value $1$ when the enclosed condition is satisfied, and $0$ otherwise. 
We define the $q$-length unit vector $\ve_a; a \in [1:q]$ as $\ve_a := \left(\indicator{\{c=a\}}\right)_{c \in [1:q]}$.
For any $\vt_1,\vt_2 \in \mathbb{Z}^q$, we define the functions 
\begin{align*}
\delta_{\vt_1}(\vt) &= \indicator\{\vt=\vt_1\}, \\
\delta_{\vt_1 \rightarrow \vt_2}(\vt) &=\delta_{\vt_2}(\vt)-\delta_{\vt_1}(\vt),
\end{align*}
for all $ \vt \in \mathbb{Z}^q$.

An $n$-length vector over $[1:q]$ is denoted by a bold-faced lowercase letter, e.g. $$\vx=(x_1, \ldots, x_n)=(x_i)_{i \in [1:n]}.$$
For any $n$-length vector $\vx$, we define the support of $\vx$ as $\supp{\vx} \coloneqq \{i \in [1:n]| x_i >0\}$.
For any $\vx\in [1:q]^n$ and $S\subseteq [1:q]$, we denote the subvector obtained by selecting only the components indexed by $S$ as $\vx_S=(x_i)_{i\in S}$.
For any $\vx\in [1:q]^n$, $a\in [1:q]$, we denote the number of components in $\vx$ with value $a$ by
\begin{align}
    N(a|\vx) = \sum_{i=1}^n \indicator_{\{x_i=a\}}.
\end{align}
The {\it composition vector}, or simply composition, of $\vx$ is defined as the $q$-length vector
\begin{align}
    \compositionof{\vx} = (N(1|\vx),N(2|\vx),\cdots,N(q|\vx)).
\end{align}

The set of compositions  of vectors in $[1:q]^n$ is denoted as $\weightqn$:
\begin{align}
    \weightqn := \left\{\compositionof{\vx} | \vx\in [1:q]^n \right\}.
\end{align}
The type of $\vx$ is the normalized composition vector, i.e., $\frac{1}{n}\compositionof{\vx}$.
We denote the set of all $n$-length  vectors over $[1:q]$ of composition $\vw$ as $\compositionclass{\vw}$, and for any $\vx \in [1:q]^n$, the set of all $n$-length vector over $[1:q]$ having the same composition/type as that of $\vx$ is denoted as $\seqofsame{\vx}$, is known as the type class of $\vx$. Note that the number of distinct compositions of vectors in $[1:q]^n$ is $\binom{n+q-1}{q-1}$.
We will use the following simple bounds (see, e.g., \cite[eq. (2)]{sarkar2024identification}) for  $|\weightqn|$:
\begin{align}
    \frac{n^{(q-1)}}{(q-1)!} \le |\weightqn| \le(2n)^{(q-1)} \quad \text{ (for $n\ge q-1$)}. \label{eq:Nbounds1}
\end{align}

For two compositions  $\vt,\vt'\in \weightqn$, we define their distance to be half of their $l_1$-distance
\begin{align}
    \dc(\vt,\vt'):=\frac{1}{2}\sum_i |t_i-t_i'|. \label{defn:dc}
\end{align}
This is also a scaled version of the total variation distance between the corresponding types $\frac{1}{n}\vt$ and $\frac{1}{n}\vt'$. That is,
\begin{align*}
    \dc(\vt,\vt'):=n\,\tv \left(\frac{1}{n}\vt,\frac{1}{n}\vt'\right).
\end{align*}

For any two distributions $\vu,\vv \in \cP([1:q])$, we define the following disjoint sets. 
\begin{align}
    \splus{\vu}{\vv}&\coloneqq \{ c \in [1:q]| u_c > v_c\},\\
    \sminus{\vu}{\vv} &\coloneqq \{ c \in [1:q]| u_c < v_c\}.
\end{align}
$\splus{\vu}{\vv}$ is the set of indices where the probability measure $\vu$ dominates $\vv$, and  $\sminus{\vu}{\vv}$ is the set of indices where $\vv$ dominates $\vu$. Since these sets are disjoint subsets of $[1:q]$, it holds that
\begin{align}
    0 \le |\splus{\vu}{\vv}|+|\sminus{\vu}{\vv}| \leq q.
\end{align}

For two sets $\cA,\cB$, $\cP(\cB|\cA)$ denotes the set of conditional distributions $P_{Y|X}$ where $X\in\cA, Y\in\cB$. A channel with input alphabet $\cA$ and output alphabet $\cB$ is specified by such a conditional distribution. All logarithms are assumed to be of base $2$, when not specified.

\subsection{Noisy permutation channel}

\begin{table}[h]
  \centering
  
  \begin{tabular}{@{}ll@{}} 
    \toprule
    \textbf{Channel Description} & \textbf{Notation} \\
    \midrule
    $n$-block noiseless permutation channel & $\upcn$ \\
    Discrete memoryless channel (DMC) & $\dmc$ \\
    $n$-fold memoryless extension of DMC & $\dmcn$ \\
    $q$-ary noisy composition channel & $\qww$ \\
    $n$-block noisy permutation channel & $\nupcn$ \\
    \bottomrule
  \end{tabular}
  \caption{Notations used for various channels}
  \label{tab:notation_channels}
\end{table}

The $n$-block permutation channel $\bupcnq$ has input and output alphabets $[1:q]^n$, and the transition probability
    \begin{align*}
    \Pperm(\vz|\vx) = \frac{1}{n!} {\sum_{\sigma \in S_n} \indicator{}\left( z_{i}=x_{\sigma^{-1}(i)}, \forall i\in \{1,\ldots, n\}\right)}.
    \end{align*}
Let $\dmc(y|x)$ denote the transition probability of a discrete memoryless channel, and let $\dmcn(\vy|\vx)$ denote the transition probability of its $n$-fold memoryless extension 
$\dmcn(\vy|\vx) = \prod_{i=1}^n \dmc(y_i|x_i)$.
We will also use $\dmcmatrix$ to denote the $q\times q$ transition probability matrix of the channel $\dmc(y|x)$: 
\begin{align}
    \dmcmatrix= \begin{pmatrix}
        u_{1,1}&u_{1,2}&\cdots & u_{1,q}\\
        u_{2,1}&u_{2,2}&\cdots & u_{2,q}\\
        \vdots&\vdots&\ddots & \vdots \\
        u_{q,1}&u_{q,2}&\cdots & u_{q,q}\\
    \end{pmatrix}=\begin{pmatrix}
        \vu_{1}\\
        \vu_{2}\\
        \vdots\\
        \vu_{q}\\
    \end{pmatrix},
\end{align}
where the $i$-th row of $\dmcmatrix$ is denoted by $\vu_i$. We assume that $u_{i,j}>0$ for all $1\leq i,j\leq q$. The distribution of the composition of the output of $\dmcn$, when $\vx\in [1:q]^n$ is transmitted over it, is denoted by $W_\vx(\vw); \vw\in\weightqn$. Note that $W_{\vx}$ depends only on the composition of $\vx$. When any $n$-length vector of composition $\vt$ is transmitted over $\dmcn$, the distribution of the composition of the output vector is also denoted, with abuse of notation, as $W_{\vt}(\vw); \vw \in \weightqn$.

We consider the $n$-block noisy permutation channel, which is a composite of two channels - $\upcn$ and a $\dmcn$ as shown in Fig.~\ref{fig:noisyperm_ch1}.
The channel randomly reorders the symbols of the transmitted vector and then the DMC acts on each symbol independently to produce the output symbol.
We denote the combined channel by $\nupcn$. 
The overall behaviour of the channel remains the same if the two blocks are interchanged as in Fig.~\ref{fig:noisyperm_ch2} (\cite{makur2020coding}). We denote this equivalent channel as $\nupcn'$.
The transition probability of the channel $\nupcn$ is given by
    \begin{align}
    \Psigma (\vy|\vx)&=\frac{1}{|\seqofsame{\vx}|}\sum_{\vz \in \seqofsame{\vx} }\dmcn(\vy | \vz) \notag \\
    & = \frac{1}{|\seqofsame{\vy}|} \sum_{\vv\in \seqofsame{\vy}}  \hspace*{-1mm} \dmcn(\vv|\vx) \hspace*{2mm}\text{ (since } \nupcn \equiv  \nupcn'\text{)} \notag \\
    &= \sum_{\vv\in [1:q]^n} \hspace*{-1mm} \dmcn(\vv|\vx) \frac{\indicator\{\vy \in \seqofsame{\vv}\}}{|\seqofsame{\vv}|} 
    \label{eq:bupcntran}
\end{align}
for all $\vx, \vy \in [1:q]^n$. 
We will call the sequence of channels $\left(\nupcn\right)_{n \in \mathbb{N}}$ as the noisy permutation channel and denote it as  $\nupcnc$. 
\begin{figure}[ht] 
\centering{}
\scalebox{0.88}{
\begin{tikzpicture}[recta/.style={rectangle, draw, minimum height=0.6cm, text width=1.5cm, align=center, minimum width=0.2cm},rect2/.style={rectangle, draw, minimum height=1cm, text width=2cm, align=center, minimum width=4.5cm}]
\node (nb1) at  (0,0)  [recta]{$\Pi_n^q$};
 \node (nbsc) at (3,0) [recta]{$\dmcn$};
\draw[thick] (-1.3,-0.5) rectangle (4.2,0.5);

\begin{scope}[line width=1.5pt,>=latex]
\draw[->] (nb1) -- node[midway,above]{$\vz$} (nbsc);
\draw[->](nbsc)-- node[midway,above]{$\vy$}(5,0);
\draw[->](-2,0)--node[above, pos=0.3]{$\vx$}(nb1);
\end{scope}
\end{tikzpicture}
}
\caption{The noisy permutation channel: $\nupcn$} \label{fig:noisyperm_ch1}
\vspace*{-0mm}
\end{figure}

\begin{figure}[ht] 
\centering{}
\scalebox{0.88}{
\begin{tikzpicture}[recta/.style={rectangle, draw, minimum height=0.6cm, text width=1.5cm, align=center, minimum width=0.2cm}]
\node (nb1) at  (3,0)  [recta]{$\Pi_n^q$};
\node (nbsc) at (0,0) [recta]{$\dmcn$};
\node (nb2) at (5,0) {};
\draw[thick] (-1.3,-0.5) rectangle (4.2,0.5);
\begin{scope}[line width=1.5pt,>=latex]
\draw[->] (nbsc) -- node[midway,above]{$\vv$} (nb1);
\draw[->](nb1)-- node[pos=0.6, above]{$\vy$}(nb2);
\draw[->](-2,0)--node[above, pos=0.3]{$\vx$}(nbsc);
\end{scope}
\end{tikzpicture}
}
\caption{A channel that is equivalent to $\nupcn$: $\nupcn'$} \label{fig:noisyperm_ch2}
\end{figure}

The constituent uniform permutation channel $\upcn$ in $\nupcn$ preserves the composition of the input sequence, i.e. $\compositionof{\vx}=\compositionof{\vz}$. Since the permutation is chosen uniformly at random from $S_n$, no other information about $\vx$ is preserved in $\vz$. To be precise, given $\vx$, the output vector $\vz$ is uniformly distributed in $\seqofsame{\vx}$. Hence $I(\vZ;\vX|\compositionof{\vX}=\vt)=0$ for any $\vt\in \weightqn$.  The DMC changes the composition of the input sequence $\vz$. Since information passes through $\nupcn$ only through the composition of the transmitted sequence, we define a new channel that characterizes the change in the composition. The $q$-ary noisy composition  channel ($\qww$) with input and output alphabet $\weightqn$  is depicted in Fig.~\ref{fig:qww-channel}. 
Its transition probability $\Pqww$ is given by
\begin{align}
    \Pqww (\vw|\vt) 
    & = \sum_{\vy \in \compositionclass{\vw}}  P_{\Sigma} (\vy|\vx) = \sum_{\vz \in \compositionclass{\vw}}  \dmcn (\vz|\vx)   \label{eq:wwch1} 
\end{align}
for all $\vt,\vw \in \weightqn$,
where $\vx\in [1:q]^n$ is any vector of composition $\vt$.
We will refer to the sequence of channels $\left(\qww\right)_{n \in \mathbb{N}}$ as the $q$-ary noisy composition channel and denote it by $\qwwu$.

An $(n,N)$ communication code over $\nupcn$ is a pair of encoding map $f:[1:N] \rightarrow [1:q]^n$ and a decoding map $g:[1:q]^n \rightarrow [1:N]$. Rate of a code is defined as $R=\frac{\log N}{\log n}$. A rate $R$ is said to be achievable for reliable communication over $\nupcn$ if there exists a sequence of $(n_i,N_i)$ codes such that the rates $R_i\geq R$ and the probability of error vanishes as $i\rightarrow \infty$. The supremum of achievable rates is defined as the reliable communication capacity. 

\begin{figure}[htbp]
\centering{}
\scalebox{0.86}{
\centering{}

\begin{tikzpicture}[recta/.style={rectangle, draw, minimum height=1cm, text width=3cm, align=center, minimum width=0.2cm}]
\node (nb1) at  (0,0)  [recta]{Pick an arbitrary vector of composition $\vt$};
\node (nbsc) at (4.5,0) [recta]{$\dmcn$};
\node (nb2) at (9,0) [recta]{Compute the composition of $\vy$};

\begin{scope}[line width=1.5pt,>=latex]
\draw[->] (nb1) -- node[above]{$\vx$} (nbsc);
\draw[->](nbsc)-- node[above]{$\vz$}(nb2);
\draw[->](-3,0)--node[above, pos=0.4]{$\vt \in \weightqn $}(nb1);
\draw[->](nb2)--node[above, pos=0.5]{$\vw=\compositionof{\vz}$}(13,0);
\end{scope}

\end{tikzpicture}}
\caption{The noisy composition channel $\qww$}\label{fig:qww-channel} \vspace*{-2mm}
\end{figure}

\subsection{Identification codes}

\begin{definition}\label{def:code}
	An $\msgsize$-sized identification (ID) code with deterministic decoders  for any channel $P\in \cP(\cB|\cA)$ is a set
	$$	\left\{(Q_i,\cD_i)\mid i=1,\ldots,\msgsize\right\}$$
	of pairs with
	$
		Q_i \in \cP(\cA),
	$  and $\cD_i \subset \cB,$
for $i=1,\ldots,\msgsize$. For such a code, a message $i$ is encoded to a symbol
$x\in\cA$ with probability $Q_i(x)$, and the decoder for message $i$ outputs $1$ (``Accept'') if and only if the received symbol
$y\in \cD_i$. If the encoding distributions $Q_i$ are uniform distributions over some support sets $\cA_i$, then the ID code  is specified by $\left\{(\cA_i,\cD_i)\mid i=1,\ldots,\msgsize\right\}$. 
\end{definition}

For an ID code, a message $ i $ is encoded as a symbol $ x \in \cA $ with probability $ Q_i(x) $. A deterministic decoder for message $ i $ outputs $ 1 $ (``Accept'') if the received symbol $ y$ is in $\cD_i$, and outputs $0$ (``Reject'') otherwise.
    
\begin{definition}\label{def:stochastic ID code}
	An $\msgsize$-sized identification (ID) code with stochastic decoders  for any channel $P\in \cP(\cB|\cA)$ is a set of pairs
	\begin{align*}
		\left\{(Q_i,P_i)\mid i=1,\ldots,\msgsize\right\}
	\end{align*}
	with
	$
		Q_i \in \cP(\cA),
	$  and $P_i \in \cP(\{0,1\}|\cB),$
for $i=1,\ldots,\msgsize$. For such a code, the decision rule $P_i$ for the message $i$ outputs $1$ with probability $P_i(1|y)$ for received symbol $y \in \cB$. Deterministic decoders are a special case with $P_i(1|y):=\indicator{\{y\in \cD_i\}}$.
\end{definition}

 A stochastic decoder $P_i$ outputs $1$ with probability $P_i(1|y)$ for received symbol $y \in \cB$. Deterministic decoders are a special case with $P_i(1|y):=\indicator{\{y\in \cD_i\}}$.

When message $i$ is encoded, the probability that a different message $j\neq i$ is accepted by the decoder is given by
\begin{align*}
    \lambda_{i \rightarrow j} &\coloneqq 
    \sum_{x \in \cA, y\in \cB} \hspace*{-2mm} Q_i(x)P(y|x)P_j(1|y)\text{ (for stochastic decoders)} \label{eq:error id at n_2a} \\
    & = \sum_{x \in \cA, y\in \cD_j}Q_i(x)P(y|x) \,\text{ (for deterministic decoders),} \notag
\end{align*}
and the missed detection probability for message $i$, i.e. the probability that the decoder rejects message $i$ while the same message was encoded, is given by
\begin{align*}
    \lambda_{i \not\rightarrow i}& \coloneqq 
    \sum_{x \in \cA, y\in \cB} \hspace*{-2mm} Q_i(x)P(y|x)P_i(0|y) \text{ (for stochastic decoders)}  \\
    & = \sum_{x \in \cA, y\in \cD^c_i}Q_i(x)P(y|x) \,\text{ (for deterministic decoders).} \notag
\end{align*}
We define the Type-I and Type-II error probabilities of an ID code as 
\begin{align}
\lambda_1 \coloneqq \max_{1 \le i\le \msgsize} \lambda_{i \not\rightarrow i}, \text{ and }
&\lambda_2 \coloneqq \max_{1\le i\neq j\le \msgsize}\lambda_{i \rightarrow j}, 
\end{align}
and the probability of error of an ID code as $\lambda \coloneqq \lambda_1+\lambda_2.$

We refer to an $\msgsize$-sized (i.e. with $\msgsize$ messages) ID code for $\nupcn$ as an $(n,\msgsize)$ ID code for $\nupcn$ or $\nupcnc$, and similarly, an $\msgsize$-sized ID code for $\qww$ is referred as an $(n,\msgsize)$ ID code for $\qww$ or $\qwwu$. When relevant, we will also specify the error probabilities and refer to such a code as an $(n,\msgsize,\lambda_1,\lambda_2)$ or $(n,\msgsize,\lambda)$ code. 

For an $(n,\msgsize)$ ID code over $\nupcn$, the probability of accepting message $j$, when $i\neq j$ is transmitted, is given by
\begin{align}
    \lambda_{i \rightarrow j} 
    &=\sum_{\vx \in [1:q]^n} Q_i(\vx) \sum_{\vy\in \cD_j} \Psigma (\vy|\vx) \nonumber\\
    &=\sum_{\vx \in [1:q]^n} Q_i(\vx)  \sum_{\vz\in [1:q]^n} U^{(n)} (\vz|\vx)\frac{|\cD_j\cap \seqofsame{\vz}|}{|\seqofsame{\vz}|} \label{eq: error i to j}
\end{align}
The last equality follows by using \eqref{eq:bupcntran}.
Similarly, the missed detection probability for message $i \in [1:M]$ is given by
\begin{align}
    \lambda_{i \not \rightarrow i}  &=\sum_{\vx \in [1:q]^n} Q_i(\vx)  \sum_{\vz\in [1:q]^n} U^{(n)} (\vz|\vx)\frac{|\cD_i^c \cap \seqofsame\vz|}{|\seqofsame{\vz}|}.
    \label{eq: error i not i}.
\end{align}

The rate of an $(n,\msgsize)$ ID code over $\nupcn$ is defined to be $\frac{\log \log \msgsize}{\log n}$. An identification rate $R$ is said to be {\it achievable} over $\nupcnc$ if there is a sequence of $(\msgsize_i,n_i,\lambda_i)$ ID codes with $n_i\rightarrow \infty$, $\frac{\log \log \msgsize_i}{\log n_i} \rightarrow R$ and $\lambda_i \rightarrow 0$. The supremum of all achievable rates over $\nupcnc$ is defined as the capacity of $\nupcnc$.

\subsection{Multinomial and related distributions}

For any positive integer $n$ and a real number $p; 0\le p \le 1$, let $\left(B_{(n,p)}(t)\right)_{t \in [0:n]}$ be the binomial distribution with number of trials $n$ and success probability $p$. The distribution is given by
\begin{align*}
    B_{(n,p)}(t)=\binom{n}{t}p^t(1-p)^{n-t}, \quad \forall t \in [0:n].
\end{align*}
For any distribution $\vv=(v_1,v_2,\cdots,v_q)$ over $[1:q]$, the multinomial distribution $\multnv(\vt)$ is given by
\begin{align*}
    \multnv(\vt)=\binom{n}{t_1,\cdots,t_q} \prod_{i=1}^q v_i^{ t_i}, \quad \forall \vt \in \weightqn.
\end{align*}
Note that for $q=2$, a multinomial distribution specializes to a binomial distribution. Specifically, we have 
\begin{align}\label{eq:binom as multinom}
    \multinomial{n}{(1-p,p)}(n-t,t)=B_{(n,p)}(t)
\end{align}

For any $a\in [1:q]$, when the vector $(a,a,\cdots,a)\in [1:q]^n$ is transmitted over $\dmcmatrix$, the output composition distribution is $\multinomial{n}{\vu_a}$, where $\vu_a$ is the $a$th row of the transition probability matrix $\dmcmatrix$. Now consider an arbitrary $\vx\in [1:q]^n$ with composition  $\vt\in \weightqn$. For $a \in [1:q]$, let $J_a:=\{j\in [1:q]|x_j=a\}$ be the locations of $\vx$ with component value $a$. When $\vx$ is transmitted, the composition of the output $\vY$ is the sum of the compositions of the subsequences corresponding to the locations $J_b;b=1,2,\cdots,q$.That is, the output composition is given by $\compositionof{\vY}=\compositionof{\vY_{J_1}}+\compositionof{\vY_{J_2}}+\cdots + \compositionof{\vY_{J_q}}$. Also note that, since $|J_a|=t_a$,  $\compositionof{\vY_{J_i}}$ has the distribution $\multinomial{t_a}{\vu_a}$. Hence the distribution of $\compositionof{\vY}$, when $\vx$ is transmitted, denoted by $W_\vx$, is given by
\begin{align}\label{eq:W_x_conv_multinomial}
    W_\vx(\vw) = \multinomial{t_1}{\vu_1}(\vw) \ast
\multinomial{t_2}{\vu_2}(\vw) \ast \cdots \ast \multinomial{t_q}{\vu_q}(\vw), \quad \forall \vw \in \weightqn.
\end{align}

For any distribution $Q$ over $\setofweights{q}{n}$, and $a,b\in [1:q]$; $a\neq b$, we define the sum of all absolute differences between successive values in the distribution $Q$ along the direction $\ve_b-\ve_a$ as \bikdjournal{Give an example to explain the definition.}
\begin{align*}
    D_{ab}(Q) &:= \sum_{\vt\in \setofweights{q}{n+1}}\left| Q(\vt-\ve_{b})- Q(\vt-\ve_{a})\right|.
\end{align*}
Here, $Q(\vt)$ is defined to be $0$ when $\vt$ has any component outside the range $[0:n]$.

For example, by considering $Q(t)=\multinomial{n}{(1-p,p)}(n-t,t)=B_{(n,p)}(t); t \in [0:n]$, we have $D_{12}(\multinomial{n}{(1-p,p)}(n-t,t))=\sum_{t=0}^n |B_{(n,p)}(t-1)-B_{(n,p)}(t)|$.
Clearly, $D_{ab}(Q)=D_{ba}(Q)$ for all $a,b\in [1:q]$; $a\neq b$. 

For $\vt \in \setofweights{q}{n+1}$ by denoting $\delta_{a,b}(\vt)=\delta_{\ve_b}(\vt)-\delta_{\ve_a}(\vt)$ we can write
\begin{align}
    D_{ab}(Q) &=
    \sum_{\vt\in \setofweights{q}{n+1}}\left| Q(\vt-\ve_{b})- Q(\vt-\ve_{a})\right|\\
    &=\sum_{\vt\in \setofweights{q}{n+1}}|((\delta_{\ve_b}-\delta_{\ve_a})*Q)(\vt)|\\
    &=\sum_{\vt\in \setofweights{q}{n+1}}|(\delta_{a,b}*Q)(\vt)| \label{eq:D_ab_W}.
\end{align}

\section{Background and known results}\label{sec:background}

We now present some known results on the noisy permutation channel. Reliable communication over a noisy permutation channel was studied in \cite{makur2020coding, tang2023capacity}.
\begin{theorem}[(\cite{makur2020coding,tang2023capacity})]\label{thm:reliable}
 The reliable communication capacity of $\nupcnc$ is at least $\frac{r-1}{2}$, where $r$ is the rank of $U$. Further, if $U$ has strictly positive entries the capacity is exactly $\frac{r-1}{2}$.
\end{theorem}

The achievability was proved in \cite{makur2020coding}. In \cite{makur2020coding} it is shown that the reliable communication capacity of $\nupcn$ is at most $\frac{\min \{r,s\}-1}{2}$ if $U$ has strictly positive entries, where $s$ denotes the number of exterior points of the convex hull of rows of $U$ and $r$ be the rank of $U$. Later \cite{tang2023capacity} improved the converse to meet the achievability.

The achievability result guarantees that there exists a sequence of codes for reliable communication over $\nupcnc$  with message size growing as $\binom{k+r-1}{r-1} \ge \frac{k^{r-1}}{(r-1)!}$, where $k=n^{\frac{1}{2}-\epsilon}$, and the probability of error going to $0$ as $n\rightarrow \infty$. A minor refinement of the achievability proof of \cite{makur2020coding} can show that there exists a sequence of codes with $N=\left( \frac{n}{c\log n}\right)^{(r-1)/2}$, where $c$ is a suitably chosen constant, with vanishing probability of error. We present this result as a lemma.

\begin{lemma}\label{lem:reliable_ach}
    Consider the channel $\nupcnc$, where $U$ is a DMC with strictly positive entries and rank $r$. There exists a sequence of $\left(n, \left( \frac{n}{c \log n}\right)^{\frac{r-1}{2}}\right)$ reliable transmission codes for $\nupcnc$, where $c$ is a suitable chosen constant, such that the probability of error goes to $0$ as $n\rightarrow \infty$.
\end{lemma}

\section{Main results}\label{sec:result}

We present our results on identification with stochastic encoding in the first subsection, and with deterministic encoding in the next subsection.

\subsection{Identification over $\nupcn$ using Stochastic encoding}

We first present our main result for identification using stochastic encoding. 

\begin{theorem}[Identification under stochastic encoding] \label{thm:stoch}
~~
\begin{enumerate}[label=(\roman*)]
\item \label{thm:ach_stoch} Achievability:
    For $q \ge 2$, let $U$ be  a $q$-ary DMC with the transition probability that has rank $r$. For any $\epsilon_n \rightarrow 0$, there exists a sequence of $\left(n,2^{\epsilon_n \left( \frac{n}{\log n} \right)^{(r-1)/2} },\lambda_{1,n},\lambda_{2,n}\right)$ ID codes for $\nupcn$ with $\lambda_{1,n}, \lambda_{2,n} \rightarrow 0$ as $n \rightarrow \infty$. 
\item \label{thm:converse_stoch} Strong converse:     For $q \ge 2$, let $U$ be a $q$-ary DMC with a transition probability matrix that has positive entries. For any sequences $\{n_i\}_{i \ge 1}$ and $\{c_i\}_{i \ge 1}$ such that $n_i, c_i \rightarrow \infty$, and any sequence of $$\left(n_i, 2^{\left(c_{i} n_i^{(q-1)/2}(\log n_i)^{1+\frac{(q-1)(q-2)}{2}}\right)}, \lambda_{1,i}, \lambda_{2,i} \right)$$ ID codes with stochastic decoders for the $q$-ary noisy permutation channel $\nupcnc$, $\liminf_{i \rightarrow \infty}(\lambda_{1,i}+\lambda_{2,i}) \ge 1$.
\end{enumerate}
\end{theorem}

We give an outline of the proof next. The formal proof of the achievability is given in Sec~\ref{sec:proof_ach_thm}, whereas the formal proof of the strong converse is given in Sec~\ref{sec:proof_of_theorem_strong_converse_qary}.

{\it Proof outline of Theorem~\ref{thm:stoch}:} Part (i) (Achievability): The proof uses the known capacity result for reliable communication over noisy permutation channels. It is known that the capacity of a noisy permutation channel is $(r-1)/2$. The achievability was shown in \cite{makur2020coding}, and the converse was shown in \cite{makur2020coding,tang2023capacity}.  The proof of part (i) then follows by using Lemma~\ref{lem:reliable_ach} along with  \cite[Proposition 1]{ahlswede1989identification} (see Proposition~\ref{prop:ahlswede}).

Part (ii) (strong converse): To prove the strong converse, we begin by showing the existence of an ID code (with stochastic decoders) for the $q$-ary noisy composition channel ($\qww$), that matches the message size and error performance of the original code for $\nupcn$. This is because the constituent permutation channel in $\nupcn$ preserves only the composition of the transmitted $q$-ary vector. As a result, information passes through $\nupcn$ only through the composition of the sequences. We employ the method of channel resolvability to prove the converse result for $\qww$. In this method, we need a good quantization scheme for input distributions for $\qww$ such that the output distributions due to any input distribution and its quantized version have a small (vanishing as $n\rightarrow \infty$) TV distance. We propose an interesting two-stage deterministic quantization scheme that uses the structure of the $\qww$. We next discuss our proposed quantization scheme.

To design a good quantization scheme, we first analyze how the output distribution of the $\qww$ changes when in the input composition, a unit mass is shifted from one symbol to another. We show that the TV distance between the corresponding output distributions over $\qww$ decays as $\frac{(\log n)^{(q-2)/2}}{\sqrt{n}}$ (up to a constant factor). Using triangular inequality, it then follows that for any two arbitrary compositions $\vt$ and $\vt'$, the TV distance between their output distributions over $\qww$ decays as $ \dc(\vt, \vt') \cdot \frac{(\log n)^{(q - 2)/2}}{\sqrt{n}} $ (up to a constant factor). 
Inspired by this, we propose a two-stage deterministic quantization scheme to approximate the distribution of input composition with an $M$-type distribution by shifting probability mass in nearby (in distance $\dc(\cdot,\cdot)$) compositions as much as possible. In the first stage, we redistribute probability mass within a distance $a = o\left( \frac{\sqrt{n}}{ (\log n)^{(q-2)/2}} \right)$ to quantize (i.e. make these multiples of $1/M$) all except a few probability masses. In the second stage, we quantize the remaining few probability masses by sequentially redistributing their probability masses. It is shown that the change, measured in TV distance, in the output distribution due quantization of an input distribution this way vanishes as $n\rightarrow \infty$.

Note that if $U$ has full-rank, i.e., if $r=q$, with strictly positive entries in the transition probability matrix, the achievability and the converse meet in rate. Hence we have the following capacity result.

\begin{corollary}\label{cor:capacity}
    The identification capacity of $\nupcnc$, where $U$ is of full rank and has strictly positive entries, is given by $\frac{q-1}{2}$.
\end{corollary}

\begin{remark} 
For reliable communication, the message size $N$ for a noisy permutation channel grows as $n^R$, and hence the rate is defined as $\frac{\log N}{\log n}$. It was proved in \cite{makur2020coding,tang2023capacity} that the capacity, the supremum of 
achievable rates, of $\nupcnc$ is $\frac{r-1}{2}$ for a $U$ with strictly positive entries and rank $r$. Corollary~\ref{cor:capacity} shows the same result for identification over $\nupcnc$ for full rank (i.e. $r=q$) $U$. We believe that for any $U$ with strictly positive entries, the identification capacity is $\frac{r-1}{2}$, consistent with the case of reliable communication. However, our converse proof does not use possible rank-deficiency of $U$, and hence shows a weaker converse for $\nupcnc$, if $U$ is rank-deficient. We believe that using the rank-deficiency of $U$ will require significant technical generalization, and is left as an open problem. However, as an aid to future research towards tightening the strong converse for rank deficient $U$, we present some of our lemmas (Lemma \ref{lemma:D_M}, Lemma \ref{lemma:upper bound mode multinomial1}, and Corollary \ref{lemma:upper bound mode multinomial}) in Sec.~\ref{sec:multinomial} for general $U$ with arbitrary rank.
\end{remark}

\subsection{Identification over $\nupcn$ using Deterministic encoding}

We now consider the deterministic encoding and present achievability, weak converse and strong converse results, which are tight up to powers of $\log n$ in message size.

\begin{theorem}[Deterministic encoding]\label{thm:det}
~~
\begin{enumerate}[label=(\roman*)]
    \item {\it Achievability:} For $q \ge 2$, let $U$ be  a $q\times q$ transition probability matrix with rank $r$. For  a suitably chosen constant $c$,  there exists a sequence of $\left(n, \left( \frac{n}{c\log n} \right)^{(r-1)/2},\lambda_{1,n},\lambda_{2,n}\right)$ ID codes for $\nupcnc$ with deterministic encoders and decoders  such that $\lambda_{1,n}, \lambda_{2,n} \rightarrow 0$ as $n \rightarrow \infty$. \label{thm:ach_det}
    \item {\it Weak Converse:} For $q \ge 2$, let $U$ be  a $q\times q$ transition probability matrix with strictly positive entries. There exists a real number $R'$ such that for all $R >R'$, any sequence $\{n_i\}_{i \ge 1}$ such that $n_i \rightarrow \infty$, and any sequence of $$\left(n_i, R n_i^{(q-1)/2}(\log n_i)^{(q-1)(q-2)/2} , \lambda_{1,i}, \lambda_{2,i} \right)$$
    ID codes with deterministic encoders and decoders for $\nupcnc$, $$\liminf_{i \rightarrow \infty}(\lambda_{1,i}+\lambda_{2,i}) > 0.$$
    \item {\it Strong Converse:}
    For $q \ge 2$, let $U$ be  a $q\times q$ transition probability matrix with strictly positive entries. For any sequences $\{n_i\}_{i \ge 1}$ and $\{c_i\}_{i \ge 1}$ such that $n_i,c_i \rightarrow \infty$, and any sequence of $$\left(n_i, c_{i} n_i^{(q-1)/2}(\log n_i)^{(q-1)(q-2)/2}, \lambda_{1,i}, \lambda_{2,i} \right)$$
    ID codes with deterministic encoders and decoders for $\nupcnc$, $$\liminf_{i \rightarrow \infty}(\lambda_{1,i}+\lambda_{2,i}) \ge 1.$$
\end{enumerate}

\end{theorem}

The detailed proof of Theorem~\ref{thm:det} is given in Sec.~\ref{sec:proof_of det_ID}. Here we provide a proof sketch.

{\it Proof sketch of Theorem~\ref{thm:det}:}
The achievability proof uses a deterministic reliable transmission code for message identification. As one can transmit $\left( \frac{n}{c\log n}\right)^{(r-1)/2}$ messages reliably over the $q$-ary noisy permutation channels (see Lemma~\ref{lem:reliable_ach}), the part (i) of Theorem~\ref{thm:det} follows directly. 
Similar to the proof of Theorem~\ref{thm:stoch}, part (ii), we begin to prove part (ii) and (iii) of Theorem~\ref{thm:det} by showing the existence of an ID code with a deterministic encoder and stochastic decoders for $\qww$, having the same message size and error performance that of the original code for $\nupcn$.  Let $d^*$ denote the minimum pairwise distance (in $\dc(\cdot,\cdot)$, see \eqref{defn:dc}) between its codewords. For any ID code with deterministic encoder and stochastic decoders having message size $ L = c_n n^{\frac{q-1}{2}}\left(\log n\right)^{\frac{(q-1)(q-2)}{2}} $, by using a packing argument (see Lemma~\ref{lemma:min_distance}), we show that $ d^* \le \frac{\sqrt{n}}{c_n^{1/(q-1)} (\log n)^{(q-2)/2}}$. We then  use the fact that for any two compositions having distance $d$, the TV distance between their corresponding output distributions over $\qww$ is upper bounded (up to a constant factor) by
$d \cdot \frac{(\log n)^{(q-2)/2}}{\sqrt{n}}$ (Lemma~\ref{lemma:weight_t_1t_2}). This yields that, for any two distinct codewords, the TV distance between their corresponding output distributions is at most $d^* \cdot \frac{(\log n)^{(q-2)/2}}{\sqrt{n}}\leq \frac{1}{c_n^{1/(q-1)}}$ (up to a constant factor). Following the standard arguments in channel resolvability, we then lower bound the sum of Type-I and Type-II error probabilities of an ID code by one minus the minimum TV distance between the output distributions corresponding to pairs of distinct codewords. For a certain value of $R'$, this gives a non-zero constant lower bound on the sum probability of error for $c_n=R$ for any $R>R'$, thus proving the weak converse. If $c_n \rightarrow \infty$ (as $n \rightarrow \infty$), this also gives the strong converse.

\section{Proof of Theorem \ref{thm:stoch}, Part (i)}\label{sec:proof_ach_thm}

Our achievability proof will use the following proposition from \cite{ahlswede1989identification}.
 \begin{proposition}\cite[Proposition 1]{ahlswede1989identification} \label{prop:ahlswede}
For any finite set $\cZ$, $\lambda\in (0,0.5)$, and $\epsilon>0$ such that
\begin{align}
    \lambda \log \left(\frac{1}{\epsilon}-1\right) >2 \text{ and } \epsilon <\frac{1}{6}, \label{eq:epscondition}
\end{align}
 there exist $\msgsize$ subsets $\cU_1,\cdots,\cU_\msgsize\subseteq \cZ$, each of size $\epsilon|\cZ|$, such that $|\cU_i\cap \cU_j|\leq \lambda\epsilon |\cZ|$ $\forall i\neq j$ and
\begin{align}
    \msgsize \geq |\cZ|^{-1} 2^{\epsilon |\cZ|-1}.
\end{align}
\end{proposition}
  
Our ID encoder is composed of two stages. First, a stochastic ID encoder encodes the message into an intermediate message. This new message is then encoded by an encoder for reliable communication.
Similarly, the decoder has two stages. First, a decoder decodes the intermediate message, and then an ID decoder outputs the final decision based on the intermediate message.

The achievability proof of \cite[Theorem $1$]{makur2020coding} (for reliable communication) proved that for any $\epsilon \in (0.0.5)$, there exists a sequence of 
codes  with message size $$\frac{1}{(r-1)!}n^{\left(\frac{r-1}{2}-\epsilon(r-1)\right)}$$ and vanishing probability of error for $\nupcnc$. It can be checked that the achievability proof can be strengthened to prove 
the existence of a sequence of codes for reliable communication with message size  
$$N=\frac{1}{(r-1)!}\left(\frac{n}{c \log n}\right)^{(r-1)/2},$$ 
where $c$ is a suitably chosen constant independent of $n$, with vanishing probability of error $P_e^{(n)}$. Let $\phi(\cdot)$ and $\psi(\cdot)$ be, respectively, the encoding and decoding functions of a code from this family. $\phi(\cdot)$ will be used as the second encoding stage, and $\psi(\cdot)$ will be used as the first decoding stage.

We define
    \begin{align}
         \epsilon_n' &=\left( (r-1)! c^{(r-1)/2} \epsilon_n+\frac{1}{N}+\frac{1}{N}\log N\right)\\
         \lambda'_{2,n} &=\frac{4}{\log\left(\frac{1}{\epsilon_n'}\right)} \label{eq:lambda_ach}\\
         \cZ&= [1:N]
    \end{align} 
    and use these in 
places of $\epsilon, \lambda, \cZ$ respectively in Proposition~\ref{prop:ahlswede}.
    It follows  that $\epsilon_n'\rightarrow 0$ as $n\rightarrow \infty$ and $\lambda'_{2,n} \rightarrow 0$ as $\epsilon'_n \rightarrow 0$, and hence for large enough $n$, $\epsilon_n'<\frac{1}{6}$.
    Also, for large enough $n$,
    \begin{align*}
    \lambda'_{2,n} \log \left(\frac{1}{\epsilon_n'}-1\right) 
    &=\lambda'_{2,n} \log \left(\frac{1}{\epsilon_n'}\left(1-\frac{1}{2} \right)+\frac{1}{2\epsilon'_n}-1\right)\\
    & \stackrel{(a)}{\geq} \lambda'_{2,n} \log \left(\frac{1}{2\epsilon_n'}\right)\\
    & = \lambda'_{2,n} \times \left(\log \left(\frac{1}{\epsilon_n'} \right)-1\right) \\
    &= \lambda'_{2,n} \times \left(\log \left(\frac{1}{\epsilon_n'} \right)\left( 1-\frac{1}{2}\right)+\frac{1}{2}\log \left(\frac{1}{\epsilon_n'} \right)-1\right)\\
    & \geq  \frac{4}{\log\left(\frac{1}{\epsilon_n'}\right)} \times \frac{1}{2}\log  \left(\frac{1}{\epsilon_n'} \right)\\
    & = 2.
    \end{align*}
    Here $(a)$ follows because for large enough $n$ $\epsilon'_n < 1/2$, this implies $(\frac{1}{2\epsilon'_n}-1)>0$ and $(b)$ follows because for large enough $n$ $\epsilon'_n < 1/4$, this implies $\frac{1}{2}\log \left(\frac{1}{\epsilon_n'} \right)-1>0$.
    Hence $\epsilon_n'$ and $\lambda'_{2,n}$ satisfy the conditions in  Proposition~\ref{prop:ahlswede}, which guarantees that there exist
    \begin{align*}
        \msgsize &\ge 2^{\left(\epsilon_n' N-1-\log N \right)}\\
        &=2^{\big\{ \epsilon_n  N (r-1)!c^{(r-1)/2}\big\}}\\
        &=2^{\epsilon_n \left(\frac{n}{\log n} \right)^{(r-1)/2}}
    \end{align*}
    number of distinct subsets $\{\cU_1, \ldots, \cU_\msgsize\}$ of $\cZ$ such that
    \begin{align*}
        |\cU_i|&=\epsilon_n' N, \,\,\forall i,\\
        |\cU_i \cap \cU_j| &\le \lambda'_{2,n} \epsilon_n' N, \,\,\forall i \neq j. 
    \end{align*}
We define the ID code $\{(Q_i,\cU_i)|i=1,\ldots,M)\}$ for $\nupcn$, where  $Q_i$ is the uniform distribution over $\cU_i$.
For a message $i$, the first stage of the encoder outputs an intermediate message $m$, chosen at random from $\cU_i$ using the distribution $Q_i$. This message is encoded by the second stage encoder to output $\phi(m)$, which is transmitted over the channel. On receiving $\vy$, decoder $j$ accepts if and only if $\psi(\vy)$ is in $\cU_j$.

An identification error of either type can be either due to the decoding error in the first stage ($\psi(.)$) or due to the identification error in the second stage when the first stage of decoding is correct.
Hence for any message $i \in [1:\msgsize]$, 
the probability of missed detection for message $i$ is given by
\begin{align}
    \lambda_{i \not \rightarrow i}
    &\leq P_{e}^{(n)} + 0 \notag \\
    & = P_{e}^{(n)}, 
\end{align}
and the probability that a different message $j \neq i$ is accepted while message $i$ is encoded is given by
\begin{align}
    \lambda_{i \rightarrow j} 
    & \leq P_{e}^{(n)} + \frac{|\cU_i \cap \cU_j|}{|\cU_i|} \notag \\
    &\le  P_{e}^{(n)} + \lambda'_{2,n}. 
\end{align}
Therefore both the Type-I error probability $\lambda_{1,n}=P_{e}^{(n)}$ and Type-II error probability $\lambda_{2,n}= \lambda_{2,n}'+P_{e}^{(n)}$ asymptotically go to $0$ as $n \rightarrow \infty$. This proves the achievability part of Theorem~\ref{thm:stoch}.

\section{Proof of Theorem~\ref{thm:stoch}, Part (ii)}\label{sec:proof_of_theorem_strong_converse_qary}

In this section, we present the proof of Theorem \ref{thm:stoch}, Part (ii). We first state some key lemmas required for the proof. The proofs of these lemmas require some technical development, and are presented in subsequent sections. The first lemma relates an ID code for $\nupcn$ with an ID code with stochastic decoders for $\qww$. We show that for any $(n,\msgsize, \lambda_{1}, \lambda_{2})$ ID code with deterministic decoders for $\nupcn$, there exists an $(n,\msgsize, \lambda_{1}, \lambda_{2})$ ID code with stochastic decoders for $\qww$. The proof of this lemma is given in Appendix \ref{app:general_to_type}.
\begin{lemma}\label{lemma: noisy permutation to noisy type}
    Given an $(n,\msgsize, \lambda_{1}, \lambda_{2})$ ID code with deterministic decoders $\{(Q_j, \mathcal{D}_j) | j=1,\ldots, \msgsize\}$ for $\nupcn$,
     there exists an $(n,\msgsize, \lambda_{1}, \lambda_{2})$ ID code $\{Q'_i,P_i| i=1,\ldots, \msgsize\}$ with stochastic decoders for the $\qww$, having the same probability of errors $\{\lambda_{i\rightarrow j}|1 \le i \neq j \le \msgsize\}$ and $\{\lambda_{i \not\rightarrow i}|i=1,\ldots,\msgsize\}$. Furthermore, if the given code has a deterministic encoder, i.e., $Q_i$ is a point mass for each $i$, then the new code $\{Q'_i,P_i| i=1,\ldots, \msgsize\}$ also has a determnistic encoder. 
\end{lemma}

For any arbitrarily transmitted composition $\vt \in \weightqn$  over $\qww$, the output composition distribution is the convolution of multinomial distributions (see \eqref{eq:W_x_conv_multinomial}), hence multinomial distributions play a crucial role in the proof of Theorem \ref{thm:stoch}, Part (ii). We now present a result on the multinomial distributions. Recall that for any distribution $Q \in \cP(\weightqn)$, and any two distinct symbols $a,b \in [1:q]$, $D_{ab}(Q)$ denotes the sum of all absolute differences between successive values in the distribution $Q$ along the direction $\ve_b-\ve_a$.
The following lemma characterizes $D_{ab}(\cdot)$ for a multinomial distribution with number of trials $n$ and success probability distribution $\vu \in \cP([1:q])$.

\begin{lemma}\label{lemma:D_M}
    For large $n$, $q \ge 2$, probability distribution $\vu \in \cP([1:q])$, and any $a \neq b \in [1:q]$, we have
    \begin{align}
D_{ab}(\multinomial{n}{\vu}) \begin{cases}=2 &\textnormal{ if } a \not \in \supp{\vu} \textnormal{ or } b \not\in \supp{\vu} \\
\le \frac{C_1(\log n)^{(|\supp{\vu}|-2)/2}}{\sqrt{n}}, &\textnormal{ if } a,b \in \supp{\vu}
\end{cases}
    \end{align}
    where $C_1$ is a constant independent of $n$. In particular, if $\vu$ has strictly positive entries, then
    \begin{align}
D_{ab}(\multinomial{n}{\vu}) 
\le \frac{C_1(\log n)^{(q-2)/2}}{\sqrt{n}},
    \end{align}
for any $a\neq b$.
\end{lemma}
The proof is given in Sec~\ref{sec:multinomial}.
In Sec.~\ref{sec:unit_change}, this lemma is generalized using \eqref{eq:W_x_conv_multinomial} and we show that $D_{ab}(W_{\vx})$ also
decays as  
$O\left(\frac{(\log n)^{(q-2)/2}}{\sqrt{n}}\right)$ for any vector $\vx\in [1:q]^n$
(see Lemma \ref{lemma:conv_W}).

We next quantify the change in the output composition distribution when one symbol in the transmitted vector is changed, suppose, from $a$ to $b$. This corresponds to 
changing the composition of the input vector by subtracting $1$ from the number of occurrences of $a$ and adding $1$ to the number of occurrences of $b$. This plays a key role in the design of a good quantization scheme for input composition distributions.

\begin{lemma}\label{lemma:dTV_singleW}
    For large $n$, $q \ge 2$, any $\vx \in [1:q]^n$, and any $a \neq b \in [1:q]$, the total variation distance between the output  distributions  $W_{(\vx,a)}$ and $W_{(\vx,b)}$ over $\qww$ corresponding to input sequences $(\vx,a)$ and $(\vx,b)$ is bounded as
    \begin{align}
        \tv(W_{(\vx,a)},W_{(\vx,b)}) \le \frac{C_3(\log n)^{(q-2)/2}}{\sqrt{n}}
    \end{align}
    for some constant $C_3$ independent of $n$.
\end{lemma}
Consider two input compositions $\vt$ and $\vt'$, where $\vt'$ is obtained from $\vt$ by shifting unit mass from a symbol $a$ to another symbol $b$. Lemma~\ref{lemma:dTV_singleW} bounds the change in the distribution of output composition due to changing the distribution of input composition from $\delta_\vt$ to $\delta_{\vt'}$. 
Note that, for $q=2$, the lemma implies that for a bit-flip  channel (e.g. a BSC), changing the input Hamming weight by one changes the output Hamming weight distribution by $O\left(\frac{1}{\sqrt{n}}\right)$. 

Note that for any two compositions $\vt$ and $\vt'$, the total amount of unit mass that needs to be shifted to get $\vt'$ from $\vt$ is $d (\vt,\vt')$.
For such $\vt,\vt'$, changing the  distribution of input composition from $\delta_\vt$ to $\delta_{\vt'}$ causes a change in the distribution of output composition by at most $d (\vt,\vt')\cdot \frac{C_3(\log n)^{(q-2)/2}}{\sqrt{n}}$. The following lemma generalizes this idea to two arbitrary distributions of input composition.

\begin{lemma}\label{lemma:weight_transfer_pairs}
    For large $n$, $q \ge 2$, a positive integer $k \ge 1$, let $\vt_1,\vt_1',\vt_2,\vt_2',\cdots,\vt_k,\vt_k'\in \weightqn$ be some compositions. If $Q_1$ and $Q_2$ are two distributions on $\weightqn$ such that
    \begin{align}
        Q_2(\vt) = Q_1(\vt)+\sum_{i=1}^k p_i \delta_{\vt_i \rightarrow \vt_i'} (\vt)\hspace*{10mm} \forall \vt\in\weightqn \label{eq:distchange}
    \end{align}
    for some $p_1,p_2,\cdots,p_k\geq 0$. If $Q_{\text{op},1}$ and $Q_{\text{op},2}$ denote the output distributions over the $\qww$ corresponding to input distributions $Q_1$ and $Q_2$ respectively, then
    \begin{align*}
        \tv (Q_{\text{op},1},Q_{\text{op},2}) \leq \frac{C_4(\log (n))^{(q-2)/2}}{\sqrt{n}} \sum_{i=1}^k p_i \cdot \tv (\vt_i,\vt_i')
    \end{align*}   
\end{lemma}
The proof of the lemma is given in Sec~\ref{sec:unit_change}.
Note that any two distributions $Q_1$ and $Q_2$ can be related as in \eqref{eq:distchange} such that $\sum_i p_i = \tv (Q_1,Q_2)$.

We use the method of channel resolvability for proving Theorem~\ref{thm:stoch}\ref{thm:converse_stoch}. In this proof, a crucial role is played by a good quantization scheme for distributions of input composition that ensures a small change in the distribution of output composition (in TV distance) due to the quantization. Since a composition distribution needs to be quantized to an $M$-type (for suitable choice of $M$) distribution, probability mass can only be shifted between compositions so as to ensure that the result is still a distribution. We propose a 2-stage deterministic quantization scheme. 

Lemma~\ref{lemma:weight_transfer_pairs} motivates us to shift probability mass between nearby (in TV distance) compositions, as much as possible.
More specifically, it is clear that if all such shifts of probability mass is between compositions within a distance $(q-1)a$, where $a=\frac{\sqrt{n}}{c_n^{1/(q-1)} (\log n)^{(q-2)/2}}$ for some $c_n\rightarrow \infty$, then the change in the output composition distribution $\frac{C_4 (\log (n))^{(q-2)/2}}{\sqrt{n}}(q-1)a=\frac{C_4(q-1)}{c_n^{1/(q-1)}}\rightarrow 0$, even under the worst case total shift of $\sum_i p_i =1$. This inspires the first stage of our quantization scheme. In the first stage, the simplex $\weightqn$ partitioned into cubic cells of side $a$, and masses are redistributed within each cell so as to make each probability mass within a cell, except possibly at one point, a multiple of $1/M$. Hence this stage involves movement of mass within a distance of at most $(q-1)a$, and it results in a distribution that is ``almost'' $M$-type, in the sense that except for one mass in each cell, all other probability masses are multiples of $1/M$.

For the second stage, note that the number of cells is upper bounded by $\left( \frac{2n}{a} \right)^{q-1}$. It is first shown that the cells have a Gray-like order. Under this ordering, any two compositions in neighbouring cells have a distance at most $\leq 2(q-1)a$. In the second stage, the unquantized masses are quantized to the largest multiple of $1/M$ below, and the residual mass is transferred to the unquantized mass in the next cell (according to a Gray-like ordering). This results in a change in the output composition distribution. This change, measured in TV distance, is bounded by 
$$2(q-1)a\times \frac{1}{M}\times \left(\frac{2n}{a}\right)^{q-1}\times \frac{C_4 (\log n)^{\frac{q-2}{2}}}{\sqrt{n}}$$ 
which also vanishes asymptotically if $M=c_n \left( \sqrt{n (\log n)^{(q-2)}} \right)^{(q-1)}$ for some $c_n \rightarrow \infty$.

\begin{proposition}\label{prop:approx_qary}
    For any integer $M,q \ge 2$, large $n$, and integer $a \in [1:n]$ consider the noisy $q$-ary composition channel $\qww$ shown in Fig.~\ref{fig:qww-channel}. For any distribution $\inpdist \in \cP(\weightqn)$ of the input $\vT$,  the   distribution $\inpapprox \in \cP(\weightqn)$ given as output by {\it Algorithm \ref{algo:quan_qary}} is $M$-type, and is such that  the variational distance between the corresponding output distributions (of $\vW$) 
    \begin{align}
        \outdist(\vw) & =\sum_{\vt} \inpdist(\vt) \Pqww (\vw|\vt),\quad\forall \vt,\vw \in \weightqn,\\
        \text{and } \outapprox(\vw) &= \sum_{\vt} \inpapprox(\vt) \Pbww(\vw|\vt),\quad\forall \vt,\vw \in \weightqn
    \end{align}
    is upper bounded as
    \begin{align}
       \tv (\outdist,\outapprox) \le 
       \frac{C_4'a(\log n)^{(q-2)/2}}{\sqrt{n}}+ \left(\frac{2n}{a}\right)^{q-1} \cdot  \frac{C_4'(\log (n))^{(q-2)/2}}{\sqrt{n}} \frac{a}{M}
    \end{align}
    where $C'_4$ is a constant independent of $n$.
\end{proposition}

For any sequence $c_n \rightarrow \infty$, taking $a= \frac{\sqrt{n}}{c_n^{1/(q-1)} (\log n)^{(q-2)/2}}$ and $M=c_n \left( \sqrt{n (\log n)^{(q-2)}} \right)^{(q-1)}$ in Proposition~\ref{prop:approx_qary} gives
    \begin{align}\label{eq:dtvgoes0}
        \tv (\outdist, \outapprox) \le \frac{C_4'}{c_n^{1/(q-1)}} + 2^{q-1}C_4' \cdot \frac{1}{c_n^{1/(q-1)}}.
    \end{align}
Hence $\tv (\outdist, \outapprox)\rightarrow 0$ as $n \rightarrow \infty$.

We now proceed to prove the strong converse. By denoting 
$$
\Gamma_i=2^{c_{i} {n_i}^{(q-1)/2}(\log n_i )^{1+\frac{(q-1)(q-2)}{2}}},
$$
we consider a given sequence of $\left(n_i,\Gamma_i, \lambda_{1,i}, \lambda_{2,i}\right)$ ID codes with deterministic decoders for  $\Sigma_{n_i,U}$  with  $n_i \rightarrow \infty$ as $i \rightarrow \infty$.

For each $i$,  by Lemma~\ref{lemma: noisy permutation to noisy type}, there exists an $\left((n_i,\Gamma_i, \lambda_{1,i}, \lambda_{2,i}\right)$ ID code $$\left\{(Q_j^{(i)},P_j^{(i)})|j=1,\ldots,\Gamma_i\right\}$$ for $\qwwi$ with stochastic decoders.  We will now show that for such a sequence of codes, $\liminf_{i \rightarrow \infty} (\lambda_{1,i}+\lambda_{2,i}) \ge 1$.

Fix an $i$. Let $\left\{(\inpdistj,P_j^{(i)})|j=1, \ldots, \Gamma_i \right\}$ be an ID code with stochastic decoder for $\qwwi$. The total variational distance between any two output distributions $\oupdistj,\oupdistk$ (for distinct messages $j$ and $k$ respectively) is related to the error probabilities  $\lambda_{1,i}, \lambda_{2,i}$ as follows.

\bikdjournal{The following set of equations is a repetition.}
    \begin{align}
        \tv(\oupdistj,\oupdistk) & = \frac{1}{2} \sum_{\vw \in \weightqn} \left|\oupdistj(\vw)-\oupdistk(\vw)\right| \notag \\
	&\ge \frac{1}{2}  \sum_{\vw \in \weightqn} P_j^{(i)}(1|\vw) \left|\oupdistj(\vw)-\oupdistk(\vw)\right| \notag \\
        &\ge \frac{1}{2}  \bigg(\sum_{\vw \in \weightqn} P_j^{(i)}(1|\vw)\oupdistj(\vw) 
        -  \sum_{\vw \in \weightqn} P_j^{(i)}(1|\vw)\oupdistk(\vw) \bigg)  \notag \\
	&=\frac{1}{2}\left(1-\lambda_{i,j\not\rightarrow j}-\lambda_{i,k\rightarrow j} \right)\\
	&\ge \frac{1}{2} \left( 1-\lambda_{1,i} - \lambda_{2,i} \right), \label{eq:rmk1_1}
    \end{align}
i.e.,
\begin{align} \label{eq: lambda_n and TV distance2}
        \lambda_{1,i}+\lambda_{2,i} \ge 1-2\tv(\oupdistj,\oupdistk), \hspace{.2cm} \forall j \neq k, 
    \end{align}

By denoting $M_i=\frac{c_{i}}{2(q-1)} \left(\sqrt{ n_i (\log n_i)^{(q-2)}}\right)^{(q-1)}$ and using Proposition~\ref{prop:approx_qary}, for every $j \in [1:\Gamma_i]$, there exists an 
$M_i$-type distribution $\inpdistapproxj$ such that
\begin{align}
    \tv (\oupdistj, \outapproxj) \le \delta_i
\end{align}
where 
\begin{align}
    \delta_i &=  \frac{C_4'}{c_{i}^{1/(q-1)}} + 2^{q-1}C_4' \cdot \frac{1}{c_{i}^{1/(q-1)}},\label{eq:delta_bound_strong_oneshot_qary}
\end{align}
and $C_4'$ is a constant,
and $\outapproxj$ is the output distribution corresponding to $\inpdistapproxj$.
Clearly we have  $\delta_{i}\rightarrow 0$ as $i \rightarrow \infty$ because $c_{i} \rightarrow \infty$ as $i \rightarrow \infty$.

The number of distinct $M_i$-type distributions on $\setofweights{q}{n_i}$ is upper bounded, by using \eqref{eq:Nbounds1}, as
\begin{align}
    |\setofweights{q}{n_i}|^{M_i} &\le \left((2n_i)^{q-1}\right)^{\frac{c_{i}}{2(q-1)} {n_i}^{(q-1)/2} \cdot  \left( \log n_i\right)^{\frac{(q-1)(q-2)}{2}}}\\
    &=2^{(1+\log n_i)\frac{c_{i}}{2} {n_i}^{(q-1)/2} \cdot  \left( \log n_i\right)^{\frac{(q-1)(q-2)}{2}}}\\
    &\le 2^{c_{i}n_i^{(q-1/2)}(\log n_i)^{1+\frac{(q-1)(q-2)}{2}}}
\end{align}

Hence, $\exists I$ such that for each $i>I$, there exist at least one pair $(j,k)$ with $j\neq k$ and $\inpdistapproxj=\inpdistapproxk$ and  we have
 \begin{align}
	\tv(\oupdistj,\oupdistk)&\le \tv\left(\oupdistj,\outapproxj\right)+\tv\left(\oupdistk,\outapproxj\right)\notag \\
	&= \tv\left(\oupdistj,\outapproxj\right)  + \tv\left(\oupdistk,\outapproxk\right)\notag \\
	&\le 2 \delta_i. \label{eq:TV_lower_bound}
\end{align}
By combining \eqref{eq: lambda_n and TV distance2} and \eqref{eq:TV_lower_bound} we have
\begin{align}\label{eq:rmk1_2}
    \lambda_{1,i}+\lambda_{2,i} \ge 1-4\delta_{i}.
\end{align}
Since $\delta_{i}\rightarrow 0$ as $i \rightarrow \infty$, it follows that
\begin{align}\label{eq:rmk1_3}
    \liminf_{i \rightarrow \infty} (\lambda_{1,i}+\lambda_{2,i}) \geq 1.
\end{align}
This completes the proof of Theorem~\ref{thm:stoch}, part (ii).

\section{Some results on multinomial distributions and proof of Lemma~\ref{lemma:D_M}}\label{sec:multinomial}

Multinomial distributions play a crucial role in the proof of Theorem~\ref{thm:stoch}, part (ii).
We now present some results on the multinomial distribution. Due to our assumption that the transition probability matrix $\dmcmatrix$ of the DMC has only non-zero entries, the relevant multinomial distributions also have full support success probability distributions. However, we prove some of the results (e.g. Lemma~\ref{lemma:D_M}) without making this assumption and are applicable for general support, i.e. without requiring the success probability distribution to have full support.  We begin with a simple upper bound on the multinomial distribution. 

\begin{lemma}\label{lemma:upper bound mode multinomial1}
Consider positive integers $n$ and $q \ge 2$, any  probability distribution $\vu=(u_1, \ldots,u_q)$ over $[1:q]$. Let
$\left(\multinomial{n}{\vu}(\vt)\right)_{\vt \in \weightqn}$ be a multinomial distribution over $\weightqn$. 
For large $n$, we have
\begin{align}
    \multinomial{n}{\vu}(\vt ) \le 2^{-n \kl\left(\frac{\vt}{n}|| \vu\right)} \times \sqrt{n} \cdot  \left(\prod_{i \in \supp{\vt}} t_i\right) ^{-\frac{1}{2}}, \quad \forall \vt \in \weightqn. 
\end{align}
\end{lemma}

As a special case, Lemma~\ref{lemma:upper bound mode multinomial1} gives a bound on the binomial distribution:
\begin{align}
        \binpt \le  \frac{\sqrt{n}}{\sqrt{t(n-t)}} \cdot 2^{-n\cD_{KL}\left(\frac{t}{n}||p)\right)} \text{ for all } t\in [0:n], 0 \le p \le 1.
\end{align}

The proof of Lemma~\ref{lemma:upper bound mode multinomial1} is presented in Appendix~\ref{app:upper_bound_multinomial}. The following corollary on the peak value of $\multinomial{n}{\vu}(\vt)$ follows from Lemma~\ref{lemma:upper bound mode multinomial1}.  

 \begin{corollary}[Peak value of a multinomial distribution]\label{lemma:upper bound mode multinomial}
For any  positive integers $n$, $q \ge 2$, $k \in [1:q]$, and any  probability vector $\vu=(u_1, \ldots,u_q)$ such that $|\supp{\vu}|=k$, the multinomial distribution $\multinomial{n}{\vu}(\vt)$ is upper bounded as
\begin{align}\label{eq:multinomial_peak}
    \multinomial{n}{\vu}(\vt)
    \le \frac{2}{n^{(k-1)/2} \sqrt{\prod_{i \in \supp{\vu}}} u_i}, \hspace{8mm} \forall \vt \in \weightqn
\end{align}
for large $n$.
\end{corollary}

The proof of the corollary is given in Appendix~\ref{app:mode_multinomial}.
As a special case of Corollary~\ref{lemma:upper bound mode multinomial}, for $p\neq 0,1$, the peak value $(\bnpmax)$ for binomial distribution can be bounded as:
\begin{align}\label{eq:binomial_peak}
    \bnpmax \le \frac{1}{\sqrt{n}}\times \frac{2}{\sqrt{p(1-p)}}.
\end{align}

If any $t_i$ significantly deviates from its mean value, as a corollary, Lemma~\ref{lemma:upper bound mode multinomial1} implies at least a power law decay of the multinomial distribution.

\begin{corollary}\label{cor:tailbound}
Consider any real number $K>1$, any  positive integer  $q \ge 2$, any  probability distribution $\vu=(u_1, \ldots,u_q)$. If $\vt\in \weightqn$ is such that $|t_i - (n+1)u_i|\geq K\sqrt{n\log n}$ for some $i$, then for large enough $n$,
\begin{align}
    \multinomial{n}{\vu}(\vt ) \le n^{-\frac{1}{2}(K^2-1)}.
\end{align}
\end{corollary}
The proof of the corollary is given in Appendix~\ref{app:tailbound}.

The following lemma provides an upper bound on the sum of the successive differences of binomial distribution, which is same as $D_{12}(\multinomial{n}{(1-p,p)})$. This is a special case of Lemma~\ref{lemma:D_M}, and it is used in the proof of Lemma~\ref{lemma:D_M}.

\begin{lemma}\label{lemma:consequtive_binomial}
For any positive integer $n \ge 1$, and a real number $0<p<1$, we have the following bound on the sum of successive difference of binomial distribution:
    \begin{align}
        \sum_{t=0}^{n+1} |B_{n,p}(t-1)-B_{n,p}(t)|
        \le  \frac{1}{\sqrt{n}} \cdot \frac{4}{\sqrt{p(1-p)}}.
    \end{align}
\end{lemma}

\begin{IEEEproof}
It is well known that binomial distributions are strongly unimodal, that is, the convolution of a binomial distribution with any unimodal distribution also gives a unimodal distribution~\cite[Lemma 1]{keilson1971some}.  The support of a unimodal distribution is the union of two intervals, with one where the distribution is monotonically non-decreasing and the other where the distribution is monotonically non-increasing. Hence, the sum of all its increments and decrements, given by $\sum_{t=0}^{n+1} |B_{n,p}(t-1)-B_{n,p}(t)|$ is twice its peak value, i.e. $2\bnpmax$. 
Therefore, by \eqref{eq:binomial_peak} we have
\begin{align}
     \sum_{t=0}^{n+1} |B_{n,p}(t-1)-B_{n,p}(t)| &\le 2\bnpmax \label{eq:mode_binomial}\\
     &= \frac{1}{\sqrt{n}} \cdot \frac{4}{\sqrt{p(1-p)}}.
\end{align}
This proves the lemma.
\end{IEEEproof}

\begin{IEEEproof}[Proof of Lemma \ref{lemma:D_M}]
First consider the case when $u_a=u_b=0$. The summand in the sum
\begin{align}\label{eq:DM1}
    D_{ab}(\multinomial{n}{\vu}) &= \sum_{\vt\in \setofweights{q}{n+1}}\left| \multinomial{n}{\vu}(\vt-\ve_{a})- \multinomial{n}{\vu}(\vt-\ve_{b})\right|
\end{align}
is non-zero only for $(t_a,t_b)=(1,0)$ or $(0,1)$. We now define $\tilde{\vt}$ as a $q-2$-length vector with component indices $[1:q]\setminus \{a,b\}$ and components $\tilde{t}_c=t_c$ for all $c\in [1:q]\setminus \{a,b\}$. Similarly $\tilde{\vu}$ is also defined to be the truncated vector obtained from $\vu$ by removing the $0$-valued components at $a,b$. We can then write
\begin{align*}
    D_{ab}(\multinomial{n}{\vu}) &= \sum_{\tilde{\vt}\in \setofweights{q-2}{n}}\multinomial{n}{\tilde{\vu}}(\tilde{\vt})\cdot 2\\
    &=2.
\end{align*}
This is obtained by observing that for a given $\tilde{\vt}$, there are two values of $\vt$, obtained by appending $(1,0)$ and $(0,1)$, for which the summand in \eqref{eq:DM1} is nonzero, and the value of the summand is exactly $\multinomial{n}{\tilde{\vu}}(\tilde{\vt})$.

We now consider the case $(u_a,u_b)\neq (0,0)$, and define $p=\frac{u_a}{u_a+u_b}$. We define a new multinomial distribution $\multinomial{n}{\vu'}$. This distribution corresponds to $n$ independent trials, each with distribution $\vu'$ over $q-1$ categories indexed with the set $\cA':=([1:q]\setminus \{a,b\})\cup \{ab\}$. This distribution represents the same experiment as that of $\multinomial{n}{\vu}$, but the categories $a,b$ are counted as a single category indexed with $ab$. The probability distribution $\vu'$ is such that $u_c'=u_c$ for $c\neq a,b$ and $u'_{ab}=u_a+u_b$.
We note that
\begin{align*}
\multinomial{n}{\vu} (\vt) & = \multinomial{n}{\vu'}(\vt')  B_{\tau,p}(t_a),
\end{align*}
where $\vt' \in \setofweights{q-1}{n+1}$ is defined to be a $q-1$ length vector with index set $\cA'$, $t_c'=t_c$ for $c\neq a,b$ and $t_{ab}=t_a+t_b$.

Now consider any $\vt \in \setofweights{q}{n+1}$. We define $\tau = t_a+t_b-1$, and $\vt' \in \setofweights{q-1}{n}$ with $t_c'=t_c$ for $c\neq a,b$ and $t'_{ab}=\tau$. Note that both $\tau$ and $\vt'$ are functions of $\vt$. 
\begin{align}
    |\multinomial{n}{\vu}(\vt -\ve_a)-\multinomial{n}{\vu}(\vt- \ve_b)|
    &=\multinomial{n}{\vu'}(\vt')|B_{\tau,p}(t_a-1)-B_{\tau,p}(t_a)|.
\end{align}
Therefore, we can write
\begin{align*}
    D_{ab}(\multinomial{n}{\vu}) &= \sum_{\vt\in \setofweights{q}{n+1}}\left| \multinomial{n}{\vu}(\vt-e_{a})- \multinomial{n}{\vu}(\vt-e_{b})\right|\\
    &=\sum_{\vt' \in \setofweights{q-1}{n}} \multinomial{n}{\vu'}(\vt') \sum_{t_a=0}^{t_{ab}'+1} |B_{t_{ab}',p}(t_a-1)-B_{t_{ab}',p}(t_a)|.
\end{align*}
If exactly one of $a$ and $b$ is in $\supp{\vu}$, then $p=0$ or $1$. In that case, $B_{t_{ab}',p}$ is a point mass, and hence the inner sum above is $2$. Hence, in this case, 
\begin{align*}
    D_{ab}(\multinomial{n}{\vu}) &= 2\sum_{\vt' \in \cW_{q-1,n}} \multinomial{n}{\vu'}(\vt') =2.
\end{align*}
Now, we consider the case when $a,b\in \supp{\vu}$. Note that this implies $ab \in \supp{\vu'}$.
Fix a constant $K= \sqrt{2q+3}$ and define a subset of $\setofweights{q-1}{n}$ as
\begin{align*}
\cN_{central} :=\{\vt'\in \setofweights{q-1}{n}||t_{c}' - (n+1)u'_c|\leq K\sqrt{n\log n} \;\;\forall c \in \supp{\vu'}, t_j=0 ~ \forall j \in \cA' \setminus \supp{\vu'}\}.
\end{align*}
Then
\begin{align}
    D_{ab}(\multinomial{n}{\vu}) 
    &=\sum_{\vt' \in \cN_{central}} \multinomial{n}{\vu'}(\vt') \sum_{t_a=0}^{t_{ab}'+1} |B_{t_{ab}',p}(t_a-1)-B_{t_{ab}',p}(t_a)| \notag  \\
    &\hspace{5mm}+ \sum_{\vt' \not\in \cN_{central}} \multinomial{n}{\vu'}(\vt') \sum_{t_a=0}^{t_{ab}'+1} |B_{t_{ab}',p}(t_a-1)-B_{t_{ab}',p}(t_a)|.\label{eq:D_M_combined}
\end{align}

We upper bound the first sum as
\begin{align*}
& \hspace*{-15mm} \sum_{\vt' \in \cN_{central}} \multinomial{n}{\vu'}(\vt') \sum_{t_a=0}^{t_{ab}'+1} |B_{t_{ab}',p}(t_a-1)-B_{t_{ab}',p}(t_a)| \\
&\stackrel{(a)}{\leq} \sum_{\vt' \in \cN_{central}} \frac{2}{n^{(|\supp{\vu}|-2)/2} \sqrt{\prod_{i \in \supp{\vu}} u_i}} \cdot \frac{1}{\sqrt{t'_{ab}}} \cdot \frac{4}{\sqrt{p(1-p)}} \quad \\
&\stackrel{(b)}{\leq} (2K\sqrt{n\log n})^{|\supp{\vu}|-2} \cdot \frac{2}{n^{(|\supp{\vu}|-2)/2} \sqrt{\prod_{i\in \supp{\vu}} u_i}} \cdot \frac{\sqrt{2}}{\sqrt{nu_{ab}'}} \cdot \frac{4}{\sqrt{p(1-p)}}\\
&=\frac{(2K)^{(|\supp{\vu}|-2)}2^{7/2}}{\sqrt{p(1-p)u'_{ab}}\sqrt{\prod_{i \in \supp{\vu}} u_i}} \cdot \frac{(\log n)^{\frac{|\supp{\vu}|-2}{2}}}{\sqrt{n}}.
\end{align*}
Here $(a)$ follows from Corollary \ref{lemma:upper bound mode multinomial} and Lemma \ref{lemma:consequtive_binomial}, $(b)$ follows because $|\cN_{central}|\leq (2K\sqrt{n\log n})^{|\supp{\vu}|-2}$ and $\vt'\in \cN_{central}$, $t_{ab}' \geq n u_{ab}'/2$ for large enough $n$.

Now for the second summand in \eqref{eq:D_M_combined}, we consider the following two sub-cases:

{\it Case 1}: We consider the set of  $\vt' \not \in \cN_{central}$ such that $t'_j=0$ for all $j \in \cA' \setminus \supp{\vu'}$ and there exists at least one  $c \in \supp{\vu'}$ such that 
\begin{align}
    |t'_c-(n+1)u'_c|> K\sqrt{n \log n}.
\end{align}

We can write
\begin{align*}
    &\sum_{\substack{\vt' \not\in \cN_{central} \\ \exists c \in \supp{\vu'}: |t'_c-(n+1)u'_c|> K\sqrt{n \log n} \\{t'_j =0 ~\forall j \in \cA'\setminus \supp{\vu'}}}} \multinomial{n}{\vu'}(\vt') \sum_{t_a=0}^{t_{ab}'+1} |B_{t_{ab}',p}(t_a-1)-B_{t_{ab}',p}(t_a)| \\
    & \quad \le \sum_{\substack{\vt' \not\in \cN_{central} \\ \exists c \in \supp{\vu'}: |t'_c-(n+1)u'_c|> K\sqrt{n \log n} \\{t'_j =0 ~\forall j \in \cA'\setminus \supp{\vu'}}}} \multinomial{n}{\vu'}(\vt') \cdot 2 &\text{(by \eqref{eq:mode_binomial} and $ \bnpmax \le 1$)}\\
&\quad \stackrel{}{\le} 2|\setofweights{q-1}{n}|\cdot n^{-\frac{1}{ 2}(K^2-1)} &\text{(by Corollary \ref{cor:tailbound})}\\
    &\quad \stackrel{(a)}{\le} 2^{q-1}\cdot n^{-\left(\frac{1}{2}(K^2-1)-(q-2)\right)}\\
    &\quad \leq \frac{1}{n^2} \quad \text{(for large $n$)}\\
    &\quad \le \frac{(\log n)^{(|\supp{\vu}|-2)/2}}{\sqrt{n}}.
\end{align*}
Here $(a)$ follows \eqref{eq:Nbounds1} which implies $ |\setofweights{q-1}{n}|\leq (2n)^{q-2}$.

{\it Case 2}: We now consider the set of $\vt' \not \in \cN_{central}$ such that there exists at least one  $j \in \cA' \setminus \supp{\vu'}$ where $t'_j \neq 0$. Then we have
$\multinomial{n}{\vu'}(\vt')=0$ for such $\vt'_j \not \in \cN_{central}$.
Hence 
\begin{align}
    \sum_{\substack{\vt' \not\in \cN_{central} \\ \exists j \in \cA' \setminus \supp{\vu'}: t'_j \neq 0 }} \multinomial{n}{\vu'}(\vt') \sum_{t_a=0}^{t_{ab}'+1} |B_{t_{ab}',p}(t_a-1)-B_{t_{ab}',p}(t_a)|&=0.
\end{align}

Hence by combining the two sums in \eqref{eq:D_M_combined}, we get
\begin{align}
    D_{ab}(\multinomial{n}{\vu}) 
    &\le \frac{(2K)^{(|\supp{\vu}|-2)}2^{7/2}}{\sqrt{p(1-p)u'_{ab}}\sqrt{\prod_{i \in \supp{\vu}} u_i}} \cdot \frac{(\log n)^{\frac{|\supp{\vu}|-2}{2}}}{\sqrt{n}}
    + \frac{(\log n)^{\frac{|\supp{\vu}|-2}{2}}}{\sqrt{n}}+0\\
    &=\frac{(\log n)^{\frac{|\supp{\vu}|-2}{2}}}{\sqrt{n}}\left(1+ \frac{(2K)^{(|\supp{\vu}|-2)}2^{7/2}}{\sqrt{p(1-p)u'_{ab}}\sqrt{\prod_{i \in \supp{\vu}} u_i}} \right)
\end{align}
This proves the lemma.
\end{IEEEproof}

\section{Effect of a change in the input composition distribution on the output composition distribution, and Proofs of Lemma \ref{lemma:dTV_singleW} and Lemma \ref{lemma:weight_transfer_pairs}}\label{sec:unit_change}

In this section, we analyze the TV distance between the output distributions over $\qww$ corresponding to two point mass input distributions on two distinct input compositions. We introduce several auxiliary lemmas to prove Lemma \ref{lemma:dTV_singleW} and Lemma \ref{lemma:weight_transfer_pairs}.
We first prove an elementary result on the convolution of a probability distribution with a finite sequence.

\begin{lemma}\label{lemma:conv_two_seq}
    For a positive integer $n$, let $q(\vt)$ be a probability mass function  and $x(\vt)$ be a sequence  over $\mathbb{Z}^n$. Define
    \begin{align*}
        y(\vt) & = x(\vt) * q(\vt).
    \end{align*}
     If $x(\vt)$ is absolutely summable, then
    \begin{align*}  
 & \sum_t |y(\vt)| \leq \sum_{t} |x(\vt)|.
    \end{align*}
\end{lemma}

\begin{IEEEproof}
By definition of convolution, we have
\begin{align*}
    \sum_{\vt} |y(\vt)| &= \sum_{\vt} \left|\sum_{\vl}  q(\vl) x(\vt-\vl)\right|\\
    & \leq \sum_\vt \sum_{\vl} q(\vl) |x(\vt-\vl)|\\
    &=\sum_{\vl} q(\vl) \sum_\vt  |x(\vt-\vl)|\\
    &=\sum_{\vl} q(\vl) \sum_\vt |x(\vt-\vl)|\\
    & \le \sum_{\vt} |x(\vt)|
\end{align*}
This proves the result. 
\end{IEEEproof}

When $x(\vt)$ is bounded, it can also be shown that 
$\max_\vt |y(\vt)| \leq \max_{\vt} |x(\vt)|$. We now use Lemma \ref{lemma:conv_two_seq} to present an upper bound for $D_{ab}(W_{\vx})$.

Recall that
\begin{align}
\dmcmatrix= \begin{pmatrix}
        u_{1,1}&u_{1,2}&\cdots & u_{1,q}\\
        u_{2,1}&u_{2,2}&\cdots & u_{2,q}\\
        \vdots&\vdots&\ddots & \vdots \\
        u_{q,1}&u_{q,2}&\cdots & u_{q,q}\\
    \end{pmatrix}=\begin{pmatrix}
        \vu_{1}\\
        \vu_{2}\\
        \vdots\\
        \vu_{q}\\
    \end{pmatrix},
\end{align}
where $\forall i,j, ~~ u_{i,j}>0$, denotes the transition probability matrix of the $q$-ary DMC. We denote the $i$-th row of the channel matrix as $\vu_i$.

\begin{lemma} \label{lemma:conv_W}
    Let $q \ge 2, n$ be two positive integers, and $\vu$ be a distribution over $[1:q]$ with strictly positive entries. For any $\vx\in [1:q]^n$ of composition $\vt \in \weightqn$, the distribution $W_{\vx}$ of the composition of the output over a $q$-ary DMC with strictly positive entries satisfies
    \begin{align}
        D_{ab}(W_{\vx}) &\le \min_{i\in [1:q]} D_{ab}(\multinomial{t_i}{\vu_i}) \label{eq:D_ab_W1}\\
        &\le \frac{C_2(\log n)^{(q-2)/2}}{\sqrt{n}}\label{eq:D_ab_W2}.
    \end{align}
\end{lemma}
\begin{IEEEproof}
Note that $W_{\vx}$ is the convolution of multinomial distributions (see \eqref{eq:W_x_conv_multinomial})
\begin{align*}
   W_{\vx}(\vw) &= \multinomial{t_1}{\vu_1} * \multinomial{t_2}{\vu_2} * \cdots * \multinomial{t_q}{\vu_q}(\vw) \quad \forall \vw \in \weightqn.
\end{align*}

By \eqref{eq:D_ab_W}, we can write
    \begin{align*}
        D_{ab}(W_{\vx}) 
    &=\sum_{\vw\in \setofweights{q}{n+1}}|(\delta_{a,b}*W_{\vx})(\vw)|\\
    &=\sum_{\vw\in \setofweights{q}{n+1}}\left|(\delta_{a,b}*\multinomial{t_1}{\vu_1} * \multinomial{t_2}{\vu_2} * \cdots * \multinomial{t_q}{\vu_q})(\vw)\right|\\
    &=\sum_{\vw\in \setofweights{q}{n+1}}\left|
    (x^{(i)} * q^{(i)})(\vw)
    \right|
    \end{align*}
    where  $x^{(i)}(\vw)=\left(\delta_{a,b}*\multinomial{t_i}{\vu_i} \right) (\vw)$ and $$q^{(i)}(\vw)= \left(\multinomial{t_1}{\vu_1} * \cdots*\multinomial{t_{i-1}}{\vu_{i-1}}* \multinomial{t_{i+1}}{\vu_{i+1}} * \cdots* \multinomial{t_{q}}{\vu_{q}}\right)(\vw).$$
    Therefore by Lemma~\ref{lemma:conv_two_seq}
    \begin{align}
        D_{ab}(W_{\vx}) \le \sum_{\vw\in \setofweights{q}{n+1}}\left|x^{(i)}(\vw)\right|
    \end{align}
    Hence
    \begin{align}
        D_{ab}(W_{\vx}) &\le \min_{i \in [1:q]} \sum_{\vw\in \setofweights{q}{n+1}}\left|
    x^{(i)}(\vw)\right|\\
    &=\min_{i \in [1:q]} \sum_{\vw\in \setofweights{q}{n+1}}\left|\left(\delta_{a,b}*\multinomial{t_i}{\vu_i} \right) 
    (\vw)\right|\\
    &= \min_{i \in [1:q]}  D_{ab}(\multinomial{t_i}{\vu_i}).
    \end{align}
    This proves the first part of the lemma, i.e. \eqref{eq:D_ab_W1}. Since $\max \{t_1,t_2,\cdots,t_q\}\geq n/q$ for large enough $n$, the second part \eqref{eq:D_ab_W2} follows from Lemma~\ref{lemma:D_M}.
\end{IEEEproof}

Recall that for two distributions $\vu,\vv$ over $[1:q]$, the sets $\splus{\vu}{\vv}$ and $\sminus{\vu}{\vv}$ are the indices where $\vu$ dominates $\vv$ and vice-versa. 
Since these sets are disjoint subsets of $[1:q]$, it holds that
\begin{align}
    0 \le |\splus{\vu}{\vv}|+|\sminus{\vu}{\vv}| \leq q.
\end{align}
Since both $\vu$ and $\vv$ are distributions, we have 
$$\sum_{c \in \splus{\vu}{\vv} } (u_{c}-v_{c})=-\sum_{c \in \sminus{\vu}{\vv} } (u_{c}-v_{c}).$$
Clearly,
\begin{align}
    \tv(\vu,\vv) & =\frac{1}{2}\sum_{c \in [1:q]} |u_{c}-v_{c}|\\
    &= \frac{1}{2} \left( \sum_{c\in \splus{\vu}{\vv}} |u_{a}- v_{c}|
    +\sum_{c\in \sminus{\vu}{\vv}} |u_{a}- v_{c}| \right)\\
    &=\sum_{c\in \splus{\vu}{\vv}} |u_{c}- v_{c}|.
\end{align}

\begin{lemma}\label{lemma:tvatom}
For $q \ge 2$ and two distributions $\vu,\vv$ over $[1:q]$, there exists $$\atomno \le |\splus{\vu}{\vv}|+|\sminus{\vu}{\vv}|-1 $$ pairs 
$$
(c_1,c'_1),(c_2,c'_2),\cdots,(c_{\atomno},c'_{\atomno})\in \splus{\vu}{\vv}\times \sminus{\vu}{\vv}
$$ 
and constants $\tvatom_1,\tvatom_2,\cdots,\tvatom_{\atomno}>0$, such that 
\begin{align}
    \vv&=\vu +\sum_{i=1}^{\atomno}\tvatom_i \delta_{\ve_{c_i} \rightarrow \ve_{c'_i}}. \label{eq:ub=ua+weight}
\end{align}
As a consequence, we have 
\begin{align}
    & \tv (\vu,\vv)  = \sum_{i=1}^{\atomno} \tvatom_i\\
    &  v_{c} =\begin{cases}u_{c}  + \sum_{i: c_i'=c} \tvatom_i &;  c\in \sminus{\vu}{\vv}\\
    u_{c}  - \sum_{i: c_i'=c} \tvatom_i &;  c\in \splus{\vu}{\vv}\\
    u_{c} &; c\not\in \sminus{\vu}{\vv}\cup \splus{\vu}{\vv}
    \end{cases}.\label{eq:sum_delta_i-}
\end{align}
Furthermore, for any positive integer $n \ge 1$, if $\vu$ and $\vv$ are $n$-type distributions, then $\tvatom_i$s are multiple of $1/n$ for all $i \in [1:\atomno]$.
\end{lemma}

The proof of Lemma~\ref{lemma:tvatom} is given in Appendix~\ref{app:atom}. We now proceed to prove Lemma~\ref{lemma:dTV_singleW}.

\begin{IEEEproof}[Proof of Lemma~\ref{lemma:dTV_singleW}]
We have
  \begin{align*}
    \tv\left(W_{(\vx,a)}, W_{(\vx,b)}\right) & = \frac{1}{2}\sum_{\vw\in \setofweights{q}{n+1}}   \left| W_{(\vx,a)}(\vw) -W_{(\vx,b)}(\vw)\right|  \\
    &=\frac{1}{2}\sum_{\vw\in \setofweights{q}{n+1}}\left|  \sum_{c\in [1:q]} u_{a,c} W_{\vx}(\vw-\ve_c)-  \sum_{c\in [1:q]} u_{b,c} W_{\vx}(\vw-\ve_c)\right|\\
    &=\frac{1}{2}\sum_{\vw\in \setofweights{q}{n+1}}\left|  \sum_{c\in [1:q]} (u_{a,c}-u_{b,c}) W_{\vx}(\vw-\ve_c)\right|\\
    &=\frac{1}{2}\sum_{\vw\in \setofweights{q}{n+1}}\left|  \sum_{c\in \splus{\vu_a}{\vu_b}} (u_{a,c}-u_{b,c}) W_{\vx}(\vw-\ve_c) + \sum_{c\in \sminus{\vu_a}{\vu_b}} (u_{a,c}-u_{b,c}) W_{\vx}(\vw-\ve_c) \right|
    \end{align*}
    Now, using \eqref{eq:sum_delta_i-}, we have
    \begin{align*}
    \sum_{c\in \splus{\vu_a}{\vu_b}} (u_{a,c}-u_{b,c}) W_{\vx}(\vw-\ve_c)
    &=  \sum_{c\in\splus{\vu_a}{\vu_b}} \left(\sum_{i: c_i=c} \tvatom_i\right) W_{\vx}(\vw-\ve_c) \hspace{5mm} \\
    &=  \sum_{c\in \splus{\vu_a}{\vu_b}} \sum_{i: c_i=c} \tvatom_i W_{\vx}(\vw-\ve_{c_i}) \\
    &=\sum_{i=1}^{\atomnoab} \tvatom_i  W_{\vx}(\vw-\ve_{c_i}),
    \end{align*}
    and
     \begin{align*}
    \sum_{c\in \sminus{\vu_a}{\vu_b}} (u_{a,c}-u_{b,c}) W_{\vx}(\vw-\ve_c)
    &=-\sum_{i=1}^{\atomnoab} \tvatom_i  W_{\vx}(\vw-\ve_{c_i'}).
    \end{align*}   
    Hence
    \begin{align*}
    \tv\left(W_{(\vx,a)}, W_{(\vx,b)}\right)
    &=\frac{1}{2}\sum_{\vw\in \setofweights{q}{n+1}}  \left|\sum_{i=1}^{\atomnoab
    } \tvatom_i  (W_{\vx}(\vw-\ve_{c_i})- W_{\vx}(\vw-\ve_{c_i'}))\right|\\
    &\leq \frac{1}{2}\sum_{\vw\in \setofweights{q}{n+1}}  \sum_{i=1}^{\atomnoab} \tvatom_i \left| W_{\vx}(\vw-\ve_{c_i})- W_{\vx}(\vw-\ve_{c_i'})\right|\\ 
    &= \frac{1}{2} \sum_{i=1}^{\atomnoab} \tvatom_i \left(\sum_{\vw\in \setofweights{q}{n+1}}\left| W_{\vx}(\vw-\ve_{c_i})- W_{\vx}(\vw-\ve_{c_i'})\right|\right).\\
    &= \frac{1}{2} \sum_{i=1}^{\atomnoab} \tvatom_i D_{c_ic_i'}(W_{\vx}).
\end{align*}  
Using Lemma~\ref{lemma:conv_W}, we get
\begin{align*}
    \tv\left(W_{(\vx,a)}, W_{(\vx,b)}\right) 
    & \le \frac{1}{2} \sum_{i=1}^{\atomnoab} \tvatom_i \frac{C_2(\log n)^{(q-2)/2}}{\sqrt{n}}\\
    & \le  \frac{C_2(\log n)^{(q-2)/2}}{2\sqrt{n}}. 
\end{align*} 
Here the last equality follows since $\sum_i \tvatom_i = \tv (\vu_a,\vu_b)\le 1$.
This completes the proof of the lemma by taking $C_3=C_2/2$.
\end{IEEEproof}

Recall that for any $a,b\in [1:q]; a\neq b$, $\delta_{\ve_a \rightarrow \ve_b}$ is defined as $\delta_{\ve_a \rightarrow \ve_b}(\vx):=\indicator_{\{\vx=\ve_b\}}-\indicator_{\{\vx=\ve_a\}}$ and for any two compositions $\vt,\vt'\in \setofweights{q}{n}$, their  distance $\dc(\vt,\vt')$ is given by
\begin{align*}
    \dc(\vt,\vt'):=\frac{1}{2}\sum_{i=1}^q \left|t_i-t_i'\right|
    =n\,\tv \left(\frac{1}{n}\vt,\frac{1}{n}\vt'\right).
\end{align*}

Two compositions $\vt,\vt'\in \setofweights{q}{n}$ are said to be neighbours if they have distance $1$. A neighbour $\vt$ can be obtained from another type $\vt'$ by moving an unit mass of from one component to another,  i.e. $\vt=\vt'+\delta_{\ve_a \rightarrow \ve_b}$ for some $a,b\in [1:q]; a\neq b$. 
For any two compositions $\vt,\vt'\in \weightqn$, we can write using Lemma \ref{lemma:tvatom},
\begin{align}
    &\frac{1}{n}\vt =\frac{1}{n}\vt' + \sum_{i=1}^{\nu\left(\frac{1}{n}\vt,\frac{1}{n}\vt' \right)}\tvatom_i\delta_{\ve_{c_i} \rightarrow \ve_{c'_i}},\\
    &\tv\left(\frac{1}{n}\vt,\frac{1}{n}\vt' \right) =\sum_{i=1}^{\nu\left(\frac{1}{n}\vt,\frac{1}{n}\vt' \right)}\tvatom_i.
\end{align}

Hence if $\vt$ and $\vt'$ have a distance $d$, then one can obtain $\vt$ from $\vt'$ by making $d$ hops/steps of changes, each of unit mass. 
In other words, for any $\vt,\vt'\in \weightqn$, there exists $d=n\tv (\frac{1}{n}\vt,\frac{1}{n}\vt')$ pairs $$(a_1,b_1),(a_2,b_2)\cdots,(a_{d},b_{d})\in [1:q]^2,$$ such that 
\begin{align}
    \vt &=\vt'+\delta_{\ve_{a_1} \rightarrow \ve_{b_1}} +  \delta_{\ve_{a_2} \rightarrow \ve_{b_2}} +\cdots+ \delta_{\ve_{a_{d}} \rightarrow \ve_{b_{d}}} \label{eq:step-change}
\end{align}
We now prove the following Lemma. 

\begin{lemma}\label{lemma:weight_t_1t_2}
    (i) Let $\vt_1,\vt_2\in \weightqn$ be neighbors, i.e. $\vt_1=\vt_2+\delta_{\ve_{a} \rightarrow \ve_{b}}$ for some $a,b\in [1:q]; a\neq b$. Let $W_{\vt_1}$ and $W_{\vt_2}$ be the output distributions over $\qww$ corresponding to the input composition distributions $\delta_{\vt_1}$ and $\delta_{\vt_2}$ respectively. Then
    \begin{align*}
        \tv (W_{\vt_1},W_{\vt_2}) &\leq \frac{C_4(\log (n))^{(q-2)/2}}{\sqrt{n}},
    \end{align*}
    where $C_4$ is a constant independent of $n$. \\(ii) For any compositions $\vt_1,\vt_2\in \weightqn$, 
    \begin{align*}
        \tv (W_{\vt_1},W_{\vt_2}) &\leq \frac{C_4(\log (n))^{(q-2)/2}}{\sqrt{n}}\cdot \dc(\vt_1,\vt_2),
    \end{align*}
    where $C_4$ is a constant independent of $n$.
\end{lemma}
\begin{IEEEproof}
(i) Since $\vt_1=\vt_2+\delta_{\ve_{a} \rightarrow \ve_{b}}$, there exists a vector $\vx\in [1:q]^{n-1}$ such that $(\vx,a)$ and $(\vx,b)$ have types $\vt_1$ and $\vt_2$ respectively. Hence by Lemma~\ref{lemma:dTV_singleW},
\begin{align*}
\tv (W_{\vt_1},W_{\vt_2}) &= \tv (W_{(\vx,a)},W_{(\vx,b)})\\
&\leq \frac{C_3(\log (n-1))^{(q-2)/2}}{\sqrt{n-1}}\\
&\leq \frac{C_4(\log (n))^{(q-2)/2}}{\sqrt{n}},
\end{align*}
for large enough $n$, for some constant $C_4$. This proves the first part. (ii) The second part follows from \eqref{eq:step-change} using the first part and triangular inequality.
\end{IEEEproof}
\begin{IEEEproof}[Proof of Lemma \ref{lemma:weight_transfer_pairs}]
First note that
\begin{align*}
    Q_{\text{op},1}(\vw)&=\sum_{\vt\in \weightqn} Q_1(\vt)W_{\vt}(\vw)\\ 
    Q_{\text{op},2}(\vw)&=\sum_{\vt\in \weightqn} Q_2(\vt)W_{\vt}(\vw)
\end{align*}
By \eqref{eq:distchange}, 
\begin{align*}
   Q_{\text{op},2}(\vw)&= Q_{\text{op},1}(\vw) + \sum_{i=1}^k p_i \left(W_{\vt_i'}(\vw)-W_{\vt_i}(\vw)\right) 
\end{align*}
Hence
\begin{align*}
    \tv (Q_{\text{op},1},Q_{\text{op},2})
    &= 
    \frac{1}{2} \sum_{\vw\in \weightqn} \left| \sum_{i=1}^k p_i \left(W_{\vt'_i}(\vw)-W_{\vt_i}(\vw)\right) \right| \\
    &\le \frac{1}{2} \sum_{\vw\in \weightqn}  \sum_{i=1}^k p_i \left|W_{\vt'_i}(\vw)-W_{\vt_i}(\vw) \right|\\
    &= \sum_{i=1}^k p_i \tv \left(W_{\vt_i},W_{\vt'_i}\right)\\
    &\le \frac{C_4(\log (n))^{(q-2)/2}}{\sqrt{n}}\cdot \sum_{i=1}^k p_i \dc(\vt_i,\vt'_i)
 \hspace*{5mm}\text{(by Lemma~\ref{lemma:weight_t_1t_2}(ii)) }
\end{align*}
    This proves the lemma.
\end{IEEEproof}

\section{Approximation of input distributions by $M$-type distributions and Proof of Proposition \ref{prop:approx_qary}}\label{sec:algo}

This section proposes a deterministic approximation scheme for input distributions over $\weightqn$ with $M$-type distributions. We first consider the special case of $q=2$ and provide a simple quantization scheme for this case in Subsection~\ref{subsec:binary_algo}. In the next subsections, we consider the general values of $q$.
We first define a ``cubic'' partition of $\weightqn$ and prove some basic results in Subsection~\ref{subsec:cell_partition}. We then show the existence of a Gray-like ordering of the cells in Subsection~\ref{subsec:gray_order}. Finally, the quantization scheme is presented for general $q$ in Subsection~\ref{subsec:algo}. \bikdjournal{We need to inspire the two-stage quantization scheme for general $q$ using examples.}

\subsection{Approximation scheme for the special case of $q=2$}\label{subsec:binary_algo}
For $q=2$, note that a composition vector of $\vx\in \{0,1\}^n$ is $(n-w, w)$, where $w$ is the Hamming weight of $\vx$. Hence for a fixed $n$, the composition can be specified just by the Hamming weight $w$. In other words, the set $\setofweights{2}{n} $ is isomorphic to $[0:n]$. Thus we consider input and output distributions over Hamming weights. We want to quantize a given input Hamming weight distribution $\inpdist$, so that the change in the output Hamming weight distribution due to the quantization, measured in TV distance is small. 

The proposed algorithm quantizes the values $\inpdist (i)$ in increasing order of $i$, starting from $i=0$. First it quantizes $\inpdist (0)$ below to $\inpapprox (0)=\lfloor \inpdist (0)  \rfloor_{1/M}$ and adds the extra mass $\inpdist (0) - \lfloor \inpdist(0)  \rfloor_{1/M}$ with $\inpdist (1)$. This updated mass at $i=1$ is denoted as $\tilde{Q}_{\text{ip}}(1)$. The modified values still form a distribution as the total mass has not changed. The value $\tilde{Q}_{\text{ip}}(1)$ is then quantized to $\inpapprox (1)=\lfloor \tilde{Q}_{\text{ip}}(1) \rfloor_{1/M}$, and the extra mass $\tilde{Q}_{\text{ip}}(1)-\lfloor \tilde{Q}_{\text{ip}}(1) \rfloor_{1/M}$ is added to $\inpdist (2)$. This continues until quantization is complete up to $n-1$. At this time, the modified last value $\inpdist (n) + (\tilde{Q}_{\text{ip}}(n-1)-\lfloor \tilde{Q}_{\text{ip}}(n-1) \rfloor_{1/M})$ is also a multiple of $1/M$, since the total mass $1$ is also a multiple of $1/M$.

\begin{algorithm}
\caption{Quantization Algorithm for Binary Alphabet}
\begin{algorithmic}[0]
\State \textbf{Input:} A distribution $\inpdist$ over $[0:n]$
\State \textbf{Output:} An $M$-type distribution $\inpapprox$ over $[0:n]$
\State \textbf{Initialization:} $\tilde{Q}_{\text{ip}}(0) := \inpdist(0)$
\For{$i = 0, 1, \dots, n-1$}
    \State {\it Step 1a:} $\inpapprox(i) \gets \lfloor \tilde{Q}_{\text{ip}}(i) \rfloor_{1/M}$
    \State \hspace{5mm} $\tilde{Q}_{\text{ip}}(i+1) \gets \inpdist(i+1) + \Bigl(\tilde{Q}_{\text{ip}}(i) - \inpapprox(i)\Bigr)$
\EndFor
\State {\it Step 2:} $\inpapprox(n) \gets \tilde{Q}_{\text{ip}}(n)$
\end{algorithmic}
\label{algo:quan_binary}
\end{algorithm}

After iteration $i$, the algorithm gives a modified input composition distribution
$$Q_{\text{ip}}^{(i)}=\left(\inpapprox (0),\inpapprox(1), \cdots, \inpapprox(i), \tilde{Q}_{\text{ip}}(i+1),\inpdist(i+2),\cdots,\inpdist(n)\right).$$ 
The first $i+1$ components of this distribution are multiples of $1/M$. We also define $\inpdist^{(-1)}:=\inpdist$.   At the end (of the final iteration $n-1$), the resulting distribution $\inpapprox = \inpdist^{(n-1)}$ is $M$-type. See Appendix~\ref{app:proof_proposition_algo} for a formal proof.

Suppose the output distribution over $\qww$ corresponding to the input distribution $Q_{\text{ip}}^{(i)}$ is denoted by $\outdist^{(i)}$. In particular $\outdist^{(-1)} =\outdist$ and $\outdist^{(n-1)}=\outapprox$ correspond to the input distributions $\inpdist$ and $\inpapprox$ respectively. By triangular inequality,
\begin{align*}
    \tv (\outdist, \outapprox) & =\tv(\outdist^{(-1)},\outdist^{(n-1)})\\
    & \leq \sum_{i=0}^{n-1} \tv (\outdist^{(i-1)}, \outdist^{(i)}).
\end{align*}
By Corollary~\ref{cor: consecutive shift of mass} (see Appendix \ref{app:proof_proposition_algo}), $\tv (\outdist^{(i-1)}, \outdist^{(i)})\leq \frac{C_4}{M\sqrt{n}} $. Hence $\tv (\outdist, \outapprox) \leq \frac{C_4 \sqrt{n}}{M}$.

\subsection{General $q$: Partitioning $\weightqn$ into cubic cells}\label{subsec:cell_partition}

For any given positive integer $a$, we define a partition $\cP_a(\weightqn)$ of $\weightqn$ into cubic cells of side $a$.  
Define $\nu=\lceil(n+1)/a\rceil$, which represents the number of cells along each dimension. Each cell is indexed by a $(q-1)$-tuple $\cellid \in [1:\nu]^{q-1}$, and is defined to be
\begin{align*}
    C_{\cellid} \coloneqq \{\vt\in \weightqn | (l_i-1)a\leq t_i \leq l_i a -1 \text{ for each }  i\in [1:q-1]\}.
\end{align*}
Each $C_{\cellid}$ is the intersection of a $q$-dimensional cube of side length $a$ with $\weightqn$. The collection of sets $C_{\cellid}; \cellid\in [1:\nu]^{q-1};C_{\cellid}\neq \emptyset$ defines a partition $\cP_a(\weightqn)$ of $\weightqn$.

We now present the necessary and sufficient condition for a cell $C_{\cellid}$ to be non-empty.

\begin{lemma}
    For any $\cellid=(l_1, \cdots,l_{q-1}) \in [1:\nu]^{(q-1)}$, the cell $C_{\cellid}$ is non empty if and only if 
\begin{align}
    q-1 \le \sum_{i=1}^{q-1}l_i \le \frac{n}{a}+(q-1).
\end{align}
\end{lemma}

\begin{IEEEproof}
    {\bf If part:} 
    As the cell $C_{\cellid}$ is non empty, there exists a $\vt=(t_1, \cdots, t_q)$ such that
    \begin{align} \label{eq:ti_cell}
        (l_i-1)a \le t_i \le l_ia-1,\quad \forall i \in [1:q-1],
    \end{align}
    and $t_q= n-\sum_{i=1}^{q-1}t_i$. 
    By summing \eqref{eq:ti_cell} over $i=1, \cdots, q-1$, and noting that $\vt \in \weightqn$, i.e., $\sum_{i=1}^{q}t_i=n$, we get
    \begin{align}
        n \ge \sum_{i=1}^{q-1}t_i &\ge a \sum_{i=1}^{q-1} l_i-a(q-1)\\
        \implies   \frac{n}{a}+(q-1) &\ge \sum_{i=1}^{q-1} l_i\\
        &\ge (q-1) \hspace{8mm} \text{as } l_i \in [1:\nu], \quad \forall i \in [1:q-1].
    \end{align}

{\bf Only if part:}
Assume for cell $C_{\cellid}$, we have
\begin{align}\label{eq:onlyif_nonempty_Cell}
    q-1 \le \sum_{i=1}^{q-1}l_i \le \frac{n}{a}+(q-1),
\end{align}
Consider the following $\vt \in \weightqn$:
\begin{align}
    t_i &= (l_i-1)a, \quad \forall i \in [1:q-1]\\
    t_q&= n-\sum_{i=1}^{q-1} t_i.
\end{align}
Therefore we have
\begin{align}
    n \ge t_q &= n -\sum_{i=1}^{q-1} t_i\\
    &=n- a\sum_{i=1}^{q-1} l_i+a(q-1)\\
    &\ge 0 \quad  \text{(by the upper bound from \eqref{eq:onlyif_nonempty_Cell}).}
\end{align}
Hence $\vt \in C_{\cellid}$ by the construction of $C_{\cellid}$. Hence, $C_{\cellid}$ is non-empty.
This completes the proof of the lemma.
\end{IEEEproof}

\subsection{Gray-like ordering for the set of all non-empty cells}\label{subsec:gray_order}

For any $m \ge 1$, and $ m \le s \le m \nu$, we define the following set 
\begin{align}\label{eq:Vms}
    V(m,s) \coloneqq \left\{ 
    \cellid=(l_1, l_2, \cdots, l_m)| l_i \in [1:\nu] \quad \forall i \in [1:m], \text{ and }\sum_{i=1}^{m}l_i  \le s 
    \right\}.
\end{align}
For $m=q-1$ and $s=(q-1)+\frac{n}{a}$, $V(m,s)$ gives the set of $\cellid$ for which $C_{\cellid}$ is nonempty. Note that
\begin{align}
    |V(m,s)|\leq \nu^m \label{eq:cellnumber}
\end{align}

We say any two distinct sequences $\cellid_1$ and $\cellid_2$ in $V(m,s)$ are \emph{adjacent} if either (i) all components in them are the same except for one, where their difference is exactly $1$, or (ii) all components in them are the same except for two components indexed by, say $k,j$, where $l_{1,k}=l_{2,k}+1$ and $l_{1,j}=l_{2,j}-1$. Equivalently,
\begin{align}
    \left| \sum_{i=1}^m (l_{1,i} -l_{2,i}) \right| &\le 1, \quad \text{and} \label{eq:ad1}\\
    \sum_{i=1}^m |l_{1,i}-l_{2,i}| &\le 2. \label{eq:ad2}
\end{align}

\begin{lemma}\label{lemma:adjacent_gray}
For any integers $n,q \ge 2$ and $1 \le a \le n$, suppose
$\cellid =(l_1, \cdots,l_{q-1})$ and $\cellid'=(l'_1, \cdots,l'_{q-1})$ are ``adjacent'' in $V\left(q-1,\frac{n}{a}+q-1 \right)$. Then any $\vt= \left( t_1,\cdots,t_q\right) \in C_{\cellid}$ and $\vt'= \left( t'_1,\cdots,t'_q\right) \in C_{\cellid'}$ satisfy
\begin{align}
    \dc(\vt_{\cellid}, \vt_{\cellid'}) \le \frac{a}{2}(2q+1).
\end{align}
\end{lemma}

The proof of Lemma is presented in Appendix~\ref{app:adjacent}.

We define an ordering  $g$ of $V(m,s)$ as a bijection
$$
g:[1:|V(m,s)|] \to V(m,s).
$$
For a given ordering $g$ of $V(m,s)$, its reverse ordering $\tilde{g}$ is an ordering of $V(m,s)$, given by $\tilde{g}(i)=g(|V(m,s)-i+1|)$.

{\it Gray-like ordering:} We say that an ordering $g$ is a Gray-like ordering if  
for every  $j\in [1: |V(m,s)|-1]$,
$g(j)$ and $g(j+1)$ are adjacent. 

Note that the reverse ordering of a Gray-like ordering is also a Gray-like ordering.
The following lemma show that there exists  Gray-like ordering for $V(m,s)$. 

\begin{lemma}\label{lemma:gray}
    For any $a,m,s \ge 1$ satisfying $m\le s \le m \lceil \frac{n+1}{a} \rceil$, there exists a \emph{Gray-like ordering} $g_{(m,s)}$
    of $V(m,s)$ satisfying
\begin{align}
    g_{(m,s)}(1) &= (\underbrace{1,1, \cdots,1}_{m \text{ times}}) \label{eq:ordering_1}\\
    g_{(m,s)}(|V(m,s)|)&= ((s-m+1), \underbrace{1,1,\cdots, 1}_{m-1 \text{ times}}).\label{eq:ordering_2}
\end{align}  
\end{lemma}

\begin{IEEEproof}
    
    We show by induction on the parameter $m$, that there exists a Gray-like ordering $g_{(m,s)}$ of $V(m,s)$ satisfying \eqref{eq:ordering_1} and \eqref{eq:ordering_2}.

First,  for  $m=1$ and any $s\geq 1$, $V(1,s)=\{1,\cdots,s\}$.  We define
\begin{align}
    g_{(1,s)}(j)=j, \quad \forall j \in [1:s].
\end{align}
Note that $g_{(1,s)}(1)=\{1\}$, and  $g_{(1,s)}(|V(m,s)|)=\{s\}$, thus $g_{(1,s)}$ satisfies \eqref{eq:ordering_1} and \eqref{eq:ordering_2}.

Now given $m$ and $s\in [m:m\nu]$, we assume that for every pair  $(m',s')$ such that 
$m' < m$, and $s' \in [m':m'\nu]$, 
$V(m',s')$ has a Gray-like ordering $g_{(m',s')}$ satisfying \eqref{eq:ordering_1} and \eqref{eq:ordering_2}. We will then show that $V(m,s)$ has a Gray-like ordering $g_{(m,s)}$.

For an $ x \in [1: s-m+1]$, we define the $x$-slice of $V(m,s)$ as
\begin{align}
    V^{(x)}(m,s)&=\left\{ (x, l_2, \cdots,l_m)|  l_i \in [1:\nu] \quad \forall i \in [2:m], \text{ and } \sum_{i=2}^{m}l_i  \le s-x   \right\}\\
    &=\left\{ (x,l_2, \cdots,l_m)|  (l_2, \cdots, l_m) \in V(m-1,s-x)   \right\}.
\end{align}
Note that $|V^{(x)}(m,s)|=|V(m-1,s-x)|$, and $V(m,s)=\cup_{x=1}^{s-m+1}V^{(x)}(m,s)$. Hence we have
\begin{align}
    |V(m,s)|=\sum_{x=1}^{(s-m+1)}|V(m-1,s-x)|.
\end{align}

By inductive hypothesis, for each $x\in [1:s-m+1]$,  there exists a Gray-like ordering $g_{(m-1,s-x)}$ of $V(m-1,s-x)$ satisfying \eqref{eq:ordering_1} and \eqref{eq:ordering_2}.

We inductively construct the Gray-like ordering $g_{(m,s)}$ of $V(m,s)$ by partitioning $V(m,s)$ into slices based on their first coordinate. 
By inductive hypothesis we know that for each slice, indexed by an $x \in [1:s-m+1]$, there exists a Gray-like ordering $g_x$ of $V(m-1,s-x)$. A Gray-like ordering $g$ is then constructed by arranging the slices in increasing order of $x$. Inside the slice $x=1$, the elements are ordered according to the Gray-like ordering $g_1$. For each subsequent slice $x$, the elements are ordered by the reverse ordering $\tilde{g}_x$ if in the previous slice $g_{x-1}$ was used, and vice versa. This ensures that adjacency is respected at the transition from one slice to the next.

Consider any $j\in [1:|V(m,s)|]$. There exists $x \in [1:s-m+1]$ such that 
\begin{align}\label{eq:gray_j}
    \sum_{x'=1}^{x-1}|V(m-1,s-x')|<j \le \sum_{x'=1}^{x}|V(m-1,s-x')|.
\end{align}
We define
\begin{align}
    g_{(m,s)}(j)=\begin{cases}
        \left(x,g_{(m-1,s-x)}\left( j- \sum_{x'=1}^{x-1}|V(m-1,s-x')|\right)\right), \quad \text{for $x$ odd}\\
        \left(x,g_{(m-1,s-x)}\left( \sum_{x'=1}^x |V(m-1,s-x')|-j+1
        \right)\right), \quad \text{for $x$ even}.
    \end{cases}
\end{align}

We will now show that for any $j\in [1:|V(m,s)|]$,
$g_{(m,s)}(j)$ and $g_{(m,s)}(j+1)$ are adjacent,

If $j$ satisfies strict inequality in \eqref{eq:gray_j}, i.e. 
\begin{align}
j < \sum_{x'=1}^{x}|V(m-1,s-x')|,
\end{align}
i.e., $j$ and $j+1$  lie in the range given by \eqref{eq:gray_j} for the same value of $x$ (i.e. they are in the same ``slice''), then $g_{(m,s)}(j)$ and $g_{(m,s)}(j+1)$ are adjacent because $g_{(m-1,s-x)}$ is a Gray-like ordering.

On the other hand, if $j$ satisfies equality in the upper bound in \eqref{eq:gray_j}, i.e., $j = \sum_{x'=1}^{x}|V(m-1,s-x')|$ for some $x \in [1:s-m+1]$, then  we consider two cases.

{\it Case I ($x$ is odd):} 
By construction, we have
\begin{align}
    g_{(m,s)}(j)&=(x, g_{(m-1,s-x)}(|V(m-1,s-x)|))\\
    &=(x,\, (s-x-m+2),\, \underbrace{1,1,\dots,1}_{m-2 \text{ times}}) \label{eq:Gms_odd1}
\end{align}
since $g_{(m-1,s-x)}$ satisfies \eqref{eq:ordering_2} by the induction hypothesis.
We also have
\begin{align}
    g_{(m,s)}(j+1)&=((x+1), g_{(m-1,s-x-1)}(1))\\
    &=((x+1),\, (s-x-m+1),\, \underbrace{1,1,\dots,1}_{m-2 \text{ times}} )\label{eq:Gms_odd2}
\end{align}
since $g_{(m-1,s-x-1)}$ satisfies \eqref{eq:ordering_1} by the induction hypothesis.
From \eqref{eq:Gms_odd1} and \eqref{eq:Gms_odd2}, clearly $g_{(m,s)}(j)$ and $g_{(m,s)}(j+1)$ are adjacent.

{\it Case II ($x$ is even):} For even $x$, similarly we have
\begin{align}
    g_{(m,s)}(j)&=(x,\, \underbrace{1,1,\dots,1}_{m-1})\\
    \text{and } g_{(m,s)}(j+1)&=((x+1),\, \underbrace{1,1,\dots,1}_{m-1}).
\end{align}
Therefore $g_{(m,s)}(j)$ and $g_{(m,s)}(j+1)$ are adjacent.

Hence $g_{(m,s)}$ is a Gray-like ordering of $V(m,s)$, and this proves the lemma.
\end{IEEEproof}

\subsection{Quantization Algorithm}\label{subsec:algo}

Recall that for a given $1\le a \le n$, we partitioned the set of all compositions $\weightqn$ into $(q-1)$-dimensional cubes of side length $a$.  Let $V=V\left(q-1, n/a+(q-1)\right)$ (we omit the parameters for brevity) be the set of all such non-empty cubes which covers $\weightqn$, i.e.,
\begin{align*}
    \weightqn=\cup_{\cellid \in V} C_{\cellid},
\end{align*}
and $g$ be a Gray-like ordering  of $V$. 

We propose an algorithm to quantize an input distribution on $\weightqn$ into an $M$-type distribution in two stages.
In the first stage, we redistribute the probability mass within each cell without changing the total probability mass in each cell, so that every probability mass within the cell becomes an integer multiple of $1/M$, except possibly at one preselected element per cell. 
In the second stage, we consider the cells in the sequence of a Gray-like ordering of $V$. In each cell, the unquantized probability mass at the selected element is quantized to the nearest multiple of $1/M$ below, while adding the difference to the probability mass of the selected element in the next cell.

{\bf Stage I: (\it Partial quantization using cell-wise redistribution)}
We now propose a quantization algorithm to generate an intermediate, partially quantized distribution $\inpapproxcell{a}$ of $\inpdist$.  We first select an arbitrary $\tvtu\in C_{\cellid}$ for each $\cellid \in V$.

Consider a given input distribution $\inpdist$ over $\weightqn$, and a partition $\cP_a(\weightqn)$ of $\weightqn$ into cubic cells of side $a$. Recall that 
\begin{align}
\lfloor x\rfloor_{1/M} & := \frac{\lfloor Mx \rfloor}{M}.
\end{align}
We define the residual mass in $\inpdist$ at each $\vt\in \weightqn$ as
\begin{align}\label{eq:Q_res}
    Q_{\text{ip,res}}(\vt) & = \inpdist (\vt)-\lfloor \inpdist (\vt)\rfloor_{1/M}.
\end{align}
The total residual mass in cell $\cellid \in V$ is defined as
\begin{align}\label{eq:Q_cell_res}
    Q_{\text{ip},a,\text{cell-res}}(\cellid) & = \sum_{\vt\in C_{\cellid}} Q_{\text{ip,res}}(\vt).
\end{align}
We move the residual mass at all weight vectors in a cell $C_{\cellid}$ to $\tvtu$. Hence we define the approximated distribution as
\begin{align}\label{eq:Q_hat_ipa}
    \inpapproxcell{a}(\vt) 
    &=\begin{cases}
       \lfloor \inpdist (\vt)\rfloor_{1/M} & \text{ if } \vt\in C_{\cellid}, \vt\neq \tvtu\\
       \lfloor \inpdist (\tvtu)\rfloor_{1/M}+Q_{\text{ip},a,\text{cell-res}}(\vu) & \text{ if } \vt\in C_{\cellid}, \vt= \tvtu.
    \end{cases}
\end{align}

We summarize this process in the following algorithm. 
\begin{algorithm}
	\caption{Partial Quantization of Distribution $\inpdist$}
	\begin{algorithmic}[0]
		\State \textbf{Input:} Distribution $\inpdist$ over $\weightqn$, Cells $C_{\cellid}; \cellid \in V$, an arbitrarily chosen element $\tvtu$ from every $C_{\cellid}$
		\State \textbf{Output:} Partially quantized distribution $\inpapproxcell{a}$ over $\weightqn$
		\ForAll{ $\cellid \in V$}
		\ForAll{$\vt \in C_{\cellid}$}
        \State  $\inpapproxcell{a}(\vt) \gets \left\lfloor \inpdist(\vt) \right\rfloor_{1/M}$
		\State  $Q_{\text{ip,res}}(\vt) \gets \inpdist(\vt) - \left\lfloor \inpdist(\vt) \right\rfloor_{1/M}$
		\EndFor
		\State  $Q_{\text{ip},a,\text{cell-res}}(\cellid) \gets \sum_{\vt \in C_{\cellid}} Q_{\text{ip,res}}(\vt)$
        \State  $\inpapproxcell{a}(\tvtu) \gets \left\lfloor \inpdist(\tvtu) \right\rfloor_{1/M} + Q_{\text{ip},a,\text{cell-res}}(\cellid)$
		\EndFor
	\end{algorithmic}
\end{algorithm}

The distribution $\inpapproxcell{a}$ obtained after the first stage is partially quantized, as summarised below.

\begin{lemma}\label{lemma:inter_dist}
The function $\inpapproxcell{a}$ obtained after the first stage is a distribution that satisfies the following properties.
\begin{enumerate}
    \item Cell-wise total mass is preserved. In each cell $C_{\cellid} \in\cP_a(\weightqn)$, $\sum_{\vt\in C_{\cellid}}\inpdist(\vt)=\sum_{\vt\in C_{\cellid}}\inpapproxcell{a}(\vt)$.
    \item In each cell, i.e. for every $\cellid \in V$, all the probability masses, except possibly at $\tvtu$, are multiples of $1/M$. That is, for all $\cellid \in V, \vt\in C_{\cellid} \setminus \{\tvtu\}$,  $\inpapproxcell{a}(\vt)$ is a multiple of $1/M$. 
\end{enumerate}
\end{lemma}

\begin{IEEEproof}
    Part 1 follows from the fact that in each cell, indexed by $\cellid$, the mass removed from any $\vt\in C_{\cellid}$, i.e. $Q_{\text{ip,res}}(\vt)$ is added to $\tvtu$ in the same cell.
    \begin{align*}
    \sum_{\vt\in C_{\cellid}}\inpapproxcell{a}(\vt) &= \sum_{\vt\in C_{\cellid}\setminus \{\tvtu\}} \lfloor \inpdist (\vt)\rfloor_{1/M} + (\lfloor \inpdist (\tvtu)\rfloor_{1/M}+Q_{\text{ip},a,\text{cell-res}}(\cellid)) &\textnormal{(by \eqref{eq:Q_hat_ipa})}\\
    &=\sum_{\vt\in C_{\cellid}} \lfloor \inpdist (\vt)\rfloor_{1/M} +\sum_{\vt\in C_{\cellid}} Q_{\text{ip,res}}(\vt) & \textnormal{(by \eqref{eq:Q_cell_res})}\\
    &=\sum_{\vt\in C_{\cellid}} \lfloor \inpdist (\vt)\rfloor_{1/M} +\sum_{\vt\in C_{\cellid}} (\inpdist (\vt)-\lfloor \inpdist (\vt)\rfloor_{1/M}) &\textnormal{(by \eqref{eq:Q_res})}\\
    &=\sum_{\vt\in C_{\cellid}}\inpdist(\vt)
    \end{align*}
    Part 2 is obvious from the definition of $\inpapproxcell{a}(\vt)$ in \eqref{eq:Q_hat_ipa}.
\end{IEEEproof}

{\it Distortion at the output due to cell-wise partial quantization of an input distribution:}
Suppose $\inpapproxcell{a}$ is the distribution obtained by cellular approximation of $\inpdist$ over the partition $\cP_a(\weightqn)$ of $\weightqn$. Let the output distributions corresponding to the input distributions $\inpdist$ and $\inpapproxcell{a}$ be respectively $\outdist$ and $\outapproxcell{a}$. Note that the change from $\inpdist$ to $\inpapproxcell{a}$ involves a shift of probability mass $Q_{\text{ip,res}}(\vt)$ from $\vt$ to $\tvtu$ for each $\cellid \in  V$, ${\vt\in C_{\cellid}\setminus \{\tvtu\}}$.
Hence we can write
\begin{align}
    \inpapproxcell{a}(\vt) 
    &=\inpdist(\vt)+ \sum_{\cellid \in  V }\sum_{\vt' \in C_{\cellid} \setminus \{\tvtu\}} Q_{\text{ip,res}}(\vt') \delta_{\vt' \rightarrow \tvtu} (\vt) \hspace*{6mm} \forall \vt\in \weightqn
\end{align}
Therefore by Lemma \ref{lemma:weight_transfer_pairs}, we have
\begin{align}
    \tv (\outdist, \outapproxcell{a}) &\le \frac{C_4(\log (n))^{(q-2)/2}}{\sqrt{n}}  \sum_{\cellid \in  V }\sum_{\vt' \in C_\vu\setminus \{\tvtu\}} Q_{\text{ip,res}}(\vt') \cdot \dc(\vt',\tvtu)\\
    &\le \frac{C_4(\log (n))^{(q-2)/2}}{\sqrt{n}} \sum_{\cellid \in  V} \sum_{\vt \in C_{\cellid}} Q_{\text{ip,res}}(\vt) \cdot \dc(\vt,\tvtu)
\end{align}
Since $\vt$ and $\tvtu$ are in the same cell, $\dc(\vt,\tvtu)\leq (q-1)a$. 
Hence the total variation distance between the corresponding output distributions is bounded as
\begin{align}
    \tv (\outdist, \outapproxcell{a}) &\leq \frac{C_4(\log n)^{(q-2)/2}}{\sqrt{n}} \sum_{\cellid \in  V}\sum_{\vt\in C_{\cellid}} Q_{\text{ip,res}}(\vt)  (q-1)a \\
    & = \frac{C_4(\log n)^{(q-2)/2}}{\sqrt{n}} \sum_{\cellid \in  V}  Q_{\text{ip},a,\text{cell-res}}(\cellid)\cdot (q-1)a \cdot. \label{eq:CellQuantTV}
\end{align}
Since $\sum_{\cellid} Q_{\text{ip},a,\text{cell-res}}(\cellid) \leq \sum_{\cellid} \inpdist (C_{\cellid}) \leq 1$, we have
\begin{align}
    \tv (\outdist, \outapproxcell{a}) & \leq  \frac{C_4(q-1)a(\log n)^{(q-2)/2}}{\sqrt{n}}. \label{eq:CellQuantTV1}
\end{align}

{\bf Stage II: (\it Sequential quantization in Gray-like order:)}
At the end of the first stage, all probability masses of the distribution $\inpapproxcell{a}$, other than those at $\{\tvtu|\cellid \in V\}$, are multiples of $1/M$. In the second stage, we redistribute the probability masses at $\{\tvtu|\cellid \in V\}$ in the distribution $\inpapproxcell{a}$, so as to make them all multiples of $1/M$. We take a Gray-like ordering $g$ of $V$ and consider the nonempty cells in $\cP_{a}(\weightqn)$ in that order. That is, we consider the cells indexed by $g(1),g(2),\cdots, g(|V|-1)$ in this order. 
In step $i$, we quantize the $i$-th probability mass (at $\tilde{t}_{g(i)}$) to the largest multiple of $1/M$ below and move the residual mass to the next point $\tilde{t}_{g(i+1)}$. We describe the process in detail below before presenting the algorithm.

First  $\inpapproxcell{a}(\tilde{\vt}_{g(1)})$ is quantized
below to $\intdist{} (\tilde{\vt}_{g(1)})=\lfloor \inpapproxcell{a}(\tilde{\vt}_{g(1)})  \rfloor_{1/M}$ and the extra mass $\inpapproxcell{a} (\tilde{\vt}_{{g(1)}}) - \lfloor \inpapproxcell{a}(\tilde{\vt}_{{g(1)}})  \rfloor_{1/M}$ is added with $\inpapproxcell{a} (\tilde{\vt}_{g(2)})$. This updated mass at $\tilde{\vt}_{g(2)}$ is denoted as $\tilde{Q}_{\text{ip}}(\tilde{\vt}_{g(2)})$. This completes the first step. The modified values still form a distribution as the total mass has not changed. In the second step, the value $\tilde{Q}_{\text{ip}}(\tilde{\vt}_{g(2)})$ is then quantized to $\intdist (\tilde{\vt}_{g(2)})=\lfloor \tilde{Q}_{\text{ip}}(\tilde{\vt}_{g(2)}) \rfloor_{1/M}$, and the extra mass $\tilde{Q}_{\text{ip}}(\tilde{\vt}_{g(2)})-\lfloor \tilde{Q}_{\text{ip}}(\tilde{\vt}_{g(2)}) \rfloor_{1/M}$ is added to $\inpapproxcell{a} (\tilde{\vt}_{g(3)})$, and the updated mass at $\tilde{\vt}_{g(3)}$ is denoted as $\tilde{Q}_{\text{ip}}(\tilde{\vt}_{g(3)})$. This completes step 2. This process continues until step $|V|-1$. In these steps, the values $\inpapprox(\tilde{\vt}_{g(1)}), \cdots,\inpapprox(\tilde{\vt}_{g(|V|-1)})$ have been computed, and are multiples of $1/M$. At the end, the modified last value $\tilde{Q}_{\text{ip}}(\tilde{\vt}_{g(|V|)})=\inpapproxcell{a} (\tilde{\vt}_{g(|V|)}) + (\tilde{Q}_{\text{ip}}(\tilde{\vt}_{g(|V|-1)})-\lfloor \tilde{Q}_{\text{ip}}(\tilde{\vt}_{g(|V|-1)}) \rfloor_{1/M})$ is also a multiple of $1/M$, since the total mass $1$ is also a multiple of $1/M$.

\begin{algorithm}
\caption{Sequential quantization in Gray-like order}
\begin{algorithmic}[0]
\State \textbf{Input:} A distribution $\inpapproxcell{a}$ over $\weightqn$, and a Gray-like ordering $g$ of $V$
\State \textbf{Output:} An $M$-type distribution $\intdist$ over $\weightqn$
\State \textbf{Initialization:} $\tilde{Q}_{\text{ip}}(\tilde{\vt}_{g(1)}) := \inpapproxcell{a}(\tilde{\vt}_{g(1)})$
\For{$i \in [1:|V|-1]$}
    \State {\it Step 1a:} $\intdist(\tilde{\vt}_{g(i)}) \gets \lfloor \tilde{Q}_{\text{ip}}(\tilde{\vt}_{g(i)}) \rfloor_{1/M}$
    \State \hspace{5mm} $\tilde{Q}_{\text{ip}}(\tilde{\vt}_{g(i+1)}) \gets \inpdist(\tilde{\vt}_{g(i+1)}) + \Bigl(\tilde{Q}_{\text{ip}}(\tilde{\vt}_{g(i)}) - \intdist(\tilde{\vt}_{g(i)})\Bigr)$
\EndFor
\State {\it Step 2:} $\intdist(\tilde{\vt}_{g(|V|)}) \gets \tilde{Q}_{\text{ip}}(\tilde{\vt}_{g(|V|)})$
\end{algorithmic}
\label{algo:quan_qary}
\end{algorithm}

Each iteration $i\in [1:|V|-1]$ in the algorithm gives a modified distribution, given by
\begin{align}
    \inpdist^{(i)}(\tilde{\vt}_{g(j)}):=\begin{cases}
       \intdist (\tilde{\vt}_{g(j)}) & ;1\leq j\leq i\\
       \tilde{Q}_{\text{ip}}(\tilde{\vt}_{g(i+1)}) & ;j=i+1\\
       \inpapproxcell{a} (\tilde{\vt}_{g(j)}) & ;i+1<j \le |V|, 
    \end{cases} 
\end{align}
and $\inpdist^{(i)}(\vt)=\inpapproxcell{a}({\vt})$ for $\vt \in \weightqn\setminus \{\vt_{\cellid}: \cellid \in V\}$.

We now show that the output of Algorithm \ref{algo:quan_qary} is an $M$-type distribution.

\begin{lemma}
     For any positive integers $n$, $M,q \ge 2$, and $1 \le a \le n$ suppose $\inpapproxcell{a}$ is the distribution obtained by ``cellular approximation'' of $\inpdist$ over the partition $\cP_a(\weightqn)$ of $\weightqn$.  The quantized distribution $\intdist \in \cP(\weightqn)$ given as output by {\it Algorithm \ref{algo:quan_qary}} for $\inpapproxcell{a}$ is an $M$-type distribution.
\end{lemma}

\begin{IEEEproof} 
First we show that $\intdist$ is a distribution. By Step 1a and Step 2 in Algorithm~\ref{algo:quan_qary}, $\intdist(\tilde{\vt}_{g(i)})\geq 0$ for each $i\in [1:V-1]$. From the second line in Step 1a, we have, for each $j \in [1 :i]$,
\begin{align}
    \tilde{Q}_{\text{ip}}(\tilde{\vt}_{g(j+1)})+\intdist(\tilde{\vt}_{g(j)}))&=\inpapproxcell{a} (\tilde{\vt}_{g(j+1)}) + \tilde{Q}_{\text{ip}}(\tilde{\vt}_{g(j)}) 
\end{align}
Summing both sides over $j=1\cdots,i$, we have
\begin{align}
    \sum_{j=1}^{i}\tilde{Q}_{\text{ip}}(\tilde{\vt}_{g(j+1)})+\sum_{j=1}^{i}\intdist(\tilde{\vt}_{g(j)}))&=\sum_{j=1}^{i}\inpapproxcell{a} (\tilde{\vt}_{g(j+1)}) + \sum_{j=1}^{i}\tilde{Q}_{\text{ip}}(\tilde{\vt}_{g(j)})\\
    \implies \sum_{j=1}^{i}\intdist(\tilde{\vt}_{g(j)}))+ \tilde{Q}_{\text{ip}}(\tilde{\vt}_{g(i+1)})&=\sum_{j=1}^{i}\inpapproxcell{a} (\tilde{\vt}_{g(j+1)})+ \tilde{Q}_{\text{ip}}(\tilde{\vt}_{g(1)})\\
    &=\sum_{j=1}^{i}\inpapproxcell{a} (\tilde{\vt}_{g(j+1)})+ Q_{\text{ip}}(\tilde{\vt}_{g(1)})\\
    &=\sum_{j=1}^{i+1}\inpapproxcell{a} (\tilde{\vt}_{g(j)})
\label{eq:step1a_algo_binary1}
\end{align}
Considering $i=|V|-1$ and noting that $\intdist (\tilde{\vt}_{g(|V|)})=\tilde{Q}_{\text{ip}}(\tilde{\vt}_{g(|V|)})$ (by Step 2) and $\tilde{Q}_{\text{ip}}(\tilde{\vt}_{g(1)}):= \inpapproxcell{a} (\tilde{\vt}_{g(1)})$ (by initialization), we have
\begin{align}
     \sum_{\vt \in \weightqn\setminus \{\vt_{\cellid}: \cellid \in V\}} \inpapproxcell{a}({\vt})+
     \sum_{j=1}^{|V|}\intdist(\tilde{\vt}_{g(j)}) &= \sum_{\vt \in \weightqn\setminus \{\vt_{\cellid}: \cellid \in V\}} \inpapproxcell{a}({\vt})+ \sum_{j=1}^{|V|}\inpapproxcell{a} (\tilde{\vt}_{g(j+1)})\\
    &=1.
\end{align}
Hence $\intdist$ is a distribution.
 
The distribution $\inpapprox$  is $M$-type. To see this, first note that $\inpapprox (\vt)=\inpapproxcell{a}(\vt)$ for $\vt\in \weightqn \setminus \{ \vt_{\cellid}|\cellid \in V\} $ by Lemma~\ref{lemma:inter_dist}, and these are multiples of $1/M$. Now, in the $i$-th iteration of Stage II, $\intdist (\tilde{\vt}_{g(i)})$ is assigned a value that is a multiple of $1/M$, and this value is not changed in the subsequent steps. After the last ($i=|V|-1$) iteration, $\intdist(\tilde{\vt}_{g(j)});j=1,\cdots,|V|-1$ have values that are multiples of $1/M$. So at the end of Stage II, all the values $\inpapprox (\vt)$ are multiples of $1/M$, except possibly the value of $\inpapprox (\tilde{\vt}_{g(|V|)})$.
Since $\inpapprox$ is a distribution, the last value $\inpapprox(\tilde{\vt}_{g(|V|)})$ must also be a multiple of $1/M$. This completes the proof.
\end{IEEEproof}

{\it Proof of Proposition \ref{prop:approx_qary}:}

From the definition of $\inpdist^{(i)}$, we have, for $i \in [2:|V|-1]$ and $j \in [1:|V|]$,
\begin{align}
    \inpdist^{(i)}(\tilde{\vt}_{g(j)}) - \inpdist^{(i-1)}(\tilde{\vt}_{g(j)}) & = \begin{cases}
       0 & j\leq i-1 \text{ and } j\geq i+2\\
       (\tilde{Q}_{\text{ip}}(\tilde{\vt}_{g(i)}) - \intdist(\tilde{\vt}_{g(i)})) & j=i+1\\
       -(\tilde{Q}_{\text{ip}}(\tilde{\vt}_{g(i)}) - \intdist(\tilde{\vt}_{g(i)})) & j=i
    \end{cases} , 
\end{align}
and $\tilde{Q}_{\text{ip}}(\tilde{\vt}_{g(i)}) - \intdist(\tilde{\vt}_{g(i)})=\tilde{Q}_{\text{ip}}(\tilde{\vt}_{g(i)})-\lfloor\tilde{Q}_{\text{ip}}(\tilde{\vt}_{g(i)})\rfloor_{1/M}\leq 1/M$.
Hence we can write 
\begin{align} \label{eq:Qip_i_qary}
     \inpdist^{(i)}(\vt)= \inpdist^{(i-1)}(\vt)+ (\tilde{Q}_{\text{ip}}(\tilde{\vt}_{g(i)}) - \intdist(\tilde{\vt}_{g(i)})) \delta_{\tilde{\vt}_{g(i-1)} \rightarrow \tilde{\vt}_{g(i)} }, \quad \forall \vt \in \weightqn.
\end{align}

 We denote the output distribution corresponding to the input distribution $\inpdist^{(i)}$ as $\outdist^{(i)}$. We also denote the output distribution corresponding to $\inpdist$ and $\inpapprox$ as $\outdist$ and $\outapprox$, respectively. By \eqref{eq:Qip_i_qary} and  Lemma  \ref{lemma:weight_transfer_pairs}, we have, for $i=2,..,|V|-1$,
\begin{align*}
        \tv(\outdist^{(i-1)},\outdist^{(i)}) &\leq \frac{C_4(q-1)(\log (n))^{(q-2)/2}}{\sqrt{n}} \frac{1}{M} \cdot \dc(\tilde{\vt}_{g(i-1)}, \tilde{\vt}_{g(i)} )
    \end{align*}
Since $g$ is a Gray-like ordering, by Lemma~\ref{lemma:adjacent_gray}, we have $\dc(\tilde{\vt}_{g(i-1)}, \tilde{\vt}_{g(i)} ) \le \frac{a}{2}(2q+1), \forall i \in [2:|V|-1]$. Hence
\begin{align}
     \tv(\outdist^{(i-1)},\outdist^{(i)}) \le  \frac{C'_4(\log (n))^{(q-2)/2}}{\sqrt{n}} \frac{a}{M},\label{eq:dtv_2nd}
\end{align}
where $C_4'=C_4(q-1)(2q+1)/2$.

Let us denote $\inpdist^{(0)}:=\inpapproxcell{a}$ and the corresponding output distribution $Q_{\text{op}}^{(0)}=\outapproxcell{a}$.
Note that $\inpdist^{({1})}=\inpapproxcell{a}$ and $\inpdist^{({|V|}-1)}=\intdist$, and hence $\outdist^{({|V|}-1)}=\outapprox$. Suppose the output distribution over $\qww$ corresponding to the input distribution $Q_{\text{ip}}^{(i)}$ is denoted by $\outdist^{(i)}$. We have 
\begin{align}
    \tv (\outdist, \outapprox) & \le \tv( \outdist, \outapproxcell{a})+\tv(\outapproxcell{a},\outapprox) \quad \text{(by triangular inequality)}\\
    &\le  \tv( \outdist, \outapproxcell{a})+\sum_{i={1}}^{{|V|}-1} \tv(\outdist^{(i)},\outdist^{(i-1)}) \quad \text{(by triangular inequality)} \label{eq:dtv_final1} \\
    & \leq  \frac{C_4'a(\log n)^{(q-2)/2}}{\sqrt{n}}+ \left(\frac{2n}{a}\right)^{q-1} \cdot  \frac{C_4'(\log (n))^{(q-2)/2}}{\sqrt{n}} \frac{a}{M} \label{eq:dtv_final2}
\end{align}
The first term in \eqref{eq:dtv_final2} follows from upper bounding the first term in \eqref{eq:dtv_final1} using \eqref{eq:CellQuantTV1}, and the second term in \eqref{eq:dtv_final2} follows from the upper bound in \eqref{eq:dtv_2nd} and 
\begin{align*}
    |V|=\left|V\left(q-1,q-1+\frac{n}{a}\right) \right| \le \left(\left\lceil\frac{n+1}{a} \right\rceil \right)^{q-1} \le 
\left(\frac{2n}{a}\right)^{q-1}.
\end{align*}
Consider any sequence $c_n \rightarrow \infty$, and for $a= \frac{\sqrt{n}}{c_n^{1/(q-1)} (\log n)^{(q-2)/2}}$ the first summand in \eqref{eq:dtv_final2} can be written as
\begin{align}
    \frac{C_4'a(\log n)^{(q-2)/2}}{\sqrt{n}}= \frac{C_4'}{c_n^{1/(q-1)}}.
\end{align}

By setting 
\begin{align}
    M&= c_n n^{(q-1)/2} \cdot  \left( \log n \right)^{\frac{(q-1)(q-2)}{2}}
\end{align}
in the second summand in \eqref{eq:dtv_final2}, we get
\begin{align}
    &\left(\frac{2n}{a}\right)^{q-1} \cdot  \frac{C_4'(\log (n))^{(q-2)/2}}{\sqrt{n}} \cdot \frac{a}{M}\\
    &= \frac{2^{q-1} n^{q-1} c_n (\log n)^{\frac{(q-1)(q-2)}{2}} }{n^{(q-1)/2}} \cdot \frac{C_4' (\log (n))^{(q-2)/2} }{\sqrt{n}} 
 \cdot \frac{\sqrt{n}}{c_n n^{(q-1)/2}\left( \log n \right)^{\frac{(q-1)(q-2)}{2}}c_n^{1/(q-1)} (\log n)^{(q-2)/2}} \\
    &= 2^{q-1}C_4' \cdot \frac{1}{c_n^{1/(q-1)}}.
\end{align}
Hence we have
\begin{align}
    \tv (\outdist, \outapprox) \le \frac{C_4'}{c_n^{1/(q-1)}} + 2^{q-1}C_4' \cdot \frac{1}{c_n^{1/(q-1)}}, 
\end{align}
the right hand side goes to $0$ as $n \rightarrow \infty$ because $c_n \rightarrow \infty$ as $n \rightarrow \infty$. \hfill{$\blacksquare$}

\section{Proof of Theorem~\ref{thm:det}}\label{sec:proof_of det_ID}
We now prove our achievability and converse results for identification with deterministic encoder and decoders over a $q$-ary noisy permutation channel.

{\it Proof of part~(i):}
    We construct an identification (ID) code with deterministic encoder and decoders, leveraging a reliable transmission code over $\nupcn$.
Define $N=\left(\frac{n}{c\log n}\right)^{(r-1)/2}$. Then, by \cite{makur2020coding}, there exists a code for reliable transmission of messages over $\nupcn$ of length $n$, message size $N$, and maximum probability of error 
$P_{\textnormal{error}}^{n} \rightarrow 0$ (see \cite[eq. 35]{makur2020coding}). 
Let $\{(\vx_i,\cD_i)|i \in [1:N]\}$ be such code where message $i$ is encoded into  $\vx_i$, and $\cD_i$ be the disjoint decoding regions.

We define the $(n,N)$ ID code with deterministic encoder and decoders over $\nupcn$ as the same code: $\{(\vx_i,\cD_i)|i \in [1:N]\}$. That is, message $i$ is encoded to the channel input vector $\vx_i$, and the $i$th decoder `Accepts' the message if the channel output vector $\vy$ lies in $\cD_i$. The missed detection probability for message $i$ is given as
\begin{align}
    \lambda_{i \not \rightarrow i}= \sum_{\vy \in \cD_i^c} \Psigma(\vy|\vx_i) \le P_{\textnormal{error}}^{n}.   
\end{align}
On the other hand, the probability that the $j$th decoder outputs `Accept', when message $i\neq j$ is encoded, is
\begin{align}
    \lambda_{i \rightarrow j} =\sum_{\vy \in \cD_j} \Psigma(\vy|\vx_i) \stackrel{(a)}{\le} \sum_{\vy \in \cD_i^c} \Psigma(\vy|\vx_i) \le P_{\textnormal{error}}^n.
\end{align}
Here (a) follows because $\cD_i, \cD_j$ are disjoint. Hence both Type-I and Type-II error probabilities go to $0$ asymptotically as $n \rightarrow \infty$. This proves the achievability.

{\it Proofs of parts~(ii),(iii):}
    
Let us consider a given sequence of $$\left(n_i, c_in_i^{(q-1)/2}(\log n_i)^{(q-1)(q-2)/2}, \lambda_{1,i}, \lambda_{2,i}\right)$$ ID codes with deterministic encoder decoders for  $\Sigma_{n_i,U}$  with  $n_i \rightarrow \infty$ as $i \rightarrow \infty$. For the weak converse, we assume that $c_i=R$ is a constant, and for the strong converse, we assume that $c_i\rightarrow \infty$.
By Lemma~\ref{lemma: noisy permutation to noisy type} (in Sec~\ref{sec:proof_of_theorem_strong_converse_qary}), for every $i$, there exists an $\left((n_i,c_in_i^{(q-1)/2}(\log n_i)^{(q-1)(q-2)/2}, \lambda_{1,i}, \lambda_{2,i}\right)$ ID code $$\left\{(\vt_j^{(i)},P_j^{(i)})|j=1,\ldots,c_in_i^{(q-1)/2}(\log n_i)^{(q-1)(q-2)/2}\right\}$$ for $\qwwi$ with a deterministic encoder and stochastic decoders. For a fixed $i$, the total variational distance between any two output distributions $\oupdistj,\oupdistk$ for any two distinct messages $j$ and $k$ respectively is bounded in terms of the error probabilities  $\lambda_{1,i}, \lambda_{2,i}$ of this code as follows.
    \begin{align}
        \tv(\oupdistj,\oupdistk) & = \frac{1}{2} \sum_{\vw \in \weightqn} \left|\oupdistj(\vw)-\oupdistk(\vw)\right| \notag \\
	&\ge \frac{1}{2}  \sum_{\vw \in \weightqn} P_j^{(i)}(1|\vw) \left|\oupdistj(\vw)-\oupdistk(\vw)\right| \notag \\
        &\ge \frac{1}{2}  \bigg(\sum_{\vw \in \weightqn} P_j^{(i)}(1|\vw)\oupdistj(\vw) 
        -  \sum_{\vw \in \weightqn} P_j^{(i)}(1|\vw)\oupdistk(\vw) \bigg)  \notag \\
	&=\frac{1}{2}\left((1-\lambda_{i,j\not\rightarrow j})-\lambda_{i,k\rightarrow j} \right)\\
	&\ge \frac{1}{2} \left( 1-\lambda_{1,i} - \lambda_{2,i} \right). \label{eq:rmk1_21}
    \end{align}
Hence we have
\begin{align} \label{eq: lambda_n and TV distance}
        \lambda_{1,i}+\lambda_{2,i} &\ge 1-2\tv(\oupdistj,\oupdistk), \quad \forall j \neq k.
\end{align}
It thus follows that
\begin{align}
     \lambda_{1,i}+\lambda_{2,i} &\ge 1- \min_{j \neq k }2\tv(\oupdistj,\oupdistk)\\
        &\ge 1- \min_{j \neq k } \frac{2C_4(\log n_i)^{(q-2)/2}}{\sqrt{n_i}}\cdot \dc(\vt^{(i)}_j,\vt^{(i)}_k) &\text{(by Lemma~\ref{lemma:weight_t_1t_2}(ii))}\\
        &=1-  \frac{2C_4(\log n_i)^{(q-2)/2}}{\sqrt{n_i}}\cdot \min_{j \neq k }\dc(\vt^{(i)}_j,\vt^{(i)}_k)\label{eq:det_er_tv}
    \end{align}  
We will now state and use a lemma. The proof of the lemma is given in Appendix~\ref{app:min_distance}.

\begin{lemma}\label{lemma:min_distance}
    For any positive integer $q \ge 2$,  $n\ge 3$, and any non-empty subset $\cA$ 
 of $\weightqn$ such that $|\cA| \ge 3^{q-1}$,
 \begin{align}
     \min_{\substack{\vt,\vt' \in \cA \\ \vt \neq \vt'}}\dc(\vt,\vt') \le 2(q-1) \frac{n}{|\cA|^{\frac{1}{q-1}}}.
 \end{align}
\end{lemma}
We use $n=n_i$, and the set of codewords of the deterministic ID code as $\cA$, i.e. $$\cA= \{ \vt^{(i)}_j | j =1, \cdots, c_i n_i^{(q-1)/2}(\log n_i)^{\frac{(q-1)(q-2)}2} \},$$ in Lemma~\ref{lemma:min_distance}.
Hence
$$
|\cA|^{\frac1{q-1}}
=\left(c_i n_i^{\frac{q-1}2} (\log n_i)^{\frac{(q-1)(q-2)}2}\right)^{\frac{1}{q-1}}
=c_i^{\frac1{q-1}} \sqrt{n_i} (\log n_i)^{\frac{(q-2)}{2}}
$$
and, by Lemma~\ref{lemma:min_distance}, we have 
\begin{align}
    \min_{\substack{j \neq k}}\dc(\vt^{(i)}_j, \vt^{(i)}_k) &\le (q-1) \frac{2n_i}{c_i^{\frac1{q-1}} \sqrt{n_i} (\log n_i)^{(q-2)/2}}\\
    &=\frac{2(q-1) \sqrt{n_i} }{c_i^{\frac{1}{q-1}}(\log n_i)^{(q-2)/2}}.
\end{align}
By \eqref{eq:det_er_tv}, we have

(i) When $c_i=R$,
\begin{align}
\liminf_{i \rightarrow \infty} \lambda_{1,i}+\lambda_{2,i}  &\ge 1- \liminf_{i \rightarrow \infty} \frac{2C_4(\log n_i)^{(q-2)/2}}{\sqrt{n_i}} \cdot \frac{2(q-1) \sqrt{n_i} }{R^{\frac{1}{q-1}}(\log n_i)^{(q-2)/2}}\\
&=1-\frac{4C_4(q-1)}{R^{\frac{1}{q-1}}}>0,
\end{align}
for any $R > (4C_4(q-1))^{(q-1)}$. This proves the weak converse.

(ii) When $c_i \rightarrow \infty$, we have using~\eqref{eq:det_er_tv},
\begin{align}
\liminf_{i \rightarrow \infty}\lambda_{1,i}+\lambda_{2,i}  &\ge 1- \liminf_{i \rightarrow \infty}\frac{2C_4(\log n_i)^{(q-2)/2}}{\sqrt{n_i}} \cdot \frac{2(q-1) \sqrt{n_i} }{c_i^{\frac{1}{q-1}}(\log n_i)^{(q-2)/2}}\\
&=1-\liminf_{i \rightarrow \infty}\frac{4C_4(q-1)}{c_i^{\frac{1}{q-1}}}=1.
\end{align}
This proves the strong converse, and completes the proof of the theorem.

\section{Conclusion}\label{sec:conclusion}
In this work, we investigated the problem of message identification over noisy permutation channels, where transmitted vectors undergo a uniform permutation followed by a $q$-ary discrete memoryless channel (DMC). We give closely matching achievability and strong converse results on the identifiable message size. Under stochastic encoding, our achievability result showed that $2^{\epsilon_n \left( \frac{n}{\log n} \right)^{(r-1)/2}}$, where $r$ denotes the rank of the DMC, messages are identifiable with vanishing error probabilities, for any sequence $\epsilon_n \rightarrow 0$. Our strong converse result showed that for any sequence of codes with message size growing as $2^{R_n n^{(q-1)/2} (\log n)^{1+\frac{(q-1)(q-2)}{2}}}$ (where $R_n \rightarrow \infty$), has sum error probability approaching one, when the constituent DMC has strictly positive entries. 
Under deterministic encoding, we showed that $\left(\frac{n}{c \log n} \right)^{(r-1)/2}$ messages can be identifiable. We proved two converses under deterministic coding: the weak converse and the strong converse. For vanishing sum error probabilities, our weak converse rules out any sequence of ID codes with size $ R n^{(q-1)/2}(\log n)^{(q-1)(q-2)/2}$ with $R>R'$ where $R'>0$ is some constant, whereas our strong converse rules out any sequence of ID code of size $R_n n^{(q-1)/2}(\log n)^{(q-1)(q-2)/2}$, where $R_n \rightarrow \infty$.
The broad proof technique for our converses uses channel resolvability.
Key innovations include a novel deterministic quantization scheme for approximating input distributions by $M$-type distributions while ensuring the total variation (TV) distance between the corresponding output distributions is small. Our achievability proof uses reliable communication code along with a result on set systems, while the converse employs careful analysis of multinomial distributions and their sensitivity to input perturbations. 
In the case of a noiseless permutation channel, the achievability and converse bounds do not have $\log n$ terms in the exponent, and hence matching more tightly. In contrast, for the noisy setting, the achievability and converse bounds differ in a power of $\log n$ in the exponent.

\bibliographystyle{unsrt}
\bibliography{bibfile_ID}

\appendices

\section{Proof of Lemma \ref{lemma: noisy permutation to noisy type}}

We will  construct an ID code $\{(Q'_i, P_i) | i=1,\ldots, M\}$ with stochastic decoders for $\qww$. 
For every $i\in [1:M]$, $\vk \in \weightqn$, we define
    \begin{align*}
        Q'_i(\vk) &\coloneqq Q_i(\compositionclass{\vk})=\sum_{\vx\in \cW_{\vk}}Q_i(\vx) ,\label{eq:det_enc}\\
        P_i(1|\vk) &\coloneqq \frac{|\mathcal{D}_i \cap \compositionclass{\vk}|}{|\compositionclass{\vk}|}.
    \end{align*}

    The probability of errors for the new code are given by
    \begin{align*}
        \tilde{\lambda}_{i \rightarrow j} &\coloneqq \sum_{\vt \in \weightqn} Q'_i(\vt)\sum_{\vw \in \weightqn} \Pqww(\vw|\vt) P_j(1|\vw)\\
        &=\sum_{\vt \in \weightqn} \sum_{\vx \in \compositionclass{\vt}} Q_i(\vx) \sum_{\vw \in \weightqn} \sum_{\vz \in \compositionclass{\vw}}  \dmcn (\vz|\vx) \frac{|\mathcal{D}_j \cap \compositionclass{\vw}|}{|\compositionclass{\vw}|} \\
        &=\sum_{\vx \in [1:q]^n } Q_i(\vx) \sum_{\vz \in [1:q]^n}  \dmcn (\vz|\vx) \frac{|\mathcal{D}_j \cap \seqofsame{\vz}|}{|\seqofsame{\vz|}
        } \\
        & = \lambda_{i\rightarrow j}, \hspace*{10mm} \text{(by \eqref{eq: error i to j})}, 
    \end{align*}
    and
    \begin{align*}
        \tilde{\lambda}_{i \not\rightarrow i} &\coloneqq \sum_{\vt \in \weightqn} Q'_i(\vt)\sum_{\vw \in \weightqn} \Pqww(\vw|\vt) P_i(0|\vw)\\
        &=\sum_{\vt \in \weightqn} \sum_{\vx \in \compositionclass{\vt}} Q_i(\vx) \sum_{\vw \in \weightqn} \Pqww(\vw|\vt) \frac{|\mathcal{D}^c_i \cap \compositionclass{\vw}|}{|\compositionclass{\vw}|}\\
        &= \lambda_{i \not\rightarrow i}. \hspace*{10mm} \text{(by \eqref{eq: error i not i})}
    \end{align*}
    This proves the the first part of the lemma. From \eqref{eq:det_enc}, it follows that if $Q_i$ is a point mass, then $Q_i'$ is also a point mass. Hence the second part also follows. This completes the proof of the lemma.

 \label{app:general_to_type}

\section{Proof of Lemma~\ref{lemma:upper bound mode multinomial1}}\label{app:upper_bound_multinomial}
    We first consider the case when $\vu$ has strictly positive entries, i.e., $u_i>0$ for $i=1,2,\cdots, q$.
    For any $(t_1, \ldots, t_q) \in \weightqn$ and $i \in [1:q]$, let us denote $\alpha_i= \frac{t_i}{n}$. Clearly
    $$
    \sum_{i=1}^q \alpha_i =1.
    $$
    We can write 
    \begin{align}
        \multinomial{n}{\vu}(t_1, \ldots, t_q)=\multinomial{n}{\vu}(\alpha_1 n, \ldots, \alpha_q n)
        &=\binom{n}{\alpha_1 n \cdots \alpha_q n} \prod_{i=1}^q u_i^{ \alpha_i n}\\
        &=\frac{n! \prod_{i=1}^q u_i^{ \alpha_i n}}{\prod_{i =1}^q(\alpha_i n)!}   \\
        &=\frac{n!\prod_{i\in \supp{\vt}}u_i^{\alpha_i n}}{\prod_{i\in \supp{\vt}}(\alpha_i n)!} . 
    \end{align}
    \textbf{Case 1 ($|\supp{\vt}| >1$:)}
    Taking $\log$, we get 
    \begin{align}
        \log \multinomial{n}{\vu}(t_1, \ldots, t_q)&= \log n! - \sum_{i \in \supp{\vt}} \log (\alpha_i n)! + \sum_{i \in \supp{\vt}} \alpha_i n \log u_i \notag \\
        &\stackrel{(a)}{\le} \left(n \log n -n \log e +\frac{1}{2} \log n+\frac{1}{2}\log (2 \pi) +\frac{1}{12n}\log e \right) \notag\\
        &\hspace{5mm}- \sum_{i \in \supp{\vt}} \left((\alpha_i n) \log (\alpha_i n)- (\alpha_i n) \log e + \frac{1}{2} \log (\alpha_i n) +\frac{1}{2}\log (2 \pi) \right) +
        \sum_{i \in \supp{\vt}} \alpha_i n \log u_i \notag\\
        &\stackrel{}{=} n \log n -n \log e +\frac{1}{2} \log n -\sum_{i \in \supp{\vt}} \left( (\alpha_i n) \log (\alpha_i n)-\alpha_i n \log e+ \frac{1}{2} \log (\alpha_i n) -\alpha_i n \log u_i\right) \notag \\
        &\hspace{5mm}- \left(  \frac{|\supp{\vt}|-1}{2}\log(2\pi)-\frac{1}{12n}\log e \right) \notag\\
        &\stackrel{(b)}{\le } \left(n \log n -\sum_{i \in \supp{\vt}}(\alpha_i n) \log (\alpha_i n)+n\sum_{i \in \supp{\vt}}\alpha_i \log u_i\right)+\log e\left(\sum_{i \in \supp{\vt}} \alpha_i n - n \right) \notag\\
        &\hspace{5mm}+ \frac{1}{2} \left( \log n-\sum_{i \in \supp{\vt}} \log (\alpha_i n) \right)  \notag 
    \end{align}
Here $(a)$ uses Stirling's upper bound and $(b)$ follows because $\frac{|\supp{\vt}|-1}{2}\log (2 \pi) - \frac{1}{12n}\log e  \ge 0$ for large enough $n$.
By considering the first term, we get
\begin{align*}
    n \log n -\sum_{i \in \supp{\vt}}(\alpha_i n) \log (\alpha_i n)+n\sum_{i \in \supp{\vt}}\alpha_i \log u_i
    &= n \log n - n  \left(\sum_{i\in \supp{\vt} }\alpha_i\right)\log n - n \sum_{i \in \supp{\vt}} \alpha_i \log \alpha_i\\
    & \quad +n\sum_{i \in \supp{\vt}}\alpha_i \log u_i\\
    &=- n\sum_{i \in \supp{\vt}} ( \alpha_i \log \alpha_i-\alpha_i \log u_i )\\
    &\stackrel{(a)}{=}- n\sum_{i \in [1:q]} ( \alpha_i \log \alpha_i-\alpha_i \log u_i )\\
    &=-n \kl (\vec{\alpha} || \vu).
\end{align*}
Here (a) follows by considering $0 \log 0  \coloneqq 0$ and
$\vec{\alpha}=(\alpha_1, \cdots, \alpha_q)$.
 Clearly, the second summand 
 \begin{align}
     \log e\left(\sum_{i \in \supp{\vt}} \alpha_i n - n \right)=0.
 \end{align}
The third term can be written as
\begin{align}
\frac{1}{2} \left( \log n-\sum_{i \in \supp{\vt}} \log (\alpha_i n) \right)
    & = \frac{1}{2} \log \frac{n}{n^{|\supp{\vt}|} \times \prod_{i \in \supp{\vt}}  \alpha_i}\\
    &=\frac{1}{2} \log \frac{n}{  \prod_{i \in \supp{\vt}} t_i }\\
    &= \log \left(\sqrt{n} \cdot\prod_{i \in \supp{\vt}} t_i\right)^{-\frac{1}{2}}.
\end{align}

Therefore we can write 
\begin{align}
    \log \multinomial{n}{\vu}(t_1 , \ldots, t_q ) \le -n \kl\left(\frac{\vt}{n}|| \vu\right) + \log \left(\sqrt{n} \cdot\prod_{i \in \supp{\vt}} t_i\right)^{-\frac{1}{2}}.
\end{align}
\textbf{Case 2 ($|\supp{\vt}|=1$):}
Suppose $\supp{\vt}=\{j\}$. Then  $t_j=n$, and  $t_i=0$ for all $i \neq j$. Hence
\begin{align}    
\kl\left(\frac{\vt}{n}|| \vu\right) &= -\log u_j\\
\text{and } \frac{\prod_{i \in\supp{\vt}} t_i}{n}&=1.
\end{align}
Hence
\begin{align}
  2^{-n \kl\left(\frac{\vt}{n}|| \vu\right)} \times \sqrt{n} \cdot  \left(\prod_{i \in\supp{\vt}} t_i \right) ^{-\frac{1}{2}}=u_j^{n}.
\end{align}
Also, clearly,
\begin{align}
\multinomial{n}{\vu}(0,\cdots,0,t_j,0,\cdots,0)&=u_j^{n}
\end{align}
This proves the lemma for the case when $\vu$ has strictly positive entries.

Now let us consider $\vu$ with some zero-valued components. Wlog, suppose $u_1,u_2,\cdots,u_k> 0$ and $u_{k+1},\cdots,u_q=0$. The resulting distribution is a multinomial distribution on $k$ categories, with the success probability distribution $\vu_{[1:k]}=(u_1,u_2,\cdots,u_k)$ having strictly positive entries. We consider two cases:

\textbf{Case I} ($t_j=0$, $\forall j \in [k+1:q]$) :
For any $(t_1, \ldots, t_q) \in \weightqn$ such that $t_j =0 $ for all $j \in [k+1:q]$, we have $\supp{\vt} \subseteq [1:k]$. Therefore, using the result for striclty positive success probabilities, we can write 
    \begin{align}
    \multinomial{n}{\vu}(\vt) &
        =\multinomial{n}{\vu_{[1:k]}}(\vt_{[1:k]})\\
        &\stackrel{(a)}{\le} 2^{-n \kl\left(\frac{\vt_{[1:k]}}{n}|| \vu_{[1:k]}\right)} \times   \sqrt{n} \cdot  \left(\prod_{i \in\supp{\vt}} t_i \right) ^{-\frac{1}{2}}\\
        &\stackrel{(b)}{=} 2^{-n \kl\left(\frac{\vt}{n}|| \vu\right)} \times   \sqrt{n} \cdot  \left(\prod_{i \in\supp{\vt}} t_i \right) ^{-\frac{1}{2}}.
    \end{align}

Here $(a)$ follows from the result (already proved) for $\vu$ with strictly positive entries, and $(b)$ is true because in the definition of KL divergence, it is assumed that $0\log \frac{0}{0} \coloneqq 0$.

\textbf{Case II} ($t_j \neq 0$ for some $j \in [k+1:q]$): Since for some $j$, $t_j\neq 0$ and $u_j=0$, we have $\multinomial{n}{\vu}(t_1, \ldots, t_q)=0$. Also, in that case, $\kl\left(\frac{\vt}{n}|| \vu\right)=\infty$. Hence the result follows.

This completes the proof of the lemma.

\section{Proof of Corollary~\ref{lemma:upper bound mode multinomial}}\label{app:mode_multinomial}
For $\vt \in \weightqn$ by Lemma~\ref{lemma:upper bound mode multinomial1}, we have
\begin{align}
    \multinomial{n}{\vu}(\vt ) \le 2^{-n \kl\left(\frac{\vt}{n}|| \vu\right)} \times   \sqrt{n} \cdot  \left(\prod_{i \in\supp{\vt}} t_i \right) ^{-\frac{1}{2}},
\end{align}
where $\supp{\vt}=\{i \in [1:q]| t_i \neq 0\}$. Wlog, let us assume that $\supp{\vu}=[1:k]$.  Consider the following subset
\begin{align}
    \cN_{central} \coloneqq  \{ \vt \in \weightqn| |t_i -nu_i| \le n^{3/4}, \forall i \in [1:k], t_j =0 ~\forall j \in [k+1:q]\}.
\end{align}

\textbf{Case I ($\vt \in \cN_{central}$):}
For large enough $n$, for every $\vt \in \cN_{central}$ we have $\supp{\vt}=[1:k]$ and $t_i \ge nu_i -n^{3/4}$ for all $i \in\supp{\vt}$. Therefore 
\begin{align}
    \frac{\prod_{i \in \supp{\vt}} t_i}{n} &\ge   n^{-1} \prod_{i \in [1:k]}nu_i  \left(1-\frac{1}{n^{1/4}u_i}\right)\\
    &\stackrel{(a)}{\ge} n^{k-1} \prod_{i \in [1:k]} \left( \frac{u_i}{4^{1/k}} \right)\\ 
    &\stackrel{}{=} n^{k-1} \frac{\prod_{i \in [1:k]} u_i}{4}.
\end{align}
Here  $(a)$ follows because $ \left(1-\frac{1}{n^{1/4}u_i}\right)>\frac{1}{4^{1/k}}$ for large enough $n$.  As $\kl(\frac{\vt}{n}||\vu) \ge 0$, we have
 \begin{align}
    \multinomial{n}{\vu}(\vt) \le \frac{2}{n^{(k-1)/2} \sqrt{\prod_{i \in [1:k]}} u_i}, \hspace{8mm} \forall \vt \in \cN_{central}.
 \end{align}

 \textbf{Case II ($\vt \not\in \cN_{central}$):}
We consider the following two sub-cases:

{\it Case IIA:}
Consider a $\vt \not\in \cN_{central}$, such that there exists $i \in\supp{\vt}$ with  $|t_i- nu_i|> n^{3/4}$ and $t_j=0$ for all $j \in [k+1:q]$.
 We now use the Pinsker's inequality \cite[Lemma 17.3.2]{cover1999elements} , i.e.,
\begin{align}\label{eq:pinsker_inq}
    \kl\left(\vu|| \vv\right) \ge \frac{2}{\ln 2} (\tv (\vu,\vv))^2,
\end{align}
for any two distributions $\vu,\vv$ over the same alphabet.

Note that
\begin{align}
    \tv\left(\frac{\vt}{n}, \vu\right) \ge \frac{1}{2} \left|\frac{t_i}{n}-u_i\right| \ge \frac{1}{2n^{1/4}}.
\end{align}
Hence
 \begin{align}
     \kl\left(\frac{\vt}{n}|| \vu\right) \ge \frac{1}{2 \ln 2\sqrt{n}}.
 \end{align}
 Therefore
 \begin{align}
     \multinomial{n}{\vu}(\vt) &\le 2^{-\sqrt{n}/(2 \ln 2)} \cdot \sqrt{n} \cdot  \left(\prod_{i \in\supp{\vt}} t_i \right) ^{-\frac{1}{2}}\\
     &\stackrel{(a)}{\le} \sqrt{n} 2^{-\sqrt{n}/(2 \ln 2)}\\
     &\stackrel{}{\le}  \frac{2}{n^{(k-1)/2} \sqrt{\prod_{i \in [1:k]} u_i}} \hspace*{8mm}  \text{ (for large enough  } n \text{)}.
 \end{align}
Here $(a)$ follows from the fact that $t_i \ge 1$ for all $i \in\supp{\vt}$.

{\it Case IIB:} Consider $\vt \not \in \cN_{central}$ such that $t_j \neq 0$ for some $j \in [k+1:q]$. Then we have $ \multinomial{n}{\vu}(t_1, \ldots, t_q)=0$, which is less than the RHS of \eqref{eq:multinomial_peak}. 
This proves the corollary.

\section{Proof of Corollary~\ref{cor:tailbound}}\label{app:tailbound}
For $\supp{\vt}=\{i \in [1:q]| t_i \neq 0\}$ let $i \in\supp{\vt}$ such that $|t_i - (n+1)u_i|\geq K\sqrt{n\log n}$. Wlog, let us assume that $\supp{\vu}=[1:k]$. If $i \in [1:k]$, then
by Lemma \ref{lemma:upper bound mode multinomial1}, we have
\begin{align}
    \multinomial{n}{\vu}(\vt) &\le 2^{-n \kl(\frac{\vt}{n}|| \vu)} \times   \left( \frac{\prod_{j \in\supp{\vt}} t_j}{n}\right) ^{-\frac{1}{2}}\\
    &\stackrel{(a)}{\le} \sqrt{n}2^{-n \kl\left(\frac{\vt}{n}|| \vu\right)}\\
    &\stackrel{(b)}{\le} \sqrt{n} 2^{-\frac{2n}{\ln 2}\left(\tv \left(\frac{\vt}{n}, \vu\right)\right)^2}\\
    &\le \sqrt{n}  2^{-\frac{n}{2\ln 2}  \left(\frac{t_i-nu_i}{n}\right)^2}\\
    &\stackrel{(c)}{\le} \sqrt{n} 2^{-\frac{n}{2}\left(\frac{t_i-(n+1)u_i}{n}\right)^2}\\
    &\stackrel{(d)}{\le} \sqrt{n}  2^{-\frac{n}{2} \cdot \left(K\sqrt{\frac{ \log n}{n}}\right)^2}\\
    &=\sqrt{n}2^{-\frac{K^2 \log n}{2}}\\
    &=n^{-\frac{1}{2}(K^2-1)}.
\end{align}
Here (a) follows because $t_i \ge 1$ for $i \in\supp{\vt}$, $(b)$ follows from Pinsker's inequality in \eqref{eq:pinsker_inq}, $(c)$ follows because 
$$
\left( \frac{t_i -nu_i}{n} \right)^2 \ge \frac{1}{\log e} \cdot \left(\frac{t_i - (n+1)u_i}{n} \right)^2,
$$
and $(d)$ follows because $|t_i - (n+1)u_i|\geq K\sqrt{n\log n}$.
Otherwise, i.e., if $i \in [k+1:q]$, we have
\begin{align}
    \multinomial{n}{\vu}(t_1, \cdots, t_k, \cdots, t_q)=0 \le  n^{-\frac{1}{2}(K^2-1)}.
\end{align}
This proves the corollary.

\section{Proof of Lemma~\ref{lemma:tvatom}}\label{app:atom}
 Clearly, if $\Ssum=0$, then $\vu=\vv$ and $\atomno=0$. The statement of the lemma is then trivially true. 
 
As $\vu$ and $\vv$ are distributions, for $\vu\neq \vv$, $\splus{\vu}{\vv}$ and $\sminus{\vu}{\vv}$ are both nonempty, hence 
$\Ssum \ge 2$. We shall prove the lemma by induction on the parameter $\Ssum$ for $\vu\neq \vv$.
First, for $\Ssum=2$, suppose $\splus{\vu}{\vv}=\{c\}$, $\sminus{\vu}{\vv}=\{c'\}$. Hence the statement is true by defining $\atomno=1$, $(c_1,c_1')=(c,c')$, $\alpha_1= u_{c}-v_{c}=v_{c'}-u_{c'}$.

Suppose for all distributions  $\vu$ and $\vv$ with $
\Ssum < s$, the statement is true.

Now consider the case $\Ssum=s>2$. Since the sets $\splus{\vu}{\vv}$ and  $\sminus{\vu}{\vv}$ are nonempty, we arbitrarily choose $c_1 \in \splus{\vu}{\vv}$ and $c_1' \in \sminus{\vu}{\vv}$, and define 
\begin{align}
    \tvatom_1=\min \{ |u_{c_1}-v_{c_1}|,|u_{c_1'}-v_{c_1'}|\}.
\end{align}
We define a new distribution
\begin{align}
    \vu'=\vu +\tvatom_1\delta_{\ve_{c_1} \rightarrow \ve_{c'_1}}.\label{eq:va =ua+weight}
\end{align}
Note that, if 
\begin{align}
    \tvatom_1= |u_{c_1}-v_{c_1}| \label{eq:atom1}
\end{align}
, we have
\begin{align}
    \splus{\vu'}{\vv} &=\splus{\vu}{\vv} \setminus \{c_1\}\\
    \sminus{\vu'}{\vv} &\subseteq \sminus{\vu}{\vv},
\end{align}
otherwise, i.e. if 
\begin{align}
    \tvatom_1= |u_{c'_1}-v_{c'_1}| \label{eq:atom2}
\end{align}
we have
\begin{align}
    \splus{\vu'}{\vv} & \subseteq \splus{\vu}{\vv}\\
    \sminus{\vu'}{\vv} &= \sminus{\vu}{\vv} \setminus \{c'_1\}.
\end{align}
Hence we can write
\begin{align}
    |\splus{\vu'}{\vv}|+|\sminus{\vu'}{\vv}| < |\splus{\vu}{\vv}|+ |\sminus{\vu}{\vv}|=s.\label{eq:Sv<Su}
\end{align}
By induction hypothesis, there exists $\atomnoi$ pairs 
$$(c_2,c'_2),(c_3,c'_3),\cdots,(c_{\atomnoi+1},c'_{\atomnoi+1})\in \splus{\vu'}{\vv}\times \sminus{\vu'}{\vv}
$$ 
and constants $ \tvatom_2, \tvatom_3, \cdots, \tvatom_{\atomnoi+1}>0$
such that
\begin{align}
    \atomnoi &\le |\splus{\vu'}{\vv}|+|\sminus{\vu'}{\vv}|-1 \le s-2 \label{eq:nu'} \\
    \vv&=\vu'+ \sum_{i=2}^{\atomnoi+1}\tvatom_i \delta_{\ve_{c_i} \rightarrow \ve_{c'_i}}.\label{eq:ub=va+weights}
\end{align}
Combining  \eqref{eq:va =ua+weight} and \eqref{eq:ub=va+weights}, we have
\begin{align}
    \vv&=\vu+\tvatom_1\delta_{\ve_{c_1} \rightarrow \ve_{c'_1}}+ \sum_{i=2}^{\atomnoi+1}\tvatom_i \delta_{\ve_{c_i} \rightarrow \ve_{c'_i}}.\label{eq:ub=ua+weights_new}
\end{align}
By defining $\atomno \coloneqq \atomnoi+1$, we have
\begin{align}
    \atomno&= \atomnoi+1 \le s-2\\ 
    \vv&=\vu+  \sum_{i=1}^{\atomno}\tvatom_i \delta_{\ve_{c_i} \rightarrow \ve_{c_i'}}, \quad \text{(by \eqref{eq:ub=ua+weights_new})}.
\end{align}
This proves the lemma.
The second part follows from the choice of $\alpha_1$ in \eqref{eq:atom1} or \eqref{eq:atom2}.

\section{Analysis of approximation scheme for $q=2$}\label{app:proof_proposition_algo}

We shall first present a corollary of Lemma~\ref{lemma:weight_transfer_pairs}, by specializing it for $q=2$. 
\begin{corollary}\label{cor: consecutive shift of mass}
    Consider the channel $2\textnormal{-NC}_{n,U}$ and two input distributions (of $T$) $\inpdist$ and $\inpdist'$. Let $\opdist$ and $\opdist'$ denote the corresponding output distributions, i.e., $\forall w \in [1:n]$,
    \begin{align}
        \opdist(w)  =\sum_{t \in [0:n]} \inpdist(t) \Pbww(w|t), \hspace{5mm}
        \text{and} \hspace{5mm} \opdist'(w) = \sum_{t \in [0:n]} \inpdist'(t) \Pbww(w|t).
    \end{align}
 For $\epsilon>0$, $k \in [0:n]$, and $\lambda \in [1:n-k]$,
   \begin{enumerate}[label=\roman*).]
        \item If $\inpdist(k)\ge \epsilon$, and  $\inpdist' =
        \inpdist + \epsilon \delta_{k\rightarrow k+\lambda}$
then
\begin{align}
    \tv(\opdist,\opdist') 
       &\le \frac{C_4\epsilon \lambda}{\sqrt{n}}.
\end{align}
        \item If $\inpdist(k+\lambda)\ge \epsilon$, and $\inpdist' =
        \inpdist - \epsilon \delta_{k\rightarrow k+\lambda}$
then
\begin{align}
    \tv(\opdist,\opdist') 
       &\le \frac{C_4\epsilon \lambda}{\sqrt{n}}.
\end{align}
   \end{enumerate}
\end{corollary}

We now show that the output of Algorithm \ref{algo:quan_binary} is an $M$-type distribution.

\begin{lemma}\label{lemma:binary_Mtype_algo}
     For any positive integers $n,M \ge 2$ and for any hamming weight distribution $\inpdist \in \cP([0:n])$,  the quantized distribution $\inpapprox \in \cP([0:n])$ given as output by {\it Algorithm \ref{algo:quan_binary}} is an $M$-type distribution.
\end{lemma}

\begin{IEEEproof} 
First we show that $\hat{Q}_{\text{ip}}$ is a distribution. By definition (Step 1a and Step 2), $\hat{Q}_{\text{ip}}(i)\geq 0$ for each $i\in [0:n]$. From the second line in Step 1a, we have, for each $j$,
\begin{align}
    \inpapprox(j)+\tilde{Q}_{\text{ip}}(j+1)=\inpdist (j+1) + \tilde{Q}_{\text{ip}}(j).
\end{align}
Summing both sides over $j=0,1,\cdots,i$, we have
\begin{align}
    \sum_{j=0}^{i}\inpapprox(j) +  \tilde{Q}_{\text{ip}}(i+1)&=\sum_{j=0}^{i}\inpdist (j+1) + \tilde{Q}_{\text{ip}}(0)\\
    &=\sum_{j=0}^{i+1} \inpdist (j) \hspace*{5mm}\text{(by initialization)} \label{eq:step1a_algo_binary}
\end{align}
Considering $i=n-1$ and noting that $\inpapprox (n)=\tilde{Q}_{\text{ip}}(n)$ (by Step 2) and $\tilde{Q}_{\text{ip}}(0):= \inpdist (0)$ (By initialization), we have
\begin{align}
    \sum_{j=0}^{n}\inpapprox(j) &=\sum_{j=0}^{n}\inpdist (j)\\
    &=1.
\end{align}
Hence $\hat{Q}_{\text{ip}}$ is a distribution.
 
The distribution $\hat{Q}_{\text{ip}}$  is $M$-type. To see this, note that in the $i$-th iteration, $\hat{Q}_{\text{ip}}$ is assigned a value that is a multiple of $1/M$, and this value is not changed in the subsequent steps. After the last ($i=n-1$) iteration, $\hat{Q}_{\text{ip}}(j);j=0,1,\cdots,n-1$ are assigned values that are multiples of $1/M$. 
Since $\inpapprox$ is a distribution, the last value $\inpapprox(n)$ must also be a multiple of $1/M$. This completes the proof.
\end{IEEEproof}

Recall that, at the end of iteration $i$, we have a distribution
 $\inpdist^{(i)}$ as given below, where the values up to $i$ are multiples of $1/M$, the value at $i+1$ is modified by adding any extra mass resulting from the quantization of $\tilde{Q}_{\text{ip}}(i)$, and the values $\inpdist (j)$ for $j>i+1$. That is,
\begin{align}
    \inpdist^{(i)}(j):=\begin{cases}
       \inpapprox (j) & j\leq i\\
       \tilde{Q}_{\text{ip}}(i+1) & j=i+1\\
       \inpdist (j) & j>i+1
    \end{cases} .
\end{align}
It follows by \eqref{eq:step1a_algo_binary} that
$\sum_j \inpdist^{(i)}(j) =\sum_j \inpdist(j) =1$, and hence
$\inpdist^{(i)}$ is a distribution for each $i$. We denote the output distribution corresponding to the input distribution $\inpdist^{(i)}$ as $\outdist^{(i)}$. We also denote the output distribution corresponding to $\inpdist$ and $\inpapprox$ as $\outdist$ and $\outapprox$, respectively.

\begin{proposition}\label{prop: approx op dist binary}
    For any positive integer $n \ge 2$ and any real number $0<p<0.5$, consider the noisy binary composition channel. For any distribution $\inpdist \in \cP([0:n])$ of the input $T$,  the quantized $M$-type  distribution $\inpapprox \in \cP([0:n])$ output by {\it Algorithm \ref{algo:quan_binary}} is such that  the variational distance between the corresponding output distributions (of $W$) 
    \begin{align}
        \outdist(w) & =\sum_t \inpdist(t) \Pbww(w|t)\\
        \text{and } \outapprox(w) &= \sum_t \inpapprox(t) \Pbww(w|t)
    \end{align}
    is upper bounded as
    \begin{align}
       \tv (\outdist,\outapprox) \le \frac{C_4 \sqrt{n}}{M}.
    \end{align}
\end{proposition}

{\it Proof of Proposition~\ref{prop: approx op dist binary}:}
By Lemma~\ref{lemma:binary_Mtype_algo}, the distribution $\inpapprox \in \cP([0:n])$ given as output by {\it Algorithm A} is $M$-type.
    Note that 
\begin{align}
    \inpdist^{(i-1)}(j) - \inpdist^{(i)}(j) & = \begin{cases}
       0 & j\leq i \text{ and } j\geq i+2\\
       (\tilde{Q}_{\text{ip}}(i) - \inpapprox(i)) & j=i\\
       -(\tilde{Q}_{\text{ip}}(i) - \inpapprox(i)) & j=i+1
    \end{cases} . 
\end{align}
Since $\tilde{Q}_{\text{ip}}(i) - \inpapprox(i)=\tilde{Q}_{\text{ip}}(i)-\lfloor\tilde{Q}_{\text{ip}}(i)\rfloor_{1/M}\leq 1/M$, we have
\begin{align}
    \outdist^{(i-1)}=\outdist^{(i)}+(\tilde{Q}_{\text{ip}}(i)-\lfloor\tilde{Q}_{\text{ip}}(i)\rfloor_{1/M}) \cdot 
    \delta_{i \rightarrow i+1},
\end{align}
and $\dc(i,i+1)=1$. Hence by Corollary~\ref{cor: consecutive shift of mass} with $\lambda=1$, we have

\begin{align*}
    \tv (\outdist^{(i-1)} \outdist^{(i)}) \leq \frac{C_4}{M \sqrt{n}}.
\end{align*}
Hence, using the triangular inequality, we have
\begin{align*}
    \tv (\outdist, \outapprox) & =\tv(\outdist^{(0)},\outdist^{(n-1)})\\
    & \leq \sum_{i=0}^{n-1} \tv (\outdist^{(i-1)}, \outdist^{(i)})\\
    &\leq \frac{n C_4 }{M \sqrt{n}}\\
    &=\frac{C_4 \sqrt{n}}{M}.
\end{align*}
This proves the proposition.
\hfill{$\blacksquare$}

\section{Proof of Lemma~\ref{lemma:adjacent_gray}}\label{app:adjacent}

Recall that, for all $j \in [1:q-1]$, 
\begin{align}
    t_j \in [a(l_j-1):l_ja],\\
    t'_j \in [a(l'_j-1):l'_ja].
\end{align}
Then we have 
\begin{align}
    \sum_{j=1}^{q-1} \left|t_j-t'_j\right| &\le \sum_{j=1}^{q-1} \left|l_ja- a(l'_j-1)\right|\\
    &=a(q-1)+a\sum_{j=1}^{q-1} \left| l_j-l'_j  \right|\\
    &\le a(q-1) + a \cdot 2 \quad \text{(by \eqref{eq:ad2})}\\
    &=a(q+1).\label{eq:ad_1}
\end{align}
We also have
\begin{align}
    \left| \sum_{j=1}^{q-1} (t_j-t'_j)\right| & \le \left| \sum_{j=1}^{q-1} (al_j-a(l'_j-1))\right|\\
    &=a(q-1)+a \left| \sum_{j=1}^{q-1} (l_j-l'_j)\right|\\
    &\le a(q-1)+a \cdot 1  \quad \text{(by \eqref{eq:ad1})}\\
    &=aq.\label{eq:ad_2}
\end{align}
Now 
\begin{align}
    2\dc(\vt_{\vu}, \vt_{\vu'})&=\sum_{j=1}^q \left| 
t_j-t'_j\right|\\
    &=\sum_{j=1}^{q-1} \left| 
t_j-t'_j\right|+\left| 
t_q-t'_q\right|\\
&=\sum_{j=1}^{q-1} \left| 
t_j-t'_j\right|+ \left| \left( n- \sum_{j=1}^{q-1} t_j\right) -  \left( n- \sum_{j=1}^{q-1} t'_j\right) \right|\\
&=\sum_{j=1}^{q-1} \left| 
t_j-t'_j\right|+ \left| 
\sum_{j=1}^{q-1}(t_j-t'_j)\right|\\
&\le a(q+1)+aq \quad \text{(by \eqref{eq:ad_1} and \eqref{eq:ad_2})}\\
&=a(2q+1).
\end{align}
This proves the lemma.
\qed

\section{Proof of Lemma~\ref{lemma:min_distance}}
By setting $\eta= \frac{2n}{|\cA|^{1/(q-1)}} $ and 
$ \zeta = \left\lceil \frac{n+1}{\eta}\right\rceil
$, we divide each interval $[0,n]$ for coordinate $t_i$ ($i=1,\dots,q-1$) into $\zeta$ intervals of length $\eta$. That is, for each $\vk=(k_1, \cdots,k_{q-1}) \in [1:\zeta]^{q-1}$, we define the cell
\begin{align}
    C_{\vk} :=\left\{(t_1,\dots,t_{q-1}) \in [0,n]^{q-1}|
(k_i-1)\eta  \le t_i < k_i \eta ,
\quad \forall i \in [1:q]
\right\}.
\end{align}
There are $\zeta^{q-1}$ cells.
Note that
\begin{align*}
    \zeta &= \left\lceil \frac{n+1}{\eta}\right\rceil\\
&=\left\lceil \frac{n+1}{2n}|\cA|^{\frac{1}{q-1}}
\right\rceil\\
&< \frac{n+1}{2n}|\cA|^{\frac{1}{q-1}}+1 \\
& =\frac{|\cA|^{\frac{1}{q-1}}}{2}\left(1 +\frac{1}{n}\right) +1\\
&=\frac{|\cA|^{\frac{1}{q-1}}}{2} \left(1+\frac{1}{n}+\frac{2}{|\cA|^{\frac{1}{q-1}}} 
\right)\\
&\stackrel{(a)}{\le} |\cA|^{\frac{1}{q-1}}.
\end{align*}
Here $(a)$ follows because $|\cA| \ge 3^{q-1}$, i.e.
\begin{align*}
    |\cA|^{\frac{1}{q-1}} & \ge 3\\
    &\ge \frac{2n}{n-1} &\text{(for $n \ge 3$)},
\end{align*}
and hence
\begin{align*}
    \frac{2}{|\cA|^{\frac{1}{q-1}}}\le 1-\frac{1}{n}.
\end{align*}
This gives $\zeta^{q-1} < |\cA|$, and hence there are at most $|\cA|$
cells.  By the pigeon-hole principle, there must be two distinct compositions $\vt, \vt' \in \cA$ that are
in the same cell. For such two compositions $\vt,\vt'$, for each $i=1,\dots,q-1$, we have $|t_i - t_i'| \le \eta$. Hence
\begin{align}
\sum_{i=1}^{q-1} |t_i - t_i'| &\le (q-1) \eta \\
&=(q-1)\frac{2n}{|\cA|^{1/(q-1)}} 
\end{align}

and

\begin{align}
t_q - t_q' = \left(n-\sum_{i=1}^{q-1}t_i\right) - \left(n-\sum_{i=1}^{q-1}t_i'\right)
&=-\sum_{i=1}^{q-1}(t_i-t_i').
\end{align}
Hence
\begin{align}
\min_{\substack{\vt,\vt' \in \cA \\ \vt \neq \vt'}}\dc(\vt,\vt') \le \dc(\vt,\vt')
&=\frac{1}{2}\sum_{i=1}^q |t_i - t_i'|\\
&=\frac{1}{2}\sum_{i=1}^{q-1} |t_i - t_i'|+\frac{1}{2}|t_q - t_q'|\\
&\le\frac{1}{2}(q-1) \frac{2n}{|{\cA}|^{1/(q-1)}}+ \frac{1}{2}(q-1) \frac{2n}{|{\cA}|^{1/(q-1)}}\\
&=(q-1) \frac{2n}{|{\cA}|^{1/(q-1)}}.
\end{align}
This proves the lemma.

\label{app:min_distance}

\end{document}